\documentclass[preprint2]{aastex}

\usepackage{aastexug}    
\usepackage{amssymb}     
\usepackage{epsfig}
 
\newfont{\sfsl}{cmssqi8 scaled 1200}	
\newfont{\sfsls}{cmssqi8 scaled 900}	
\newfont{\sfslms}{cmssqi8 scaled 1000}	
\newcommand{\gcs}{{\sfsl HIFLUGCS}}
\newcommand{\gcss}{{\sfslms HIFLUGCS}}

\newcommand{\ro}{{\rm ROSAT}}
\newcommand{\ra}{{\rm RASS}}
\newcommand{\ps}{{\rm PSPC}}
\newcommand{\as}{{\rm ASCA}}
\newcommand{\xs}{{\rm XSPEC}}
\newcommand{\mpc}{h_{50}^{-1}\,{\rm Mpc}}
\newcommand{\kpc}{h_{50}^{-1}\,{\rm kpc}}
\newcommand{\sx}{S_{\rm X}}
\newcommand{\fx}{f_{\rm X}}
\newcommand{\fxl}{f_{\rm X,lim}}
\newcommand{\cx}{C_{\rm X}}
\newcommand{\lx}{L_{\rm X}}
\newcommand{\lbol}{L_{\rm Bol}}

\newcommand{\ncl}{N_{\rm Cl}}
\newcommand{\rx}{r_{\rm X}}
\newcommand{\rc}{r_{\rm c}}
\newcommand{\rs}{r_{\rm s}}
\newcommand{\rch}{R_{\rm ch}}
\newcommand{\rab}{r_{\rm A}}
\newcommand{\mab}{M_{\rm A}}
\newcommand{\msu}{h_{50}^{-1}\,M_{\odot}}
\newcommand{\mpr}{m_{\rm p}}
\newcommand{\om}{\Omega_{\rm m}}
\newcommand{\ok}{\Omega_{\rm k}}
\newcommand{\ol}{\Omega_{\Lambda}}
\newcommand{\ob}{\Omega_{\rm b}}
\newcommand{\oc}{\Omega_{\rm Cluster}}
\newcommand{\og}{\Omega_{\rm Group}}
\newcommand{\ocdm}{\Omega_{\rm CDM}}
\newcommand{\fg}{f_{\rm gas}}
\newcommand{\rog}{\rho_{\rm gas}}
\newcommand{\rot}{\rho_{\rm tot}}
\newcommand{\roc}{\rho_{\rm c}}
\newcommand{\rov}{\rho_{\rm vir}}
\newcommand{\nh}{n_{\rm H}}
\newcommand{\tg}{T_{\rm gas}}
\newcommand{\tx}{T_{\rm X}}
\newcommand{\mt}{M_{\rm tot}}
\newcommand{\mti}{M_{\rm tot,min}}
\newcommand{\mtz}{M_{200}}
\newcommand{\mtzi}{M_{200,i}}
\newcommand{\mtf}{M_{500}}
\newcommand{\mg}{M_{\rm gas}}
\newcommand{\kev}{\rm keV}
\newcommand{\eh}{0.5-2.0\,{\rm keV}}
\newcommand{\eb}{0.1-2.4\,{\rm keV}}
\newcommand{\ek}{0.64-2.36\,{\rm keV}}
\newcommand{\ebol}{0.01-40\,{\rm keV}}
\newcommand{\cts}{{\rm cts\,s^{-1}}}
\newcommand{\esc}{\times10^{-11}\,{\rm ergs\,s^{-1}\,cm^{-2}}}
\newcommand{\escl}{\times10^{-12}\,{\rm ergs\,s^{-1}\,cm^{-2}}}
\newcommand{\esct}{10^{-11}\,{\rm ergs\,s^{-1}\,cm^{-2}}}
\newcommand{\escc}{\times10^{-10}\,{\rm ergs\,s^{-1}\,cm^{-2}}}
\newcommand{\esl}{ h_{50}^{-2}\, 10^{40}\,{\rm ergs\,s^{-1}}}
\newcommand{\esll}{ h_{50}^{-2}\, 10^{44}\,{\rm ergs\,s^{-1}}}
\newcommand{\esls}{h_{50}^{-2}\, 10^{43}\,{\rm ergs\,s^{-1}}}
\newcommand{\vmax}{V_{\rm max}}
\newcommand{\zmax}{z_{\rm max}}
\newcommand{\vmaxi}{V_{{\rm max},i}}
\newcommand{\dc}{\delta_{\rm c}}
\newcommand{\dcv}{\delta_{\rm c}^{\rm v}}
\newcommand{\bii}{b_{\rm II}}
\newcommand{\lii}{l_{\rm II}}
\newcommand{\sll}{\sigma_{\log\lx}}
\newcommand{\slm}{\sigma_{\log\mt}}

\shorttitle{The Mass Function of Galaxy Clusters}

\begin{document} 
 
\title{The Mass Function of an X-Ray Flux-Limited Sample
of Galaxy Clusters} 

\author{\anchor{http://www.xray.mpe.mpg.de/~reiprich/}{Thomas H.
Reiprich\altaffilmark{1}} and Hans B\"ohringer}
\affil{\anchor{http://www.mpe.mpg.de/}{Max-Planck-Institut f\"ur
extraterrestrische Physik}}
\affil{P.O. Box 1312, 85741 Garching, Germany}
\email{reiprich@mpe.mpg.de, hxb@mpe.mpg.de}
\altaffiltext{1}{Present address:
Department of Astronomy,
University of Virginia,
PO Box 3818,
Charlottesville, VA 22903-0818,
USA; reiprich@virginia.edu.}
\begin{abstract}
A new X-ray selected and X-ray flux-limited galaxy cluster sample is
presented. Based on the \ro\ All-Sky Survey the 63 brightest clusters
with galactic latitude $\vert \bii \vert \geq 20$\,deg and flux $\fx(\eb)\ge
2\esc$ have been compiled. Gravitational masses have been determined
utilizing intracluster gas density profiles, derived mainly from \ro\ \ps\
pointed observations, and gas temperatures, as published mainly from
\as\ observations, assuming hydrostatic equilibrium.
This sample and an extended sample of 106 galaxy clusters is used to
establish the X-ray luminosity--gravitational mass relation.
From the complete sample the galaxy cluster mass
function is determined and used to constrain the mean cosmic matter density 
and the amplitude of mass fluctuations.
Comparison to Press--Schechter type model mass functions in the framework
of Cold Dark Matter cosmological models and a
Harrison--Zeldovich initial density fluctuation spectrum
yields the constraints
$\om=0.12^{+0.06}_{-0.04}$ and
$\sigma_8=0.96^{+0.15}_{-0.12}$ (90\,\% c.l.).
Various possible systematic uncertainties are
quantified. Adding all identified systematic uncertainties to the
statistical uncertainty in a worst case fashion results in an upper limit
$\om<0.31$.
For comparison to previous results a relation
$\sigma_8=0.43\,\om^{-0.38}$ is derived.
The mass function is integrated
to show that the contribution of mass bound within virialized cluster
regions to the total matter density is small, i.e.,
$\oc = 0.012^{+0.003}_{-0.004}$
for cluster masses larger than $6.4^{+0.7}_{-0.6}\times
10^{13}\,h_{50}^{-1}\,M_{\odot}$.
\end{abstract}


\section{Introduction}\label{intro}

The galaxy cluster mass function is the most fundamental statistic of
the galaxy cluster population. 
It is determined by the initial conditions of the mass distribution
set in the early universe in a relatively straightforward way,
since the evolution of the large-scale matter
distribution on scales comparable to the size of clusters and larger is linear
and since the
formation of clusters is governed by essentially only gravitational
processes. Choosing a specific cosmological scenario provides definitive
predictions about these initial conditions in a statistical sense. The overall
process of the gravitational growth of the density
fluctuations and the development of gravitational instabilities leading to cluster
formation is comparatively easy to understand. It has
been well described by analytical models (e.g., \citealt{ps74}, \citealt{bce91},
\citealt{lc93}, \citealt{ks96}, \citealt{sba00}) and simulated in
numerical gravitational $N$-body calculations
(e.g., \citealt{efw88,lc94}). Even though the slight
deviations from the theoretical prescription found in recent
simulations (e.g., \citealt{gbq99,jfw01})
seem to require further refinement in the theoretical framework,
these deviations are too small to be significant for the current
investigation, as will be shown later. For the precision needed here
these large simulations therefore support the validity of the model
predictions.
Within this framework the observed cluster mass
function provides
the opportunity to test different cosmological models. The tests are particularly
sensitive to the amplitude of the cosmic matter density fluctuations (at a scale of
the order of 10\,Mpc) as well as the normalized total matter density, $\om$
(e.g., \citealt{ha91,bc92}). 

In addition to its importance in testing cosmological models the integral of the mass
function yields the interesting information on the fractional amount
of matter contained 
in gravitationally bound large-scale structures. Using one of the first attempts to
construct a mass function over
the mass range from giant ellipticals to massive clusters
by \citet{bc93} \citet{fhp98} obtain a mass fraction
$\og=0.12\pm0.02$
for $2\times 10^{12}\le M \le 2\times
10^{14}\,h_{50}^{-1}\,M_{\odot}$
(where the mass fraction is expressed
here in units of the critical density of the universe, $\roc$). This result is
already close to the total matter density in some of the proposed cosmological
scenarios.
Therefore a precise observational
determination of the integral mass function is a very important task
for astronomy.

Unfortunately the galaxy cluster mass is not an easily and directly observable
quantity (except for measurements of the gravitational lensing effect of
clusters which may play a large role in the construction of mass functions in
the future) and one has to resort to the observation of other cluster parameters
from which the cluster masses can be deduced.

X-ray astronomy has provided an ideal tool to first detect and select massive
clusters by measuring their X-ray luminosity and to secondly perform mass
determinations on individual clusters through X-ray imaging and X-ray
spectroscopy.

In this paper we report the first rigorous application of these
two approaches for the construction of the cluster mass function. Building on
the \ro\ All-Sky X-ray Survey (\ra ) \citep{t93,vab99}, which has been well
studied in the search of the brightest galaxy clusters through several survey
projects (see refs in Sect.~\ref{sample}), we have compiled a new, highly
complete sample of the X-ray brightest galaxy clusters (\gcs , the HIghest X-ray
FLUx Galaxy Cluster Sample).

Thanks to the numerous detailed galaxy cluster observations performed with the
\ro\ and \as\ \citep{tih94} satellite observatories and accumulated in the archives we can perform
a detailed mass determination for most of these clusters and obtain a good mass
estimate for the few remaining objects. From these data we first establish a
(tight) correlation of the measured X-ray luminosity and the cluster mass. This
relation assures that we have essentially sampled the most massive clusters in
the nearby universe, which forms the basis of the construction of the cluster
mass function.

Previous local galaxy cluster mass functions have been derived by
\citet{bc93}, \citet{bgg93}, \citet{gbg98}, and \citet[for galaxy groups]{gg00}.
\citet{bc93} used the galaxy richness to relate to cluster
masses for optical observations and an X-ray temperature--mass relation
to convert the temperature function given by \citet{ha91} to a mass
function. \citet{bgg93}, \citet{gbg98}, and \citet{gg00} used velocity dispersions
for optically selected samples to determine the mass function. Here we
use a different approach and construct
the first mass function
for X-ray selected galaxy clusters based on the \ra\ using
individually determined cluster masses. The mass function of
this cluster sample is then used to determine the mass fraction in bound objects
with masses above a minimum mass and to derive tight constraints on
cosmological scenarios. 

The paper is organized as follows. In Sect.~\ref{sample} the sample
selection is described. The details of the determination of the
observational quantities are given in Sect.~\ref{data}. The results are
presented in Sect.~\ref{resul} and discussed in Sect.~\ref{discu}. The
conclusions are summarized in Sect.~\ref{conclu}.

Throughout a Hubble constant $H_0=50\,h_{50}\,\rm
km\,s^{-1}\,Mpc^{-1}$, $h_{50}=1$, 
$\om=1$, a
normalized  cosmological constant $\ol\equiv \Lambda/(3\,H_0^2)=0$,
and a normalized curvature index $\ok\equiv 1-\om-\ol=0$ is used if not stated
otherwise (present day quantities).
We note that the determination of physical cluster parameters has a
negligible dependence on $\om$ and $\ol$ for the small redshift range
used here, as will be shown later. Therefore it is justified to
determine the parameters for an Einstein--de Sitter model but to
discuss the results also in the context of other models.

\section{Sample}\label{sample}

The mass function measures
the cluster number density as a function of mass. Therefore any
cluster fulfilling the selection criteria and
not included in the sample distorts the result systematically.
It is then obvious that for the construction of the mass function
it is vital to use a
homogeneously selected and highly complete sample of objects,
and additionally the selection must be closely related to cluster
mass. In this work the \ra , where one single instrument
has surveyed the whole sky, has been chosen as the basis for the
sample construction. Using the X-ray emission from the hot
intracluster medium for cluster selection minimizes projection
effects and the tight correlation between X-ray luminosity and
gravitational mass convincingly demonstrates that X-ray cluster
surveys have the important property of being mass selective
(Sect.~\ref{relat}). 

 Several cluster catalogs have already been constructed
from the \ra\ with high completeness down to low flux limits (see refs
below). These we have utilized for the selection of candidates. Low
thresholds have been  set for selection in order not to miss any cluster due
to measurement uncertainties. These candidates
have been homogeneously reanalyzed, using higher quality \ro\ \ps\
pointed observations whenever possible (Sect.~\ref{data}). A flux
limit well above the limit for candidate selection has then been
applied to define the new flux-limited sample of the brightest
clusters in the sky. 

In detail the candidates
emerged from the following input catalogs.
Table~\ref{tbl:cand} lists the selection criteria and the number of
clusters selected from each of the catalogs, that are contained in the
final sample.  
\\
1) The REFLEX (\ro -ESO Flux-Limited X-ray) galaxy cluster survey
\citep{bsg01} covers the
southern hemisphere (declination $\delta \leq$
+2.5\,deg; galactic latitude $\vert \bii \vert \geq 20.0$\,deg) with a
flux limit  $\fxl(\eb) = 3.0\,\escl$.
\\
2) The NORAS (Northern \ro\ All-Sky) galaxy cluster survey \citep{bvh00}
contains clusters showing extended emission in the RASS in the
northern hemisphere  ($\delta \geq$
0.0\,deg; $\vert \bii \vert \geq 20.0$\,deg) with
count rates $\cx(\eb)\geq 0.06\, \cts$. 
\\
3) NORAS II (J. Retzlaff et al., in preparation)\ is the continuation of the
NORAS survey project. It includes point like sources and aims for a
flux limit $\fxl(\eb) = 2.0\,\escl$.
\\
4) The BCS (\ro\ Brightest Cluster Sample) \citep{eeb98} covers the northern
hemisphere ($\delta \geq$
0.0\,deg; $\vert \bii \vert \geq 20.0$\,deg) with
$\fxl(\eb) = 4.4\,\escl$ and redshifts $z\leq 0.3$.
\\
5) The \ra\ 1 Bright Sample of Clusters of Galaxies \citep{gbg99} covers the
south galactic cap region in the southern hemisphere ($\delta <$
+2.5\,deg; $\bii < -20.0$\,deg) with an effective flux limit $\fxl(\eh)$
between $\sim$\,3 and $4\,\escl$. 
\\
6) XBACs (X-ray Brightest Abell-type Clusters of galaxies) \citep{evb96} is an
all-sky sample of \citet{a58}/ACO \citep{aco89} clusters limited to high
galactic latitudes $\vert \bii \vert \geq 20.0$\,deg with nominal ACO
redshifts $z\leq 0.2$ and X-ray fluxes $\fx(\eb) > 5.0\,\escl$.
\\
7) An all-sky list of Abell/ACO/ACO-supplementary clusters
(H. B\"ohringer 1999, private communication) with count rates
$\cx(\eh)\ge  0.6\,\cts$.
\\
8) Early type galaxies with measured \ra\ count rates from a magnitude
limited sample of \citet{bdb99} have been been checked in order not to
miss any X-ray faint groups. 
\\
9) All clusters from the sample of \citet{lef89} and \citet{esf90},
where clusters had been compiled from various X-ray missions, have
been checked.


\begin{deluxetable}{lcccccccc}
\tabletypesize{\footnotesize}
\tablecaption{Selection of candidates \label{tbl:cand}}
\tablewidth{0pt}
\tablehead{
\colhead{Catalog}    &  \multicolumn{3}{c}{$\cx\ [\cts]$} &   \colhead{}   &
\multicolumn{2}{c}{$\fx\ [\esct]$} & \colhead{$\ncl$} & \colhead{Ref.}\\
\cline{2-4} \cline{6-7} \\
\colhead{} & \colhead{$\eb$}   &
\colhead{$\eh$}   & \colhead{$\ek$} & 
\colhead{} &\colhead{$\eb$} &\colhead{$\eh$} &
\colhead{} & \colhead{}
}
\startdata
REFLEX       & \nodata & 0.9 & \nodata &  & 1.7 & \nodata  & 33  &1\\
NORAS        & \nodata & 0.7 & \nodata &  & 1.7 & \nodata  & 25  &2\\
NORAS II     & \nodata & 0.7 & \nodata &  & 1.7 & \nodata  & 4   &3\\
BCS          & 1.0 & \nodata & \nodata &  & 1.7 & \nodata  & 1   &4\\
\ra\ 1       & \nodata & \nodata & \nodata &  & \nodata & 1.0 & 0 &5\\
XBACs        & 1.0 & \nodata & \nodata &  & 1.7 &  \nodata  & 0  &6\\
Abell/ACO    & \nodata & 0.7 & \nodata &  & \nodata & \nodata & 0 &7\\
Early type gx  & \nodata & \nodata & 0.7 &  & \nodata & \nodata & 0 &8\\
Previous sat\tablenotemark{a} & \nodata & \nodata & \nodata &  & \nodata & \nodata  & 0 &9\\
 \enddata

\tablenotetext{a}{All clusters from this catalog have been flagged as candidates.}

\tablecomments{Only one of the criteria, count rate or flux, has to be
met for a cluster to be selected as candidate. The catalogs are
listed in search sequence, therefore $\ncl$ gives the number of
candidates additionally selected from the current catalog and
contained in the final flux-limited sample. So in the case of NORAS a
cluster is selected as candidate if it fulfils $\cx(\eh)\geq 0.7\,\cts$
or $\fx(\eb)\geq 1.7\esc$ and has not already been selected from
REFLEX. This candidate is counted under $\ncl$ if it meets the
selection criteria for \gcss .}

\tablerefs{
(1) \citealt{bsg01}. (2) \citealt{bvh00}. (3) J. Retzlaff et al.,
in preparation. (4) \citealt{eeb98}. (5) \citealt{gbg99}. (6)
\citealt{evb96}. (7) H. B\"ohringer 1999, private
communication. (8) \citealt{bdb99}. (9) \citealt{lef89,esf90}.}

\end{deluxetable}


The main criterion for candidate selection, a flux threshold $1.7\esc$,
has been chosen
to allow for measurement uncertainties in the input catalogs. E.g., for REFLEX clusters with
$1.5 \leq \fx \leq 2.5\esc$ the mean statistical flux error is less than
8\,\%.
With an additional mean systematic error of 6\,\%, caused by
underestimation of fluxes due to the comparatively low \ra\ exposure
times\footnote{This has been measured by comparing the count rates
determined using pointed observations of clusters in this work to
count rates for the same clusters determined in REFLEX and NORAS. If
count rates are compared also for fainter clusters, not relevant for
the present work, the mean systematic
error increases to about 9\,\% \citep{bvh00}.},
the flux threshold $1.7\esc$ for candidate selection then ensures
that no clusters are
missed for a final flux limit $\fxl = 2.0\esc$.

Almost none of the fluxes given in the input catalogs have been
calculated using a measured X-ray temperature, but mostly using
gas temperatures estimated from an $\lx$--$\tx$ relation.
In order to be independent of this additional
uncertainty clusters have also been
selected as candidates if they exceed a count rate threshold which
corresponds to $\fx(\eb)=2.0\esc$ for a typical cluster temperature,
$\tg=4\,\rm keV$, and redshift, $z=0.05$, and for an exceptionally
high column density, e.g., in the NORAS case $\nh=1.6\,\times
10^{21}\rm cm^{-2}$.

Most of the samples mentioned above excluded the area on the sky close to the
galactic plane as well as the area of the Magellanic Clouds. In order to
construct a highly complete sample from the candidate list we applied the
following selection criteria that successful clusters must fulfil:\\
1) \emph{redetermined} flux $\fx(\eb)\ge 2.0\esc$,
\\
2) galactic latitude $\vert \bii\vert \ge 20.0$\,deg,
\\
3) projected position outside the excluded 324\,deg$^2$
area of the Magellanic Clouds (see Tab.~\ref{tab:mc}),
\\
4) projected position outside the excluded 98\,deg$^2$
region of the Virgo galaxy cluster (see Tab.~\ref{tab:mc}).\footnote{The large scale X-ray
background of the irregular and very extended X-ray emission of the
Virgo cluster makes the undiscriminating detection/selection of
clusters in this area difficult.
Candidates excluded due to this criterium are Virgo, M86, and M49.}
 
\begin{deluxetable}{lccc}
\tabletypesize{\footnotesize}
\tablewidth{0pt}
\tablecaption{Regions of the sky not sampled in \gcss \label{tab:mc}} 
\tablehead{ {\rm Region}& {\rm R.A. Range }  & {\rm Dec.\ Range} & {\rm
Area (sr)}} 
\startdata
{\rm LMC 1}  & 58  $\to 103^o$  & $-63 \to -77^o$   & 0.0655 \\                     
{\rm LMC 2}  & 81  $\to 89^o $ & $-58 \to -63^o$   & 0.0060 \\                      
{\rm LMC 3}  & 103 $\to 108^o$  & $-68 \to -74^o$   & 0.0030 \\                     
{\rm SMC 1}  &358.5 $\to 20^o$  & $-67.5 \to -77^o$ & 0.0189 \\                     
{\rm SMC 2}  &356.5 $\to 358.5^o$ & $-73 \to -77^o$   & 0.0006 \\                   
{\rm SMC 3}  & 20  $\to 30^o$   & $-67.5 \to -72^o$ & 0.0047 \\                     
{\rm Virgo}  & 182.7  $\to 192.7^o$   & $7.4 \to 17.4^o$ &  0.0297\\                     
{\rm Milky Way}\tablenotemark{a}  & 0  $\to 360^o$ ($\lii$)  & $-20 \to 20^o$ ($\bii$)  & 4.2980 \\                     
\enddata
\tablenotetext{a}{Galactic coordinates.}
\tablecomments{Excised areas for the Magellanic Clouds are the same as
in \citet{bsg01}, because REFLEX forms the basic input catalog in the
southern hemisphere.}
\end{deluxetable}


These selection criteria are fulfilled by 63 candidates. The
advantages of the redetermined fluxes over the fluxes from the input
catalogs are summarized at the end of Sect.~\ref{fluxd}.
In Tab.~\ref{tbl:cand} one notes that 98\,\% of all clusters in
\gcs\ have been flagged as candidates in REFLEX, NORAS, or in the
candidate list for NORAS II; these surveys are not only all based on
the \ra\ but all use the same algorithm for the
count rate determination, further substantiating the homogeneous
candidate selection for \gcs .

The fraction of available \ro\ \ps\ pointed observations for clusters
included in \gcs\ equals 86\,\%. The actually used fraction is
slightly reduced to 75\,\% because some clusters appear extended
beyond the \ps\ field of view and therefore \ra\ data have been
used. The fraction of clusters with published ASCA temperatures equals 87\,\%.
If a lower flux limit had been chosen the fraction of available
\ps\ pointed observations and published ASCA temperatures
would have been decreased thereby increasing the uncertainties in the derived
cluster parameters.
Furthermore this value for the flux limit ensures that no 
corrections, due to low exposure in the \ra\ or high galactic hydrogen column
density, need to be applied for the effective area covered. This can
be seen by the
effective sky coverage in the REFLEX survey area for a flux limit
$\fxl(\eb) = 2.0\,\esc$ and a minimum of 30 source counts, which
amounts to 99\,\%.
The clear advantage is that the \gcs\ catalog can be used in a
straightforward manner in statistical analyses, because the effective
area is the same for all clusters and simply equals the covered solid
angle on the sky.

\begin{figure}[thbp]
\psfig{file=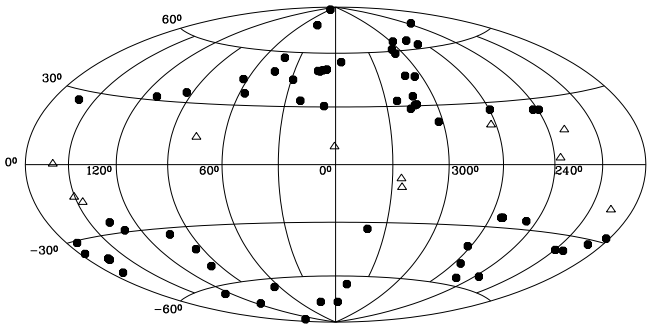,width=7.5cm,angle=270,bbllx=1pt,bblly=120pt,bburx=570pt,bbury=840pt,clip=}
\caption{Aitoff projection of the 63 \gcs\ galaxy clusters in galactic coordinates
(filled circles). Additionally shown are 11 clusters above the flux limit
but with $\vert \bii\vert < 20.0\,\rm deg$ (open triangles).}\label{aito}
\end{figure}

The distribution of clusters included in \gcs\ projected onto the
sky is shown in Fig.~\ref{aito}.
The sky coverage for the cluster sample equals 26\,721.8\,deg$^2$
(8.13994\,sr),  about two thirds of the sky.
The cluster names, coordinates and redshifts
are listed in Tab.~\ref{tbl:data1}.
Further properties of the cluster sample are discussed
in Sect.~\ref{sample_c}. 

For later analyses which do not necessarily require a complete sample,
e.g., correlations between physical parameters, 43 clusters (not
included in \gcs ) from the candidate list have been combined with
\gcs\ to form an `extended sample' of 106 clusters.
 
\section{Data Reduction and Analysis}\label{data}

This Section describes the derivation of the basic quantities in this work,
e.g., count rates, fluxes,
luminosities, and mass estimates for the galaxy clusters.
These and other relevant cluster parameters are tabulated along
with their uncertainties.

\subsection{Flux Determination}\label{fluxd}

Measuring the count rate of galaxy clusters is an important step in constructing a
flux-limited cluster sample. The count rate determination performed here is
based on the growth curve analysis method \citep{bvh00}, with
modifications adapted to the higher photon statistics available here. The main
features of the method
as well as the modifications are outlined below.

The instrument used is the \ro\ \ps\ \citep{pbh87}, with a low internal
background ideally suited for this study which needs good signal to
noise of the outer, low surface brightness regions of the
clusters. Mainly pointed observations from the
\anchor{http://www.xray.mpe.mpg.de/cgi-bin/rosat/seq-browser}{public archive at
MPE} have been 
used. If the cluster is extended beyond the \ps\ field of view
making a proper background 
determination difficult or if there is no pointed \ps\ observation available,
\ra\ data have been used.
The \ro\ hard energy band (channels $52-201\approx \eh$) has been used for all count
rate measurements
because of the higher background in the soft band.

Two X-ray cluster centers are
determined by finding the two-dimensional `center of mass' of the photon
distribution iteratively for an aperture radius of 3 and 7.5\,arcmin around the
starting position. The small aperture yields the center representing the
position of the cluster's peak emission and therefore probably indicates the
position where the cluster's potential well is deepest. This center is used for
the regional selection, e.g.\ $\vert \bii\vert \ge 20.0$. The more
globally defined center with 
the larger aperture is used for the subsequent analysis tasks since for the
mass determination it is most important to have a good estimate of the slope of
the surface brightness profile in the outer parts of the cluster.

\begin{figure}[thbp]
\psfig{file=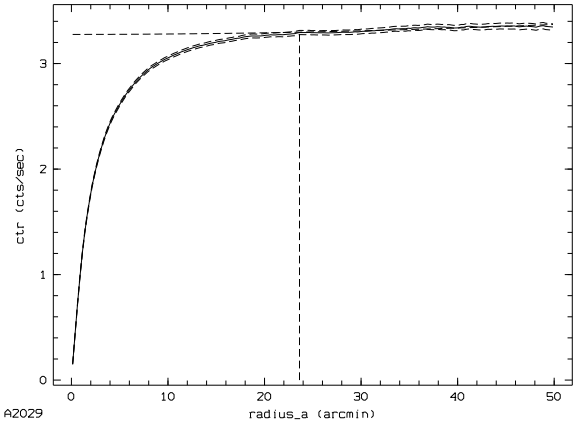,width=6cm,angle=270,clip=}
\caption{Cumulative source count rate as a function of radius (solid line) for the cluster
A2029 (pointed observation). The vertical dashed line indicates the
outer significance radius, $\rx$. The dashed lines just above and
below the source count rate indicate the 1-$\sigma$ Poissonian error
bars.}\label{a2029}
\end{figure}

The background surface brightness is determined in a
ring outside the cluster emission. To minimize the influence of
discrete sources the ring is subdivided into twelve parts of equal area and a
sigma clipping is performed. To determine the count rate the area around the
global center is divided into concentric rings. For pointed observations 200
rings with a width of 15\,arcsec each are used. Due to the lower photon statistics
a width of 30\,arcsec is used for \ra\ data and the number of rings
depends on the field size extracted ($100 - 300$ rings for field sizes of
$2\times 2\,\rm deg^2 - 8\times 8\,\rm deg^2$). Each photon is divided by the
vignetting and deadtime corrected exposure time of the skypixel where it has
been
detected and these ratios are summed up in each ring yielding the ring count
rate. From this value the background count rate for the respective ring area is
subtracted yielding a source ring count rate. These individual source ring count
rates are integrated with increasing radius yielding the (cumulative)
source count rate for a
given radius (Fig.~\ref{a2029}). Obvious contaminating point sources
have been excluded manually.
The cut-out regions have then been assigned the average surface
brightness of the ring.
If a cluster has been found to be clearly made up of
two components, for instance A3395n/s,
these components have been treated separately.
This procedure ensures that double clusters are not treated as a
single entity for which spherical symmetry is assumed.
For the same reason strong
substructure has been excluded in the same manner as contaminating
point sources.
In this work the aim is to characterize all cluster
properties consistently and homogeneously. Therefore if strong substructure
is identified then it is excluded for the
flux/luminosity \emph{and} mass determination.

\begin{figure}[thbp]
\psfig{file=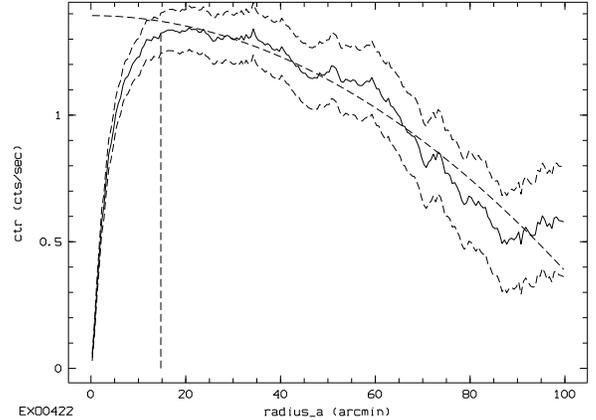,width=6cm,angle=270,clip=}
\caption{Cumulative source count rate as a function of radius for the cluster
EXO0422, shown as an extreme example (\ra\ data). The parabolic dashed
line indicates the best fit parabola for count rates larger than
$\rx$.}\label{exo}
\end{figure}

\begin{figure}[thbp]
\psfig{file=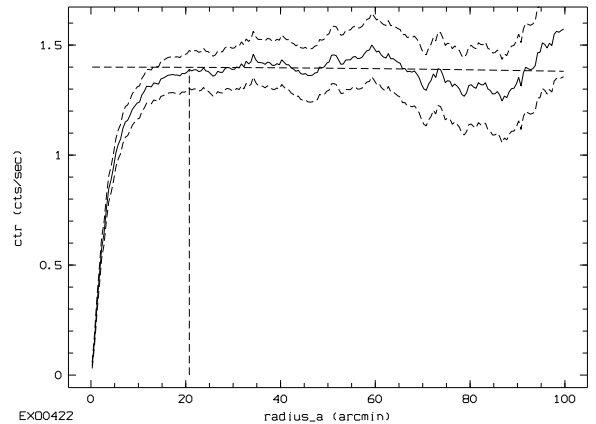,width=6cm,angle=270,clip=}
\caption{Corrected cumulative source count rate as a function of radius for the cluster
EXO0422. The count rate correction is less than 5\,\%.}\label{exoc}
\end{figure}

An outer 
significance radius of the cluster, $\rx$, is determined at the position from
where on the Poissonian
1-$\sigma$ error rises faster than the source count rate. Usually the source count
rate settles into a nearly horizontal line for radii larger than $\rx$. We have
found, however, that in some cases the source count rate
seems to increase or
decrease roughly quadratically for radii larger than $\rx$ indicating a possibly
under- or overestimated background (Fig.~\ref{exo}). We therefore
fitted a parabola of the form 
$y=mx^2+b$ to the source count rate for radii larger than $\rx$ and corrected
the measured background.
An example for a corrected source count rate profile is shown in Fig.~\ref{exoc}.

\begin{figure}[thbp]
\psfig{file=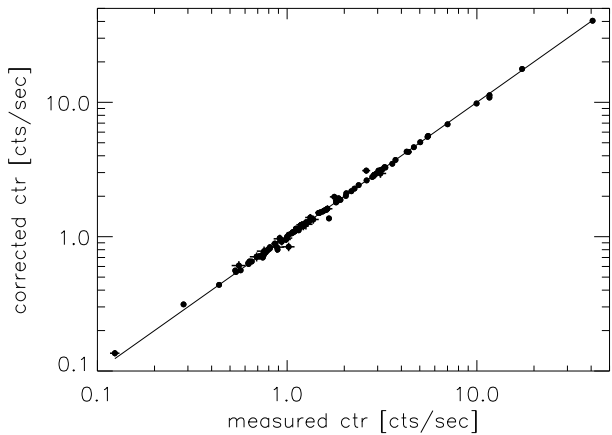,width=9cm,angle=0,clip=}
\caption{Comparison of measured and corrected source count rates for
the extended sample of 106 galaxy clusters. The
solid line indicates equality.}\label{counts}
\end{figure}

Figure~\ref{counts} shows for the extended sample (106 clusters) that
the difference between measured and 
corrected source count rate is generally very small. Nevertheless an
inspection of each count rate profile has been performed, to decide
whether the measured or corrected count rate is adopted as the final
count rate, to avoid artificial corrections due to large scale
variations of the background (especially in the large \ra\
fields). The count rates are given in Tab.~\ref{tbl:data1}. 

The conversion factor for the count rate to flux conversion depends on the
hydrogen column density, $\nh$, on the cluster gas temperature, $\tg$, on the
cluster gas metallicity, on the cluster redshift, $z$, and on the respective
detector responses for the two different \ps s used. The $\nh$ value is taken as
the value inferred from 21\,cm radio measurements for our galaxy at the
projected cluster position (\citealt{dl90};
included in the EXSAS 
software package, \citealt{zbb98}; photoelectric absorption cross
sections are taken from \citealt{mm83}).
Gas temperatures have been estimated by compiling X-ray temperatures, $\tx$, 
from the literature, giving preference to temperatures measured
with the \as\ satellite. 
For clusters where no \as\ measured temperature has been
available,
$\tx$ measured with previous X-ray satellites have been used.
The X-ray temperatures and corresponding references are given in Tab.~\ref{tbl:data2}.
For two
clusters included in \gcs\ no measured temperature has been found in the
literature and the $\lx (<2\,\mpc)-\tx$ relation of \citet{m98} has
been used. The relation 
for non cooling flow corrected luminosities and cooling flow
corrected/emission 
weighted temperatures has been chosen. Since the conversion from count
rate to flux depends only weakly on $\tg$
in the \ro\ energy band for the relevant temperature
range a cluster temperature $k\tg = 4\,\kev$ has been assumed in a first step
to determine $\lx (<2\,\mpc)$ for the clusters where no gas
temperature has been found in the literature. With this luminosity the
gas temperature has been estimated. 
The metallicity is set to 0.35 times the solar value for all
clusters \citep[e.g.,][]{arb92}.
The redshifts have been compiled from the literature and are given in
Tab.~\ref{tbl:data1} together with the corresponding references. With these quantities
and the count rates given in Tab.~\ref{tbl:data1} fluxes in the observer rest frame
energy range $\eb$ have
been calculated applying a modern version of a Raymond-Smith spectral code
\citep{rs77}.  
The results are listed in Tab.~\ref{tbl:data1}. The flux calculation has
also been checked using \xs\ \citep{a96}
by folding the model spectrum created with the
parameters given above with the detector response and adjusting the
normalization to reproduce the observed count rate.
It is found that for 90\,\% of the clusters the deviation between the two
results for the flux measurement is less than 1\,\%.
Luminosities in the source rest frame energy range $\eb$
have then been calculated within \xs\ by adjusting the normalization to
reproduce the initial flux measurements.

The improvements of the flux determination performed here compared to
the input catalogs in general are now summarized.
1) Due to the use of a high fraction of pointed observations the
photon statistics is on average much better, e.g., for the 33 clusters
contained in REFLEX and \gcs\ one finds a mean of 841 and 19580 source photons,
respectively. Consequently the cluster emission has been traced out to
larger radii for \gcs . 
2) The higher photon statistics has allowed a proper exclusion of
contaminating point sources (stars, AGN, etc.) and substructure, and the
separation of double clusters.
3) An iterative background correction has been performed.
4) A measured X-ray temperature has been used for the flux calculation in
most cases.

Simulations have shown that even for the
\gcs\ clusters with the lowest number of photons the determined flux
shows no significant trend with redshift in the relevant redshift range
\citep{irb01}.

\begin{deluxetable}{lrrrrrrrrrrcc}
\tablecolumns{13} 
\tabletypesize{\footnotesize}
\tablecaption{Cluster properties \label{tbl:data1}}
\tablewidth{0pt}
\tablehead{
\colhead{Cluster}	& \colhead{R.A.}	& \colhead{Dec.}	&
\colhead{$z$}	& \colhead{$\nh$}	& \colhead{$\cx$}	&
\colhead{$\Delta$}	& \colhead{$\rx$}	& \colhead{$\fx$}	&
\colhead{$\lx$}	& \colhead{$\lbol$}	& \colhead{Obs}	&
\colhead{Ref}	\\
\colhead{(1)}	& \colhead{(2)}	& \colhead{(3)}	&
\colhead{(4)}	& \colhead{(5)}	& \colhead{(6)}	&
\colhead{(7)}	& \colhead{(8)}	& \colhead{(9)}	&
\colhead{(10)}	& \colhead{(11)}	& \colhead{(12)}	&
\colhead{(13)}
}
\startdata
A0085	&  10.4632 & $ -9.3054$ & 0.0556 &  3.58 &  3.488 &  0.6 & 2.13 &   7.429 &  9.789 &  24.448 & P & 2 \\ 
A0119	&  14.0649 & $ -1.2489$ & 0.0440 &  3.10 &  1.931 &  0.9 & 2.68 &   4.054 &  3.354 &   7.475 & P & 1 \\ 
A0133	&  15.6736 & $-21.8806$ & 0.0569 &  1.60 &  1.058 &  0.8 & 1.52 &   2.121 &  2.944 &   5.389 & P & 3 \\ 
NGC507	&  20.9106 & $ 33.2553$ & 0.0165 &  5.25 &  1.093 &  1.3 & 0.88 &   2.112 &  0.247 &   0.326 & P & 5 \\ 
A0262	&  28.1953 & $ 36.1528$ & 0.0161 &  5.52 &  4.366 &  3.8 & 1.48 &   9.348 &  1.040 &   1.533 & R & 6 \\ 
A0400	&  44.4152 & $  6.0170$ & 0.0240 &  9.38 &  1.146 &  1.1 & 1.85 &   2.778 &  0.686 &   1.033 & P & 8 \\ 
A0399	&  44.4684 & $ 13.0462$ & 0.0715 & 10.58 &  1.306 &  5.4 & 3.18 &   3.249 &  7.070 &  17.803 & R & 6 \\ 
A0401	&  44.7384 & $ 13.5796$ & 0.0748 & 10.19 &  2.104 &  1.1 & 3.81 &   5.281 & 12.553 &  34.073 & P & 6 \\ 
A3112	&  49.4912 & $-44.2367$ & 0.0750 &  2.53 &  1.502 &  1.1 & 2.18 &   3.103 &  7.456 &  16.128 & P & 1 \\ 
FORNAX	&  54.6686 & $-35.3103$ & 0.0046 &  1.45 &  5.324 &  5.6 & 0.53 &   9.020 &  0.082 &   0.107 & P+R & 4 \\ 
2A0335	&  54.6690 & $  9.9713$ & 0.0349 & 18.64 &  3.028 &  0.8 & 1.54 &   9.162 &  4.789 &   7.918 & P & 10 \\ 
IIIZw54	&  55.3225 & $ 15.4076$ & 0.0311 & 16.68 &  0.708 &  7.7 & 1.27 &   2.001 &  0.831 &   1.226 & R & 11 \\ 
A3158	&  55.7282 & $-53.6301$ & 0.0590 &  1.06 &  1.909 &  1.5 & 1.94 &   3.794 &  5.638 &  12.779 & P & 1 \\ 
A0478	&  63.3554 & $ 10.4661$ & 0.0900 & 15.27 &  1.827 &  0.6 & 3.12 &   5.151 & 17.690 &  49.335 & P & 6 \\ 
NGC1550	&  64.9066 & $  2.4151$ & 0.0123 & 11.59 &  1.979 &  5.4 & 0.71 &   4.632 &  0.302 &   0.407 & R & 13 \\ 
EXO0422	&  66.4637 & $ -8.5581$ & 0.0390 &  6.40 &  1.390 &  6.2 & 1.32 &   3.085 &  2.015 &   3.283 & R & 10 \\ 
A3266	&  67.8410 & $-61.4403$ & 0.0594 &  1.48 &  2.879 &  0.7 & 2.99 &   5.807 &  8.718 &  23.663 & P & 4 \\ 
A0496	&  68.4091 & $-13.2605$ & 0.0328 &  5.68 &  3.724 &  0.7 & 1.78 &   8.326 &  3.837 &   7.306 & P & 8 \\ 
A3376	&  90.4835 & $-39.9741$ & 0.0455 &  5.01 &  1.115 &  1.4 & 2.86 &   2.450 &  2.174 &   4.077 & P & 4 \\ 
A3391	&  96.5925 & $-53.6938$ & 0.0531 &  5.42 &  0.999 &  1.9 & 1.98 &   2.225 &  2.681 &   5.857 & P & 4 \\ 
A3395s	&  96.6920 & $-54.5453$ & 0.0498 &  8.49 &  0.836 &  3.8 & 1.45 &   2.009 &  2.131 &   4.471 & P & 4 \\ 
A0576	& 110.3571 & $ 55.7639$ & 0.0381 &  5.69 &  1.374 &  6.8 & 2.32 &   3.010 &  1.872 &   3.518 & R & 6 \\ 
A0754	& 137.3338 & $ -9.6797$ & 0.0528 &  4.59 &  1.537 &  1.6 & 1.91 &   3.366 &  3.990 &  11.967 & P & 6 \\ 
HYDRA-A	& 139.5239 & $-12.0942$ & 0.0538 &  4.86 &  2.179 &  0.6 & 1.66 &   4.776 &  5.930 &  11.520 & P & 13 \\ 
A1060	& 159.1784 & $-27.5212$ & 0.0114 &  4.92 &  4.653 &  3.3 & 0.95 &   9.951 &  0.554 &   0.945 & R & 6 \\ 
A1367	& 176.1903 & $ 19.7030$ & 0.0216 &  2.55 &  2.947 &  0.8 & 1.55 &   6.051 &  1.206 &   2.140 & P & 8 \\ 
MKW4	& 181.1124 & $  1.8962$ & 0.0200 &  1.86 &  1.173 &  1.7 & 1.23 &   2.268 &  0.390 &   0.543 & P & 10 \\ 
ZwCl1215	& 184.4220 & $  3.6604$ & 0.0750 &  1.64 &  1.081 &  1.3 & 2.55 &   2.183 &  5.240 &  11.656 & P & 19 \\ 
NGC4636	& 190.7084 & $  2.6880$ & 0.0037 &  1.75 &  3.102 &  7.2 & 0.39 &   4.085 &  0.023 &   0.027 & R & 13 \\ 
A3526	& 192.1995 & $-41.3087$ & 0.0103 &  8.25 & 11.655 &  2.2 & 1.64 &  27.189 &  1.241 &   2.238 & R & 15 \\ 
A1644	& 194.2900 & $-17.4029$ & 0.0474 &  5.33 &  1.853 &  5.1 & 1.85 &   4.030 &  3.876 &   7.882 & R & 8 \\ 
A1650	& 194.6712 & $ -1.7572$ & 0.0845 &  1.54 &  1.218 &  6.6 & 3.17 &   2.405 &  7.308 &  17.955 & R & 6 \\ 
A1651	& 194.8419 & $ -4.1947$ & 0.0860 &  1.71 &  1.254 &  1.2 & 2.03 &   2.539 &  8.000 &  18.692 & P & 22 \\ 
COMA	& 194.9468 & $ 27.9388$ & 0.0232 &  0.89 & 17.721 &  1.4 & 4.04 &  34.438 &  7.917 &  22.048 & R & 8 \\ 
NGC5044	& 198.8530 & $-16.3879$ & 0.0090 &  4.91 &  3.163 &  0.5 & 0.56 &   5.514 &  0.193 &   0.246 & P & 24 \\ 
A1736	& 201.7238 & $-27.1765$ & 0.0461 &  5.36 &  1.631 &  6.3 & 2.47 &   3.537 &  3.223 &   5.682 & R & 25 \\ 
A3558	& 201.9921 & $-31.5017$ & 0.0480 &  3.63 &  3.158 &  0.5 & 2.11 &   6.720 &  6.615 &  14.600 & P & 1 \\ 
A3562	& 203.3984 & $-31.6678$ & 0.0499 &  3.91 &  1.367 &  0.9 & 2.01 &   2.928 &  3.117 &   6.647 & P & 4 \\ 
A3571	& 206.8692 & $-32.8553$ & 0.0397 &  3.93 &  5.626 &  0.7 & 2.35 &  12.089 &  8.132 &  20.310 & P & 21 \\ 
A1795	& 207.2201 & $ 26.5944$ & 0.0616 &  1.20 &  3.132 &  0.3 & 2.14 &   6.270 & 10.124 &  27.106 & P & 6 \\ 
A3581	& 211.8852 & $-27.0153$ & 0.0214 &  4.26 &  1.603 &  3.2 & 0.64 &   3.337 &  0.657 &   0.926 & P & 28 \\ 
MKW8	& 220.1596 & $  3.4717$ & 0.0270 &  2.60 &  1.255 &  8.4 & 1.90 &   2.525 &  0.789 &   1.355 & R & 29 \\ 
A2029	& 227.7331 & $  5.7450$ & 0.0767 &  3.07 &  3.294 &  0.6 & 2.78 &   6.938 & 17.313 &  50.583 & P & 6 \\ 
A2052	& 229.1846 & $  7.0211$ & 0.0348 &  2.90 &  2.279 &  1.0 & 1.14 &   4.713 &  2.449 &   4.061 & P & 6 \\ 
MKW3S	& 230.4643 & $  7.7059$ & 0.0450 &  3.15 &  1.578 &  1.0 & 1.39 &   3.299 &  2.865 &   5.180 & P & 10 \\ 
A2065	& 230.6096 & $ 27.7120$ & 0.0721 &  2.84 &  1.227 &  6.1 & 3.09 &   2.505 &  5.560 &  12.271 & R & 6 \\ 
A2063	& 230.7734 & $  8.6112$ & 0.0354 &  2.92 &  2.038 &  1.3 & 2.13 &   4.232 &  2.272 &   4.099 & P & 8 \\ 
A2142	& 239.5824 & $ 27.2336$ & 0.0899 &  4.05 &  2.888 &  0.9 & 3.09 &   6.241 & 21.345 &  64.760 & P & 6 \\ 
A2147	& 240.5628 & $ 15.9586$ & 0.0351 &  3.29 &  2.623 &  3.2 & 1.87 &   5.522 &  2.919 &   6.067 & P & 8 \\ 
A2163	& 243.9433 & $ -6.1436$ & 0.2010 & 12.27 &  0.773 &  1.5 & 3.15 &   2.039 & 34.128 & 123.200 & P & 31 \\ 
A2199	& 247.1586 & $ 39.5477$ & 0.0302 &  0.84 &  5.535 &  1.8 & 2.37 &  10.642 &  4.165 &   7.904 & R & 8 \\ 
A2204	& 248.1962 & $  5.5733$ & 0.1523 &  5.94 &  1.211 &  1.6 & 3.29 &   2.750 & 26.938 &  68.989 & P & 6 \\ 
A2244	& 255.6749 & $ 34.0578$ & 0.0970 &  2.07 &  1.034 &  2.1 & 2.64 &   2.122 &  8.468 &  21.498 & P & 6 \\ 
A2256	& 255.9884 & $ 78.6481$ & 0.0601 &  4.02 &  2.811 &  1.4 & 3.09 &   6.054 &  9.322 &  22.713 & P & 6 \\ 
A2255	& 258.1916 & $ 64.0640$ & 0.0800 &  2.51 &  0.976 &  1.2 & 3.22 &   2.022 &  5.506 &  13.718 & P & 6 \\ 
A3667	& 303.1362 & $-56.8419$ & 0.0560 &  4.59 &  3.293 &  0.7 & 2.81 &   7.201 &  9.624 &  24.233 & P & 1 \\ 
S1101	& 348.4941 & $-42.7268$ & 0.0580 &  1.85 &  1.237 &  0.9 & 1.64 &   2.485 &  3.597 &   5.939 & P & 35 \\ 
A2589	& 350.9868 & $ 16.7753$ & 0.0416 &  4.39 &  1.200 &  1.3 & 1.46 &   2.591 &  1.924 &   3.479 & P & 37 \\ 
A2597	& 351.3318 & $-12.1246$ & 0.0852 &  2.50 &  1.074 &  1.2 & 1.43 &   2.213 &  6.882 &  13.526 & P & 6 \\ 
A2634	& 354.6201 & $ 27.0269$ & 0.0312 &  5.17 &  1.096 &  1.6 & 1.79 &   2.415 &  1.008 &   1.822 & P & 6 \\ 
A2657	& 356.2334 & $  9.1952$ & 0.0404 &  5.27 &  1.148 &  0.9 & 1.52 &   2.535 &  1.771 &   3.202 & P & 8 \\ 
A4038	& 356.9322 & $-28.1415$ & 0.0283 &  1.55 &  2.854 &  1.3 & 1.35 &   5.694 &  1.956 &   3.295 & P & 4 \\ 
A4059	& 359.2541 & $-34.7591$ & 0.0460 &  1.10 &  1.599 &  1.3 & 1.72 &   3.170 &  2.872 &   5.645 & P & 36 \\ 
\cutinhead{Clusters from the extended sample not included in \gcss .} 
A2734	&   2.8389 & $-28.8539$ & 0.0620 &  1.84 &  0.710 &  2.5 & 1.74 &   1.434 &  2.365 &   4.357 & P & 1 \\ 
A2877	&  17.4796 & $-45.9225$ & 0.0241 &  2.10 &  0.801 &  1.2 & 1.06 &   1.626 &  0.405 &   0.714 & P & 4 \\ 
NGC499	&  20.7971 & $ 33.4587$ & 0.0147 &  5.25 &  0.313 &  2.5 & 0.30 &   0.479 &  0.045 &   0.051 & P & 5 \\ 
AWM7	&  43.6229 & $ 41.5781$ & 0.0172 &  9.21 &  7.007 &  2.0 & 1.58 &  16.751 &  2.133 &   3.882 & R & 7 \\ 
PERSEUS	&  49.9455 & $ 41.5150$ & 0.0183 & 15.69 & 40.723 &  0.8 & 3.30 & 113.731 & 16.286 &  40.310 & R & 9 \\ 
S405	&  58.0078 & $-82.2315$ & 0.0613 &  7.65 &  0.781 &  8.2 & 2.14 &   1.800 &  2.899 &   5.574 & R & 12 \\ 
3C129	&  72.5602 & $ 45.0256$ & 0.0223 & 67.89 &  1.512 &  5.6 & 1.61 &  10.566 &  2.242 &   4.996 & R & 10 \\ 
A0539	&  79.1560 & $  6.4421$ & 0.0288 & 12.06 &  1.221 &  1.3 & 1.37 &   3.182 &  1.135 &   1.935 & P & 14 \\ 
S540	&  85.0265 & $-40.8431$ & 0.0358 &  3.53 &  0.788 &  5.0 & 0.84 &   1.611 &  0.887 &   1.353 & R & 4 \\ 
A0548w	&  86.3785 & $-25.9340$ & 0.0424 &  1.79 &  0.136 &  5.4 & 0.73 &   0.234 &  0.183 &   0.240 & P & 15 \\ 
A0548e	&  87.1596 & $-25.4692$ & 0.0410 &  1.88 &  0.771 &  1.8 & 2.12 &   1.551 &  1.117 &   1.870 & P & 15 \\ 
A3395n	&  96.9005 & $-54.4447$ & 0.0498 &  5.42 &  0.699 &  3.9 & 1.37 &   1.555 &  1.650 &   3.461 & P & 4 \\ 
UGC03957	& 115.2481 & $ 55.4319$ & 0.0340 &  4.59 &  0.936 &  6.0 & 0.94 &   1.975 &  0.980 &   1.531 & R & 16 \\ 
PKS0745	& 116.8837 & $-19.2955$ & 0.1028 & 43.49 &  1.268 &  1.0 & 2.44 &   6.155 & 27.565 &  70.604 & P & 17 \\ 
A0644	& 124.3553 & $ -7.5159$ & 0.0704 &  5.14 &  1.799 &  1.0 & 4.02 &   3.994 &  8.414 &  22.684 & P & 6 \\ 
S636	& 157.5151 & $-35.3093$ & 0.0116 &  6.42 &  3.102 &  4.9 & 1.18 &   5.869 &  0.341 &   0.446 & R & 18 \\ 
A1413	& 178.8271 & $ 23.4051$ & 0.1427 &  1.62 &  0.636 &  1.6 & 2.39 &   1.289 & 11.090 &  28.655 & P & 6 \\ 
M49	& 187.4437 & $  7.9956$ & 0.0044 &  1.59 &  1.259 &  1.0 & 0.27 &   1.851 &  0.015 &   0.019 & P & 15 \\ 
A3528n	& 193.5906 & $-29.0130$ & 0.0540 &  6.10 &  0.560 &  2.3 & 1.51 &   1.263 &  1.581 &   2.752 & P & 1 \\ 
A3528s	& 193.6708 & $-29.2254$ & 0.0551 &  6.10 &  0.756 &  1.6 & 1.35 &   1.703 &  2.224 &   3.746 & P & 20 \\ 
A3530	& 193.9211 & $-30.3451$ & 0.0544 &  6.00 &  0.438 &  2.8 & 1.55 &   0.987 &  1.252 &   2.317 & P & 21 \\ 
A3532	& 194.3375 & $-30.3698$ & 0.0539 &  5.96 &  0.797 &  1.8 & 1.64 &   1.797 &  2.235 &   4.483 & P & 21 \\ 
A1689	& 197.8726 & $ -1.3408$ & 0.1840 &  1.80 &  0.712 &  1.1 & 2.36 &   1.454 & 20.605 &  60.707 & P & 23 \\ 
A3560	& 203.1119 & $-33.1355$ & 0.0495 &  3.92 &  0.714 &  2.5 & 2.00 &   1.519 &  1.601 &   2.701 & P & 26 \\ 
A1775	& 205.4582 & $ 26.3820$ & 0.0757 &  1.00 &  0.654 &  1.8 & 2.02 &   1.290 &  3.175 &   5.735 & P & 27 \\ 
A1800	& 207.3408 & $ 28.1038$ & 0.0748 &  1.18 &  0.610 &  7.9 & 1.98 &   1.183 &  2.840 &   5.337 & R & 28 \\ 
A1914	& 216.5035 & $ 37.8268$ & 0.1712 &  0.97 &  0.729 &  1.4 & 2.35 &   1.454 & 17.813 &  56.533 & P & 6 \\ 
NGC5813	& 225.2994 & $  1.6981$ & 0.0064 &  4.19 &  0.976 &  6.7 & 0.17 &   1.447 &  0.025 &   0.029 & R & 13 \\ 
NGC5846	& 226.6253 & $  1.6089$ & 0.0061 &  4.25 &  0.569 &  2.3 & 0.21 &   0.851 &  0.014 &   0.016 & P & 13 \\ 
A2151w	& 241.1465 & $ 17.7252$ & 0.0369 &  3.36 &  0.754 &  1.9 & 1.46 &   1.568 &  0.917 &   1.397 & P & 8 \\ 
A3627	& 243.5546 & $-60.8430$ & 0.0163 & 20.83 &  9.962 &  3.0 & 2.20 &  31.084 &  3.524 &   8.179 & R & 30 \\ 
TRIANGUL	& 249.5758 & $-64.3557$ & 0.0510 & 12.29 &  4.294 &  0.7 & 2.54 &  11.308 & 12.508 &  37.739 & P & 32 \\ 
OPHIUCHU	& 258.1115 & $-23.3634$ & 0.0280 & 20.14 & 11.642 &  2.0 & 2.29 &  35.749 & 11.953 &  37.391 & R & 33 \\ 
ZwCl1742	& 266.0623 & $ 32.9893$ & 0.0757 &  3.56 &  0.889 &  4.4 & 1.83 &   1.850 &  4.529 &   9.727 & R & 34 \\ 
A2319	& 290.2980 & $ 43.9484$ & 0.0564 &  8.77 &  5.029 &  1.0 & 3.57 &  12.202 & 16.508 &  47.286 & P & 6 \\ 
A3695	& 308.6991 & $-35.8135$ & 0.0890 &  3.56 &  0.836 &  9.2 & 2.58 &   1.739 &  5.882 &  12.715 & R & 1 \\ 
IIZw108	& 318.4752 & $  2.5564$ & 0.0494 &  6.63 &  0.841 &  7.3 & 2.20 &   1.884 &  1.969 &   3.445 & R & 5 \\ 
A3822	& 328.5438 & $-57.8668$ & 0.0760 &  2.12 &  0.964 &  7.3 & 3.18 &   1.926 &  4.758 &   9.877 & R & 1 \\ 
A3827	& 330.4869 & $-59.9641$ & 0.0980 &  2.84 &  0.953 &  5.8 & 1.78 &   1.955 &  7.963 &  20.188 & R & 1 \\ 
A3888	& 338.6255 & $-37.7343$ & 0.1510 &  1.20 &  0.546 &  2.4 & 1.52 &   1.096 & 10.512 &  30.183 & P & 23 \\ 
A3921	& 342.5019 & $-64.4286$ & 0.0936 &  2.80 &  0.626 &  1.7 & 2.43 &   1.308 &  4.882 &  11.023 & P & 12 \\ 
HCG94	& 349.3041 & $ 18.7060$ & 0.0417 &  4.55 &  0.820 &  1.0 & 2.09 &   1.775 &  1.324 &   2.319 & P & 36 \\ 
RXJ2344	& 356.0723 & $ -4.3776$ & 0.0786 &  3.54 &  0.653 &  1.4 & 1.61 &   1.385 &  3.661 &   7.465 & P & 12 \\ 
\enddata

\tablecomments{
Column (1) lists the cluster name. Names have
been truncated to at most eight characters to preserve the compactness
of the table. Columns (2)
and (3) give the equatorial coordinates of the cluster center used for the
regional selection for the epoch J2000 in decimal degrees. Column (4)
gives the heliocentric cluster redshift. Column (5) lists the column density
of neutral galactic hydrogen in units of $10^{20}\,\rm
atoms\,cm^{-2}$.
Column (6) gives the count rate in the
channel range 52--201 which corresponds to about (the energy
resolution of the \ps\ is limited) the energy range
$\eh$ in units of $\cts$.
Column (7) lists the relative 1-$\sigma$ Poissonian error of the count rate,
the flux, and the luminosity in
percent. Column (8) gives the significance radius in $\mpc$. Column (9)
lists the flux in the energy range $\eb$ in units of $\esct$. Column
(10) gives the luminosity in the energy range $\eb$ in units of $\esll$.
 Column (11) gives the bolometric luminosity (energy range $\ebol$) in
units of $\esll$. Column (12) indicates whether a \ra\ (R) or a
pointed (P) \ro\ \ps\ observation has been used. Column (13) lists the
code for the redshift reference decoded below.
}

\tablerefs{
(1) \citealt{1996A&A...310....8K}.
(2) \citealt{mkd96}. 
(3) Median of 9 galaxy redshifts compiled from \citealt{1989spce.book.....L,1991AJ....101..783M,1996ApJS..107..201L,1998ApJ...502..134W}.
(4) \citealt{aco89}.
(5) \citealt{1999ApJS..121..287H}.
(6) \citealt{sr87}.
(7) \citealt{1994AJ....107..427D}.
(8) \citealt{1993AJ....106.1273Z}.
(9) \citealt{1992A&AS...95..129P}.
(10) \citealt{1992NEDR....1....1N}.
(11) \citealt{bvh00}.
(12) \citealt{gbg99}.
(13) \citealt{1991RC3...C......0D}.
(14) \citealt{1990ApJS...74....1Z}.
(15) \citealt{1996MNRAS.279..349d}.
(16) \citealt{1988PASP..100.1423M}.
(17) \citealt{1995ApJ...454...44Y}.
(18) \citealt{1995A&A...297...56G}.
(19) \citealt{eeb98}.
(20) Median of 8 galaxy redshifts compiled from \citealt{1991RC3...C......0D,1995AJ....110..463Q,1998A&AS..129..399K}.
(21) \citealt{1990AJ.....99.1709V}.
(22) \citealt{1992MNRAS.259...67A}.
(23) \citealt{1990ApJS...72..715T}.
(24) \citealt{1998AJ....116....1D}.
(25) \citealt{1988AJ.....95..985D}.
(26) \citealt{1987RMxAA..14...72M}.
(27) Median of 13 galaxy redshifts compiled from \citealt{1983AJ.....88.1285K,1990ApJS...74....1Z,1992NEDR....1....1N,1995A&AS..110...19D,1995AJ....110...32O}.
(28) \citealt{1992ApJ...384..404P}.
(29) \citealt{1994AJ....108..361A}.
(30) \citealt{1996Natur.379..519K}.
(31) \citealt{1995A&A...293..337E}.
(32) \citealt{es91a}.
(33) \citealt{lef89}.
(34) \citealt{1976ApJ...206..364U}.
(35) \citealt{1991ApJS...76..813S}.
(36) \citealt{1992ApJ...399..353H}.
(37) \citealt{1991AJ....102.1581B}.
}

\end{deluxetable}

\subsection{Mass Determination}\label{massd}

A parametric description of the cluster gas density profile has been
derived using the standard
$\beta$ model \citep[e.g.,][]{cf76,gft78,jf84}. Assuming spherical symmetry the
model
\begin{equation}
\sx(R)=S_0\left(1+\frac{R^{2}}{\rc^{2}}\right)^{-3\beta+\frac{1}{2}}+B
\label{eq:sx}
\end{equation}
is fitted to the measured surface brightness profile (ring count
rates per ring area), where $R$ denotes the projected distance from the
cluster center. This yields values for the core radius, $\rc$, the $\beta$
parameter, and the normalization,
$S_0$ (and also a fitted value for the background surface brightness, $B$, since 
the fit is performed on the non background subtracted data). 
The fit values have been used to construct the radial gas density distribution
\begin{equation}
\rog(r)=\rho_0\left(1+\frac{r^{2}}{\rc^{2}}\right)^{-\frac{3}{2}\beta}.
\label{eq:rog}
\end{equation}

The gravitational cluster mass, $\mt$, has been determined assuming
the intracluster gas to be in hydrostatic equilibrium and isothermal.
Using (\ref{eq:rog}) and the ideal gas equation under these
assumptions leads to
\begin{equation}
\mt(<r)=\frac{3k\tg r^{3}\beta}{\mu \mpr G}\left(\frac{1}{r^2+\rc^2}\right),
\label{eq:mt}
\end{equation}
where $\mu\,(=0.61)$ represents the mean molecular weight, $\mpr$ the proton
mass, and $G$ the gravitational constant.
Combined $N$-body/hydrodynamic cluster simulations have
shown that this method generally gives unbiased results with an uncertainty
of 14--29\,\% \citep[e.g.,][]{s96,emn96}.
Currently there are contradictory measurements concerning the general presence of
gas temperature gradients in clusters
\citep[e.g.,][]{f97,mfs98,ibe99,w00,ib00}. If there is
a systematic trend in the sense that the gas temperature decreases
with increasing radius 
in the outer cluster parts similar to that found by \citeauthor{mfs98} then the
isothermal assumption leads to an overestimation of the cluster mass of about
30\,\% at about 6 core radii \citep{mfs98}.
\citet{frb00} determined masses by employing the assumption
of isothermality and also using measured cluster gas
temperature profiles. A comparison for 38 clusters
included in their sample indicates that the latter masses are on average a
factor of 0.80 smaller than the isothermal masses within $r_{200}$ (this
radius is defined in the next paragraph).
Until the final answer on this issue is
given by XMM-Newton\footnote{First results indicate that apart from
the central regions the
intracluster gas is isothermal out to at least $1/2\,r_{200}$
(e.g., \citealt{r01m}; M.\ Arnaud 2001, private
communication).}, we retain the isothermal assumption.
The influence of a possible overestimation of the cluster mass
on the determination of cosmological parameters is
investigated in Sect.~\ref{func_pred}.

Having determined the integrated mass as a function of radius, a
physically meaningful fiducial radius for the mass measurement has to
be defined.
The radii commonly used are either the Abell
radius, $r_{200}$, or $r_{500}$. The Abell radius is fixed at
$\rab\equiv 3\,\mpc$.
The radius $r_{200}$ ($r_{500}$) is the radius within which the mean
gravitational mass density
$\langle \rot \rangle = 200\, (500)\,\roc$. The
critical cosmic matter density is defined as $\roc \equiv 3\,H^2/(8\,\pi\,G)$, where
$H^2=H^2_0\,E(z)^2$ and $E(z)=[\om (1+z)^3+\ok (1+z)^2+\ol]^{1/2}$.
It has been shown that a correction for redshift
is not necessary for the nearby clusters included in \gcs\
\citep{frb00} and we use the zero redshift value for all calculations,
i.e.\ $\roc=4.6975\times 10^{-30}\rm g\,cm^{-3}$, unless noted otherwise.
Nevertheless the
influence of this approximation is tested in Sect.~\ref{func_pred} for
the model ($\om=1,\ol=0$), where evolution is strong. 

In order to treat clusters of different size in a homogeneous way we determine
the cluster mass at a characteristic density but also give the mass determined
formally at a fixed radius for comparison. Spherical collapse models
predict a cluster virial density
$\langle \rov \rangle \approx 178\,\roc$ for ($\om=1,\ol=0$), so a
pragmatic approximation to the virial mass is to use
$r_{200}$ as the outer boundary. Simulations performed by
\citet{emn96} have shown, however, that isothermal X-ray mass
measurements may be biased 
towards high masses for $ r > r_{500}$. Furthermore for
most of the clusters in \gcs\ (86\,\%) up to $r_{500}$ no extrapolation outside
the significantly 
detected cluster emission is necessary, i.e.\ $r_{500} < \rx$, whereas
the fraction is lower
for $r_{200}$ (25\,\%) and $\rab$ (17\,\%).
In summary the most accurate results are expected for $\mt (<r_{500}) \equiv
M_{500}$, but for a comparison to predicted mass functions $M_{200}$ is
the more appropriate value (Sect.~\ref{func_pred}).
Results for all determined masses and their corresponding radii are
given in Tab.~\ref{tbl:data2}. Masses for the cluster gas will be given in a
subsequent paper.

A major source of uncertainty comes from the temperature
measurements. However, this (statistical) error is less than 5\,\% for
one third of the clusters, therefore also other sources of error have
to be taken into account, in particular one cannot neglect the
uncertainties of the fit parameter values when assessing the statistical errors
of the mass measurements.
Therefore mass errors have been calculated by varying the
fit parameter values, $\beta$ and $\rc$, along their 68\,\% confidence level error
ellipse and using the upper and lower bound of the quoted temperature
ranges. The statistical mass error range has then been defined 
between the maximum and minimum mass.
Note that a simple error propagation applied to (\ref{eq:mt}) would
underestimate the uncertainty of $\mtz$ and $\mtf$, since $r_{200}$
and $r_{500}$ also
depend on $\tg$, $\beta$, and (weakly) $\rc$.
The individual mass errors have been used in
subsequent calculations, unless noted otherwise.\footnote{In log space errors are
transformed as 
$\Delta\log x=\log (e)\, (x^+-x^-)/(2\, x)$, where $x^+$ and 
$x^-$ denote the upper and lower boundary of the quantity's error range,
respectively.}
A mean statistical error of 23\,\% for clusters
included in \gcs\ and a mean error of 27\,\% for the extended
sample has been found.

\begin{deluxetable}{lccccccccc}
\tablecolumns{10} 
\tabletypesize{\footnotesize}
\tablecaption{Cluster properties \label{tbl:data2}}
\tablewidth{0pt}
\tablehead{
\colhead{Cluster}	& 
\colhead{$\beta$}	& \colhead{$\rc$}		& \colhead{$\tx$}		&
 \colhead{$\mtf$}	& \colhead{$r_{500}$}	&
\colhead{$\mtz$}	& \colhead{$r_{200}$}	& \colhead{$\mab$}		&  
\colhead{Ref}	\\
\colhead{(1)}	& \colhead{(2)}	& \colhead{(3)}	&
\colhead{(4)}	& \colhead{(5)}	& \colhead{(6)}	&
\colhead{(7)}	& \colhead{(8)}	& \colhead{(9)}	&
\colhead{(10)}
}
\startdata
A0085	& $0.532^{+0.004}_{-0.004}$ & $  83^{+  3}_{-  3}$ & $ 6.90^{+0.40}_{-0.40}$ & $ 6.84^{+ 0.66}_{- 0.66}$ & $1.68^{+0.05}_{-0.06}$ & $10.80^{+ 1.12}_{- 1.04}$ & $2.66^{+0.09}_{-0.09}$ & 12.21 & 1 \\[0.9mm] 
A0119	& $0.675^{+0.026}_{-0.023}$ & $ 501^{+ 28}_{- 26}$ & $ 5.60^{+0.30}_{-0.30}$ & $ 6.23^{+ 0.92}_{- 0.76}$ & $1.63^{+0.08}_{-0.07}$ & $10.76^{+ 1.50}_{- 1.39}$ & $2.66^{+0.11}_{-0.13}$ & 12.24 & 1 \\[0.9mm] 
A0133	& $0.530^{+0.004}_{-0.004}$ & $  45^{+  2}_{-  2}$ & $ 3.80^{+2.00}_{-0.90}$ & $ 2.78^{+ 2.51}_{- 0.95}$ & $1.24^{+0.30}_{-0.16}$ & $ 4.41^{+ 4.00}_{- 1.52}$ & $1.97^{+0.47}_{-0.27}$ & 6.71 & 9 \\[0.9mm] 
NGC507	& $0.444^{+0.005}_{-0.005}$ & $  19^{+  1}_{-  1}$ & $ 1.26^{+0.07}_{-0.07}$ & $ 0.41^{+ 0.04}_{- 0.04}$ & $0.66^{+0.02}_{-0.02}$ & $ 0.64^{+ 0.07}_{- 0.06}$ & $1.04^{+0.04}_{-0.04}$ & 1.86 & 2 \\[0.9mm] 
A0262	& $0.443^{+0.018}_{-0.017}$ & $  42^{+ 12}_{- 10}$ & $ 2.15^{+0.06}_{-0.06}$ & $ 0.90^{+ 0.10}_{- 0.09}$ & $0.86^{+0.03}_{-0.03}$ & $ 1.42^{+ 0.15}_{- 0.13}$ & $1.35^{+0.05}_{-0.04}$ & 3.17 & 2 \\[0.9mm] 
A0400	& $0.534^{+0.014}_{-0.013}$ & $ 154^{+  9}_{-  9}$ & $ 2.31^{+0.14}_{-0.14}$ & $ 1.28^{+ 0.17}_{- 0.15}$ & $0.96^{+0.04}_{-0.04}$ & $ 2.07^{+ 0.30}_{- 0.25}$ & $1.53^{+0.08}_{-0.06}$ & 4.10 & 2 \\[0.9mm] 
A0399	& $0.713^{+0.137}_{-0.095}$ & $ 450^{+132}_{-100}$ & $ 7.00^{+0.40}_{-0.40}$ & $10.00^{+ 3.73}_{- 2.48}$ & $1.91^{+0.21}_{-0.18}$ & $16.64^{+ 6.61}_{- 4.32}$ & $3.07^{+0.36}_{-0.30}$ & 16.24 & 1 \\[0.9mm] 
A0401	& $0.613^{+0.010}_{-0.010}$ & $ 246^{+ 11}_{- 10}$ & $ 8.00^{+0.40}_{-0.40}$ & $10.27^{+ 1.08}_{- 0.93}$ & $1.92^{+0.07}_{-0.05}$ & $16.59^{+ 1.62}_{- 1.62}$ & $3.07^{+0.09}_{-0.10}$ & 16.21 & 1 \\[0.9mm] 
A3112	& $0.576^{+0.006}_{-0.006}$ & $  61^{+  3}_{-  3}$ & $ 5.30^{+0.70}_{-1.00}$ & $ 5.17^{+ 1.17}_{- 1.45}$ & $1.53^{+0.11}_{-0.16}$ & $ 8.22^{+ 1.79}_{- 2.31}$ & $2.43^{+0.16}_{-0.25}$ & 10.16 & 1 \\[0.9mm] 
FORNAX	& $0.804^{+0.098}_{-0.084}$ & $ 174^{+ 17}_{- 15}$ & $ 1.20^{+0.04}_{-0.04}$ & $ 0.87^{+ 0.22}_{- 0.16}$ & $0.84^{+0.07}_{-0.06}$ & $ 1.42^{+ 0.36}_{- 0.27}$ & $1.35^{+0.11}_{-0.09}$ & 3.20 & 2 \\[0.9mm] 
2A0335	& $0.575^{+0.004}_{-0.003}$ & $  33^{+  1}_{-  1}$ & $ 3.01^{+0.07}_{-0.07}$ & $ 2.21^{+ 0.10}_{- 0.09}$ & $1.15^{+0.02}_{-0.02}$ & $ 3.51^{+ 0.16}_{- 0.15}$ & $1.83^{+0.03}_{-0.03}$ & 5.76 & 2 \\[0.9mm] 
IIIZw54	& $0.887^{+0.320}_{-0.151}$ & $ 289^{+124}_{- 73}$ & ($ 2.16^{+0.35}_{-0.30}$) & $ 2.36^{+ 2.22}_{- 0.90}$ & $1.18^{+0.29}_{-0.17}$ & $ 3.93^{+ 3.83}_{- 1.54}$ & $1.89^{+0.48}_{-0.29}$ & 6.32 & 11 \\[0.9mm] 
A3158	& $0.661^{+0.025}_{-0.022}$ & $ 269^{+ 20}_{- 19}$ & $ 5.77^{+0.10}_{-0.05}$ & $ 7.00^{+ 0.52}_{- 0.42}$ & $1.69^{+0.04}_{-0.03}$ & $11.29^{+ 0.95}_{- 0.68}$ & $2.69^{+0.08}_{-0.06}$ & 12.61 & 3 \\[0.9mm] 
A0478	& $0.613^{+0.004}_{-0.004}$ & $  98^{+  2}_{-  2}$ & $ 8.40^{+0.80}_{-1.40}$ & $11.32^{+ 1.78}_{- 2.81}$ & $1.99^{+0.10}_{-0.18}$ & $17.89^{+ 2.95}_{- 4.35}$ & $3.13^{+0.18}_{-0.27}$ & 17.12 & 1 \\[0.9mm] 
NGC1550	& $0.554^{+0.049}_{-0.037}$ & $  45^{+ 15}_{- 11}$ & $ 1.43^{+0.04}_{-0.03}$ & $ 0.69^{+ 0.12}_{- 0.09}$ & $0.78^{+0.04}_{-0.04}$ & $ 1.09^{+ 0.20}_{- 0.14}$ & $1.23^{+0.07}_{-0.06}$ & 2.64 & 5 \\[0.9mm] 
EXO0422	& $0.722^{+0.104}_{-0.071}$ & $ 142^{+ 40}_{- 30}$ & $ 2.90^{+0.90}_{-0.60}$ & $ 2.89^{+ 2.39}_{- 1.14}$ & $1.26^{+0.28}_{-0.19}$ & $ 4.63^{+ 3.84}_{- 1.82}$ & $2.00^{+0.44}_{-0.31}$ & 6.96 & 9 \\[0.9mm] 
A3266	& $0.796^{+0.020}_{-0.019}$ & $ 564^{+ 21}_{- 20}$ & $ 8.00^{+0.50}_{-0.50}$ & $14.17^{+ 1.94}_{- 1.84}$ & $2.14^{+0.09}_{-0.10}$ & $23.76^{+ 3.23}_{- 2.91}$ & $3.45^{+0.15}_{-0.14}$ & 20.47 & 1 \\[0.9mm] 
A0496	& $0.484^{+0.003}_{-0.003}$ & $  30^{+  1}_{-  1}$ & $ 4.13^{+0.08}_{-0.08}$ & $ 2.76^{+ 0.11}_{- 0.11}$ & $1.24^{+0.02}_{-0.02}$ & $ 4.35^{+ 0.18}_{- 0.17}$ & $1.96^{+0.03}_{-0.03}$ & 6.66 & 2 \\[0.9mm] 
A3376	& $1.054^{+0.101}_{-0.083}$ & $ 755^{+ 69}_{- 60}$ & $ 4.00^{+0.40}_{-0.40}$ & $ 6.32^{+ 2.11}_{- 1.59}$ & $1.64^{+0.17}_{-0.15}$ & $11.96^{+ 3.82}_{- 2.98}$ & $2.75^{+0.26}_{-0.25}$ & 13.20 & 1 \\[0.9mm] 
A3391	& $0.579^{+0.026}_{-0.024}$ & $ 234^{+ 24}_{- 22}$ & $ 5.40^{+0.60}_{-0.60}$ & $ 5.18^{+ 1.31}_{- 1.08}$ & $1.53^{+0.12}_{-0.11}$ & $ 8.41^{+ 2.13}_{- 1.81}$ & $2.44^{+0.19}_{-0.19}$ & 10.35 & 1 \\[0.9mm] 
A3395s	& $0.964^{+0.275}_{-0.167}$ & $ 604^{+173}_{-118}$ & $ 5.00^{+0.30}_{-0.30}$ & $ 8.82^{+ 4.79}_{- 2.61}$ & $1.83^{+0.29}_{-0.20}$ & $15.34^{+ 8.79}_{- 4.74}$ & $2.99^{+0.49}_{-0.35}$ & 15.42 & 1 \\[0.9mm] 
A0576	& $0.825^{+0.432}_{-0.185}$ & $ 394^{+221}_{-125}$ & $ 4.02^{+0.07}_{-0.07}$ & $ 5.36^{+ 4.42}_{- 1.66}$ & $1.55^{+0.34}_{-0.18}$ & $ 8.96^{+ 8.01}_{- 2.91}$ & $2.50^{+0.60}_{-0.31}$ & 10.86 & 3 \\[0.9mm] 
A0754	& $0.698^{+0.027}_{-0.024}$ & $ 239^{+ 17}_{- 16}$ & $ 9.50^{+0.70}_{-0.40}$ & $16.37^{+ 2.91}_{- 1.84}$ & $2.25^{+0.13}_{-0.09}$ & $26.19^{+ 4.45}_{- 2.95}$ & $3.57^{+0.18}_{-0.15}$ & 21.94 & 1 \\[0.9mm] 
HYDRA-A	& $0.573^{+0.003}_{-0.003}$ & $  50^{+  1}_{-  1}$ & $ 4.30^{+0.40}_{-0.40}$ & $ 3.76^{+ 0.58}_{- 0.55}$ & $1.38^{+0.07}_{-0.07}$ & $ 5.94^{+ 0.91}_{- 0.84}$ & $2.17^{+0.11}_{-0.10}$ & 8.21 & 1 \\[0.9mm] 
A1060	& $0.607^{+0.040}_{-0.034}$ & $  94^{+ 15}_{- 13}$ & $ 3.24^{+0.06}_{-0.06}$ & $ 2.66^{+ 0.34}_{- 0.28}$ & $1.23^{+0.05}_{-0.04}$ & $ 4.24^{+ 0.55}_{- 0.47}$ & $1.95^{+0.08}_{-0.08}$ & 6.54 & 2 \\[0.9mm] 
A1367	& $0.695^{+0.035}_{-0.032}$ & $ 383^{+ 24}_{- 22}$ & $ 3.55^{+0.08}_{-0.08}$ & $ 3.34^{+ 0.36}_{- 0.32}$ & $1.32^{+0.05}_{-0.04}$ & $ 5.69^{+ 0.63}_{- 0.56}$ & $2.14^{+0.08}_{-0.07}$ & 8.08 & 2 \\[0.9mm] 
MKW4	& $0.440^{+0.004}_{-0.005}$ & $  11^{+  1}_{-  1}$ & $ 1.71^{+0.09}_{-0.09}$ & $ 0.64^{+ 0.06}_{- 0.06}$ & $0.76^{+0.02}_{-0.03}$ & $ 1.00^{+ 0.10}_{- 0.09}$ & $1.20^{+0.04}_{-0.03}$ & 2.51 & 2 \\[0.9mm] 
ZwCl1215	& $0.819^{+0.038}_{-0.034}$ & $ 431^{+ 28}_{- 25}$ & ($ 5.58^{+0.89}_{-0.78}$) & $ 8.79^{+ 3.00}_{- 2.29}$ & $1.83^{+0.19}_{-0.18}$ & $14.52^{+ 4.92}_{- 3.67}$ & $2.93^{+0.30}_{-0.27}$ & 14.91 & 11 \\[0.9mm] 
NGC4636	& $0.491^{+0.032}_{-0.027}$ & $   6^{+  3}_{-  2}$ & $ 0.76^{+0.01}_{-0.01}$ & $ 0.22^{+ 0.03}_{- 0.02}$ & $0.53^{+0.02}_{-0.02}$ & $ 0.35^{+ 0.04}_{- 0.04}$ & $0.85^{+0.03}_{-0.03}$ & 1.24 & 4 \\[0.9mm] 
A3526	& $0.495^{+0.011}_{-0.010}$ & $  37^{+  5}_{-  4}$ & $ 3.68^{+0.06}_{-0.06}$ & $ 2.39^{+ 0.15}_{- 0.13}$ & $1.18^{+0.02}_{-0.02}$ & $ 3.78^{+ 0.23}_{- 0.18}$ & $1.87^{+0.04}_{-0.03}$ & 6.07 & 2 \\[0.9mm] 
A1644	& $0.579^{+0.111}_{-0.074}$ & $ 300^{+128}_{- 92}$ & $ 4.70^{+0.90}_{-0.70}$ & $ 4.10^{+ 2.64}_{- 1.41}$ & $1.42^{+0.26}_{-0.18}$ & $ 6.73^{+ 4.54}_{- 2.38}$ & $2.27^{+0.43}_{-0.31}$ & 8.98 & 10 \\[0.9mm] 
A1650	& $0.704^{+0.131}_{-0.081}$ & $ 281^{+104}_{- 71}$ & $ 6.70^{+0.80}_{-0.80}$ & $ 9.62^{+ 4.91}_{- 2.92}$ & $1.88^{+0.28}_{-0.21}$ & $15.60^{+ 8.08}_{- 4.85}$ & $3.01^{+0.45}_{-0.35}$ & 15.56 & 1 \\[0.9mm] 
A1651	& $0.643^{+0.014}_{-0.013}$ & $ 181^{+ 10}_{- 10}$ & $ 6.10^{+0.40}_{-0.40}$ & $ 7.45^{+ 1.00}_{- 0.95}$ & $1.73^{+0.07}_{-0.08}$ & $11.91^{+ 1.60}_{- 1.52}$ & $2.75^{+0.12}_{-0.13}$ & 13.01 & 1 \\[0.9mm] 
COMA	& $0.654^{+0.019}_{-0.021}$ & $ 344^{+ 22}_{- 21}$ & $ 8.38^{+0.34}_{-0.34}$ & $11.99^{+ 1.28}_{- 1.29}$ & $2.03^{+0.07}_{-0.08}$ & $19.38^{+ 2.08}_{- 1.97}$ & $3.22^{+0.11}_{-0.11}$ & 18.01 & 2 \\[0.9mm] 
NGC5044	& $0.524^{+0.002}_{-0.003}$ & $  11^{+  1}_{-  1}$ & $ 1.07^{+0.01}_{-0.01}$ & $ 0.41^{+ 0.01}_{- 0.01}$ & $0.66^{+0.01}_{-0.01}$ & $ 0.65^{+ 0.01}_{- 0.01}$ & $1.04^{+0.01}_{-0.01}$ & 1.87 & 2 \\[0.9mm] 
A1736	& $0.542^{+0.147}_{-0.092}$ & $ 374^{+178}_{-130}$ & $ 3.50^{+0.40}_{-0.40}$ & $ 2.19^{+ 1.23}_{- 0.74}$ & $1.15^{+0.18}_{-0.15}$ & $ 3.78^{+ 2.41}_{- 1.34}$ & $1.87^{+0.34}_{-0.25}$ & 6.22 & 1 \\[0.9mm] 
A3558	& $0.580^{+0.006}_{-0.005}$ & $ 224^{+  5}_{-  5}$ & $ 5.50^{+0.40}_{-0.40}$ & $ 5.37^{+ 0.70}_{- 0.64}$ & $1.55^{+0.07}_{-0.06}$ & $ 8.64^{+ 1.12}_{- 1.03}$ & $2.46^{+0.10}_{-0.10}$ & 10.56 & 1 \\[0.9mm] 
A3562	& $0.472^{+0.006}_{-0.006}$ & $  99^{+  5}_{-  5}$ & $ 5.16^{+0.16}_{-0.16}$ & $ 3.68^{+ 0.24}_{- 0.23}$ & $1.37^{+0.03}_{-0.03}$ & $ 5.83^{+ 0.38}_{- 0.36}$ & $2.16^{+0.05}_{-0.04}$ & 8.10 & 3 \\[0.9mm] 
A3571	& $0.613^{+0.010}_{-0.010}$ & $ 181^{+  7}_{-  7}$ & $ 6.90^{+0.20}_{-0.20}$ & $ 8.33^{+ 0.56}_{- 0.53}$ & $1.79^{+0.04}_{-0.04}$ & $13.31^{+ 0.90}_{- 0.85}$ & $2.85^{+0.06}_{-0.06}$ & 14.04 & 1 \\[0.9mm] 
A1795	& $0.596^{+0.003}_{-0.002}$ & $  78^{+  1}_{-  1}$ & $ 7.80^{+1.00}_{-1.00}$ & $ 9.75^{+ 2.01}_{- 1.90}$ & $1.89^{+0.12}_{-0.14}$ & $15.39^{+ 3.17}_{- 2.92}$ & $2.99^{+0.19}_{-0.20}$ & 15.46 & 1 \\[0.9mm] 
A3581	& $0.543^{+0.024}_{-0.022}$ & $  35^{+  5}_{-  4}$ & $ 1.83^{+0.04}_{-0.04}$ & $ 0.96^{+ 0.09}_{- 0.09}$ & $0.87^{+0.02}_{-0.03}$ & $ 1.52^{+ 0.16}_{- 0.13}$ & $1.38^{+0.05}_{-0.04}$ & 3.30 & 5 \\[0.9mm] 
MKW8	& $0.511^{+0.098}_{-0.059}$ & $ 107^{+ 70}_{- 42}$ & $ 3.29^{+0.23}_{-0.22}$ & $ 2.10^{+ 0.86}_{- 0.52}$ & $1.14^{+0.13}_{-0.10}$ & $ 3.33^{+ 1.45}_{- 0.83}$ & $1.79^{+0.24}_{-0.17}$ & 5.60 & 5 \\[0.9mm] 
A2029	& $0.582^{+0.004}_{-0.004}$ & $  83^{+  2}_{-  2}$ & $ 9.10^{+1.00}_{-1.00}$ & $11.82^{+ 2.14}_{- 1.99}$ & $2.01^{+0.11}_{-0.12}$ & $18.79^{+ 3.40}_{- 3.17}$ & $3.20^{+0.18}_{-0.19}$ & 17.62 & 1 \\[0.9mm] 
A2052	& $0.526^{+0.005}_{-0.005}$ & $  37^{+  2}_{-  2}$ & $ 3.03^{+0.04}_{-0.04}$ & $ 1.95^{+ 0.07}_{- 0.06}$ & $1.10^{+0.02}_{-0.01}$ & $ 3.10^{+ 0.09}_{- 0.11}$ & $1.75^{+0.01}_{-0.02}$ & 5.30 & 3 \\[0.9mm] 
MKW3S	& $0.581^{+0.008}_{-0.007}$ & $  66^{+  3}_{-  3}$ & $ 3.70^{+0.20}_{-0.20}$ & $ 3.06^{+ 0.32}_{- 0.30}$ & $1.29^{+0.05}_{-0.04}$ & $ 4.84^{+ 0.51}_{- 0.47}$ & $2.03^{+0.07}_{-0.07}$ & 7.16 & 1 \\[0.9mm] 
A2065	& $1.162^{+0.734}_{-0.282}$ & $ 690^{+360}_{-186}$ & $ 5.50^{+0.40}_{-0.40}$ & $13.44^{+16.12}_{- 5.17}$ & $2.10^{+0.63}_{-0.31}$ & $23.37^{+29.87}_{- 9.42}$ & $3.43^{+1.09}_{-0.54}$ & 20.21 & 1 \\[0.9mm] 
A2063	& $0.561^{+0.011}_{-0.011}$ & $ 110^{+  7}_{-  6}$ & $ 3.68^{+0.11}_{-0.11}$ & $ 2.84^{+ 0.23}_{- 0.19}$ & $1.25^{+0.04}_{-0.03}$ & $ 4.54^{+ 0.36}_{- 0.31}$ & $1.99^{+0.06}_{-0.04}$ & 6.86 & 2 \\[0.9mm] 
A2142	& $0.591^{+0.006}_{-0.006}$ & $ 154^{+  6}_{-  6}$ & $ 9.70^{+1.50}_{-1.10}$ & $13.29^{+ 3.45}_{- 2.41}$ & $2.10^{+0.17}_{-0.14}$ & $21.04^{+ 5.46}_{- 3.69}$ & $3.31^{+0.26}_{-0.20}$ & 19.05 & 1 \\[0.9mm] 
A2147	& $0.444^{+0.071}_{-0.046}$ & $ 238^{+103}_{- 65}$ & $ 4.91^{+0.28}_{-0.28}$ & $ 2.99^{+ 0.92}_{- 0.63}$ & $1.28^{+0.12}_{-0.10}$ & $ 4.84^{+ 1.64}_{- 1.03}$ & $2.03^{+0.21}_{-0.15}$ & 7.21 & 2 \\[0.9mm] 
A2163	& $0.796^{+0.030}_{-0.028}$ & $ 519^{+ 31}_{- 29}$ & $13.29^{+0.64}_{-0.64}$ & $31.85^{+ 4.24}_{- 3.74}$ & $2.81^{+0.12}_{-0.11}$ & $51.99^{+ 6.96}_{- 6.13}$ & $4.49^{+0.19}_{-0.18}$ & 34.18 & 3 \\[0.9mm] 
A2199	& $0.655^{+0.019}_{-0.021}$ & $ 139^{+ 10}_{- 10}$ & $ 4.10^{+0.08}_{-0.08}$ & $ 4.21^{+ 0.33}_{- 0.29}$ & $1.43^{+0.04}_{-0.03}$ & $ 6.73^{+ 0.52}_{- 0.51}$ & $2.27^{+0.06}_{-0.06}$ & 8.92 & 2 \\[0.9mm] 
A2204	& $0.597^{+0.008}_{-0.007}$ & $  67^{+  3}_{-  3}$ & $ 7.21^{+0.25}_{-0.25}$ & $ 8.67^{+ 0.67}_{- 0.57}$ & $1.82^{+0.05}_{-0.04}$ & $13.79^{+ 0.96}_{- 1.00}$ & $2.89^{+0.06}_{-0.08}$ & 14.34 & 3 \\[0.9mm] 
A2244	& $0.607^{+0.016}_{-0.015}$ & $ 126^{+ 11}_{- 10}$ & $ 7.10^{+5.00}_{-2.20}$ & $ 8.65^{+11.47}_{- 3.89}$ & $1.82^{+0.59}_{-0.33}$ & $13.78^{+18.02}_{- 6.20}$ & $2.89^{+0.92}_{-0.52}$ & 14.33 & 10 \\[0.9mm] 
A2256	& $0.914^{+0.054}_{-0.047}$ & $ 587^{+ 40}_{- 37}$ & $ 6.60^{+0.40}_{-0.40}$ & $12.83^{+ 2.38}_{- 2.00}$ & $2.07^{+0.12}_{-0.11}$ & $21.81^{+ 4.07}_{- 3.54}$ & $3.36^{+0.19}_{-0.20}$ & 19.34 & 1 \\[0.9mm] 
A2255	& $0.797^{+0.033}_{-0.030}$ & $ 593^{+ 35}_{- 32}$ & $ 6.87^{+0.20}_{-0.20}$ & $10.90^{+ 1.15}_{- 0.95}$ & $1.96^{+0.07}_{-0.05}$ & $18.65^{+ 2.01}_{- 1.67}$ & $3.18^{+0.11}_{-0.09}$ & 17.54 & 3 \\[0.9mm] 
A3667	& $0.541^{+0.008}_{-0.008}$ & $ 279^{+ 10}_{- 10}$ & $ 7.00^{+0.60}_{-0.60}$ & $ 6.88^{+ 1.08}_{- 1.02}$ & $1.68^{+0.08}_{-0.09}$ & $11.19^{+ 1.76}_{- 1.65}$ & $2.69^{+0.13}_{-0.14}$ & 12.50 & 1 \\[0.9mm] 
S1101	& $0.639^{+0.006}_{-0.007}$ & $  56^{+  2}_{-  2}$ & $ 3.00^{+1.20}_{-0.70}$ & $ 2.58^{+ 1.76}_{- 0.88}$ & $1.22^{+0.23}_{-0.16}$ & $ 4.08^{+ 2.78}_{- 1.39}$ & $1.92^{+0.36}_{-0.25}$ & 6.38 & 9 \\[0.9mm] 
A2589	& $0.596^{+0.013}_{-0.012}$ & $ 118^{+  8}_{-  7}$ & $ 3.70^{+2.20}_{-1.10}$ & $ 3.14^{+ 3.44}_{- 1.35}$ & $1.29^{+0.37}_{-0.22}$ & $ 5.01^{+ 5.41}_{- 2.15}$ & $2.06^{+0.56}_{-0.35}$ & 7.33 & 9 \\[0.9mm] 
A2597	& $0.633^{+0.008}_{-0.008}$ & $  58^{+  2}_{-  2}$ & $ 4.40^{+0.40}_{-0.70}$ & $ 4.52^{+ 0.72}_{- 1.11}$ & $1.47^{+0.07}_{-0.14}$ & $ 7.14^{+ 1.14}_{- 1.72}$ & $2.31^{+0.11}_{-0.20}$ & 9.27 & 1 \\[0.9mm] 
A2634	& $0.640^{+0.051}_{-0.043}$ & $ 364^{+ 44}_{- 39}$ & $ 3.70^{+0.28}_{-0.28}$ & $ 3.15^{+ 0.78}_{- 0.60}$ & $1.29^{+0.10}_{-0.09}$ & $ 5.35^{+ 1.34}_{- 1.04}$ & $2.10^{+0.17}_{-0.14}$ & 7.77 & 2 \\[0.9mm] 
A2657	& $0.556^{+0.008}_{-0.007}$ & $ 119^{+  5}_{-  5}$ & $ 3.70^{+0.30}_{-0.30}$ & $ 2.83^{+ 0.43}_{- 0.39}$ & $1.25^{+0.06}_{-0.06}$ & $ 4.52^{+ 0.68}_{- 0.62}$ & $1.99^{+0.10}_{-0.09}$ & 6.84 & 1 \\[0.9mm] 
A4038	& $0.541^{+0.009}_{-0.008}$ & $  59^{+  4}_{-  4}$ & $ 3.15^{+0.03}_{-0.03}$ & $ 2.16^{+ 0.09}_{- 0.08}$ & $1.14^{+0.02}_{-0.02}$ & $ 3.41^{+ 0.14}_{- 0.11}$ & $1.80^{+0.03}_{-0.01}$ & 5.67 & 3 \\[0.9mm] 
A4059	& $0.582^{+0.010}_{-0.010}$ & $  90^{+  5}_{-  5}$ & $ 4.40^{+0.30}_{-0.30}$ & $ 3.95^{+ 0.52}_{- 0.48}$ & $1.40^{+0.06}_{-0.06}$ & $ 6.30^{+ 0.83}_{- 0.76}$ & $2.22^{+0.09}_{-0.09}$ & 8.52 & 1 \\
\cutinhead{Clusters from the extended sample not included in \gcss .} 
A2734	& $0.624^{+0.034}_{-0.029}$ & $ 212^{+ 26}_{- 23}$ & ($ 3.85^{+0.62}_{-0.54}$) & $ 3.49^{+ 1.25}_{- 0.89}$ & $1.34^{+0.15}_{-0.12}$ & $ 5.67^{+ 1.98}_{- 1.48}$ & $2.14^{+0.22}_{-0.21}$ & 7.97 & 11 \\[0.9mm] 
A2877	& $0.566^{+0.029}_{-0.025}$ & $ 190^{+ 19}_{- 17}$ & $ 3.50^{+2.20}_{-1.10}$ & $ 2.61^{+ 3.32}_{- 1.24}$ & $1.22^{+0.39}_{-0.23}$ & $ 4.24^{+ 5.28}_{- 2.00}$ & $1.95^{+0.60}_{-0.38}$ & 6.57 & 10 \\[0.9mm] 
NGC499	& $0.722^{+0.034}_{-0.030}$ & $  24^{+  2}_{-  2}$ & $ 0.72^{+0.03}_{-0.02}$ & $ 0.36^{+ 0.05}_{- 0.04}$ & $0.63^{+0.03}_{-0.02}$ & $ 0.58^{+ 0.08}_{- 0.06}$ & $1.00^{+0.04}_{-0.04}$ & 1.73 & 4 \\[0.9mm] 
AWM7	& $0.671^{+0.027}_{-0.025}$ & $ 173^{+ 18}_{- 15}$ & $ 3.75^{+0.09}_{-0.09}$ & $ 3.79^{+ 0.38}_{- 0.32}$ & $1.38^{+0.05}_{-0.04}$ & $ 6.08^{+ 0.62}_{- 0.52}$ & $2.19^{+0.08}_{-0.06}$ & 8.35 & 2 \\[0.9mm] 
PERSEUS	& $0.540^{+0.006}_{-0.004}$ & $  64^{+  2}_{-  2}$ & $ 6.79^{+0.12}_{-0.12}$ & $ 6.84^{+ 0.29}_{- 0.26}$ & $1.68^{+0.02}_{-0.02}$ & $10.80^{+ 0.46}_{- 0.41}$ & $2.66^{+0.04}_{-0.04}$ & 12.20 & 2 \\[0.9mm] 
S405	& $0.664^{+0.263}_{-0.133}$ & $ 459^{+262}_{-159}$ & ($ 4.21^{+0.67}_{-0.59}$) & $ 3.91^{+ 3.56}_{- 1.57}$ & $1.40^{+0.33}_{-0.22}$ & $ 6.75^{+ 6.80}_{- 2.81}$ & $2.27^{+0.60}_{-0.37}$ & 9.09 & 11 \\[0.9mm] 
3C129	& $0.601^{+0.260}_{-0.131}$ & $ 318^{+178}_{-107}$ & $ 5.60^{+0.70}_{-0.60}$ & $ 5.68^{+ 5.58}_{- 2.29}$ & $1.58^{+0.40}_{-0.25}$ & $ 9.30^{+ 9.51}_{- 3.85}$ & $2.53^{+0.67}_{-0.42}$ & 11.08 & 9 \\[0.9mm] 
A0539	& $0.561^{+0.020}_{-0.018}$ & $ 148^{+ 13}_{- 12}$ & $ 3.24^{+0.09}_{-0.09}$ & $ 2.33^{+ 0.21}_{- 0.19}$ & $1.18^{+0.03}_{-0.03}$ & $ 3.74^{+ 0.35}_{- 0.34}$ & $1.87^{+0.05}_{-0.06}$ & 6.04 & 2 \\[0.9mm] 
S540	& $0.641^{+0.073}_{-0.051}$ & $ 130^{+ 38}_{- 29}$ & ($ 2.40^{+0.38}_{-0.34}$) & $ 1.83^{+ 0.83}_{- 0.54}$ & $1.08^{+0.14}_{-0.12}$ & $ 2.93^{+ 1.34}_{- 0.87}$ & $1.72^{+0.23}_{-0.19}$ & 5.13 & 11 \\[0.9mm] 
A0548w	& $0.666^{+0.194}_{-0.111}$ & $ 198^{+ 90}_{- 62}$ & ($ 1.20^{+0.19}_{-0.17}$) & $ 0.63^{+ 0.48}_{- 0.23}$ & $0.76^{+0.16}_{-0.11}$ & $ 1.06^{+ 0.84}_{- 0.41}$ & $1.23^{+0.26}_{-0.19}$ & 2.64 & 11 \\[0.9mm] 
A0548e	& $0.480^{+0.013}_{-0.013}$ & $ 118^{+ 12}_{- 11}$ & $ 3.10^{+0.10}_{-0.10}$ & $ 1.74^{+ 0.15}_{- 0.15}$ & $1.07^{+0.03}_{-0.04}$ & $ 2.77^{+ 0.27}_{- 0.23}$ & $1.68^{+0.06}_{-0.05}$ & 4.95 & 3 \\[0.9mm] 
A3395n	& $0.981^{+0.619}_{-0.244}$ & $ 672^{+383}_{-203}$ & $ 5.00^{+0.30}_{-0.30}$ & $ 8.70^{+ 9.53}_{- 3.19}$ & $1.82^{+0.51}_{-0.26}$ & $15.47^{+18.82}_{- 6.07}$ & $2.99^{+0.92}_{-0.46}$ & 15.55 & 1 \\[0.9mm] 
UGC03957	& $0.740^{+0.133}_{-0.086}$ & $ 142^{+ 45}_{- 33}$ & ($ 2.58^{+0.41}_{-0.36}$) & $ 2.51^{+ 1.50}_{- 0.83}$ & $1.20^{+0.21}_{-0.15}$ & $ 4.02^{+ 2.41}_{- 1.33}$ & $1.91^{+0.33}_{-0.23}$ & 6.35 & 11 \\[0.9mm] 
PKS0745	& $0.608^{+0.006}_{-0.006}$ & $  71^{+  2}_{-  2}$ & $ 7.21^{+0.11}_{-0.11}$ & $ 8.88^{+ 0.35}_{- 0.28}$ & $1.83^{+0.03}_{-0.01}$ & $14.12^{+ 0.56}_{- 0.53}$ & $2.91^{+0.04}_{-0.04}$ & 14.58 & 3 \\[0.9mm] 
A0644	& $0.700^{+0.011}_{-0.011}$ & $ 203^{+  7}_{-  7}$ & $ 7.90^{+0.80}_{-0.80}$ & $12.50^{+ 2.29}_{- 2.11}$ & $2.06^{+0.12}_{-0.12}$ & $19.83^{+ 3.79}_{- 3.23}$ & $3.24^{+0.21}_{-0.17}$ & 18.33 & 1 \\[0.9mm] 
S636	& $0.752^{+0.217}_{-0.123}$ & $ 344^{+130}_{- 86}$ & ($ 1.18^{+0.19}_{-0.17}$) & $ 0.61^{+ 0.44}_{- 0.22}$ & $0.75^{+0.15}_{-0.10}$ & $ 1.16^{+ 0.90}_{- 0.44}$ & $1.26^{+0.27}_{-0.18}$ & 2.93 & 11 \\[0.9mm] 
A1413	& $0.660^{+0.017}_{-0.015}$ & $ 179^{+ 12}_{- 11}$ & $ 7.32^{+0.26}_{-0.24}$ & $10.20^{+ 0.93}_{- 0.82}$ & $1.92^{+0.05}_{-0.05}$ & $16.29^{+ 1.49}_{- 1.31}$ & $3.05^{+0.09}_{-0.08}$ & 16.03 & 3 \\[0.9mm] 
M49	& $0.592^{+0.007}_{-0.007}$ & $  11^{+  1}_{-  1}$ & $ 0.95^{+0.02}_{-0.01}$ & $ 0.41^{+ 0.02}_{- 0.01}$ & $0.66^{+0.01}_{-0.01}$ & $ 0.65^{+ 0.04}_{- 0.02}$ & $1.04^{+0.02}_{-0.01}$ & 1.87 & 4 \\[0.9mm] 
A3528n	& $0.621^{+0.034}_{-0.030}$ & $ 178^{+ 17}_{- 16}$ & $ 3.40^{+1.66}_{-0.64}$ & $ 2.89^{+ 2.84}_{- 0.94}$ & $1.26^{+0.32}_{-0.15}$ & $ 4.65^{+ 4.54}_{- 1.48}$ & $2.00^{+0.51}_{-0.23}$ & 7.00 & 8 \\[0.9mm] 
A3528s	& $0.463^{+0.013}_{-0.012}$ & $ 101^{+  9}_{-  8}$ & $ 3.15^{+0.89}_{-0.59}$ & $ 1.69^{+ 0.87}_{- 0.50}$ & $1.05^{+0.16}_{-0.12}$ & $ 2.70^{+ 1.39}_{- 0.80}$ & $1.67^{+0.25}_{-0.19}$ & 4.86 & 8 \\[0.9mm] 
A3530	& $0.773^{+0.114}_{-0.085}$ & $ 421^{+ 75}_{- 61}$ & $ 3.89^{+0.27}_{-0.25}$ & $ 4.52^{+ 1.52}_{- 1.05}$ & $1.47^{+0.15}_{-0.13}$ & $ 7.64^{+ 2.72}_{- 1.80}$ & $2.36^{+0.26}_{-0.20}$ & 9.82 & 7 \\[0.9mm] 
A3532	& $0.653^{+0.034}_{-0.029}$ & $ 282^{+ 27}_{- 24}$ & $ 4.58^{+0.19}_{-0.17}$ & $ 4.77^{+ 0.70}_{- 0.52}$ & $1.49^{+0.07}_{-0.05}$ & $ 7.79^{+ 1.16}_{- 0.91}$ & $2.38^{+0.12}_{-0.10}$ & 9.88 & 7 \\[0.9mm] 
A1689	& $0.690^{+0.011}_{-0.011}$ & $ 163^{+  7}_{-  6}$ & $ 9.23^{+0.28}_{-0.28}$ & $15.49^{+ 1.18}_{- 1.00}$ & $2.20^{+0.06}_{-0.05}$ & $24.68^{+ 1.70}_{- 1.76}$ & $3.50^{+0.07}_{-0.10}$ & 21.13 & 3 \\[0.9mm] 
A3560	& $0.566^{+0.033}_{-0.029}$ & $ 256^{+ 30}_{- 27}$ & ($ 3.16^{+0.51}_{-0.44}$) & $ 2.16^{+ 0.79}_{- 0.56}$ & $1.14^{+0.12}_{-0.11}$ & $ 3.59^{+ 1.30}_{- 0.95}$ & $1.84^{+0.20}_{-0.18}$ & 5.92 & 11 \\[0.9mm] 
A1775	& $0.673^{+0.026}_{-0.023}$ & $ 260^{+ 19}_{- 18}$ & $ 3.69^{+0.20}_{-0.11}$ & $ 3.61^{+ 0.50}_{- 0.34}$ & $1.36^{+0.06}_{-0.05}$ & $ 5.91^{+ 0.83}_{- 0.56}$ & $2.17^{+0.09}_{-0.07}$ & 8.21 & 3 \\[0.9mm] 
A1800	& $0.766^{+0.308}_{-0.139}$ & $ 392^{+223}_{-132}$ & ($ 4.02^{+0.64}_{-0.56}$) & $ 4.75^{+ 4.64}_{- 1.85}$ & $1.49^{+0.38}_{-0.23}$ & $ 7.97^{+ 8.31}_{- 3.17}$ & $2.39^{+0.65}_{-0.37}$ & 10.08 & 11 \\[0.9mm] 
A1914	& $0.751^{+0.018}_{-0.017}$ & $ 231^{+ 11}_{- 10}$ & $10.53^{+0.51}_{-0.50}$ & $21.43^{+ 2.39}_{- 2.16}$ & $2.46^{+0.09}_{-0.08}$ & $33.99^{+ 4.06}_{- 3.43}$ & $3.88^{+0.16}_{-0.13}$ & 26.20 & 3 \\[0.9mm] 
NGC5813	& $0.766^{+0.179}_{-0.103}$ & $  25^{+  9}_{-  6}$ & ($ 0.52^{+0.08}_{-0.07}$) & $ 0.24^{+ 0.17}_{- 0.08}$ & $0.55^{+0.11}_{-0.07}$ & $ 0.38^{+ 0.27}_{- 0.13}$ & $0.87^{+0.16}_{-0.12}$ & 1.32 & 11 \\[0.9mm] 
NGC5846	& $0.599^{+0.016}_{-0.015}$ & $   7^{+  1}_{-  1}$ & $ 0.82^{+0.01}_{-0.01}$ & $ 0.33^{+ 0.02}_{- 0.02}$ & $0.61^{+0.01}_{-0.01}$ & $ 0.53^{+ 0.03}_{- 0.03}$ & $0.97^{+0.02}_{-0.01}$ & 1.63 & 4 \\[0.9mm] 
A2151w	& $0.564^{+0.014}_{-0.013}$ & $  68^{+  5}_{-  5}$ & $ 2.40^{+0.06}_{-0.06}$ & $ 1.52^{+ 0.12}_{- 0.10}$ & $1.02^{+0.03}_{-0.02}$ & $ 2.42^{+ 0.18}_{- 0.18}$ & $1.61^{+0.03}_{-0.04}$ & 4.51 & 3 \\[0.9mm] 
A3627	& $0.555^{+0.056}_{-0.044}$ & $ 299^{+ 56}_{- 49}$ & $ 6.02^{+0.08}_{-0.08}$ & $ 5.63^{+ 0.95}_{- 0.68}$ & $1.57^{+0.09}_{-0.06}$ & $ 9.20^{+ 1.61}_{- 1.16}$ & $2.51^{+0.14}_{-0.10}$ & 11.03 & 3 \\[0.9mm] 
TRIANGUL	& $0.610^{+0.010}_{-0.010}$ & $ 279^{+ 11}_{- 10}$ & $ 9.60^{+0.60}_{-0.60}$ & $13.42^{+ 1.70}_{- 1.55}$ & $2.10^{+0.09}_{-0.09}$ & $21.54^{+ 2.73}_{- 2.36}$ & $3.34^{+0.14}_{-0.11}$ & 19.35 & 1 \\[0.9mm] 
OPHIUCHU	& $0.747^{+0.035}_{-0.032}$ & $ 279^{+ 23}_{- 22}$ & $10.26^{+0.32}_{-0.32}$ & $20.25^{+ 2.51}_{- 2.10}$ & $2.41^{+0.10}_{-0.08}$ & $32.43^{+ 4.05}_{- 3.38}$ & $3.83^{+0.16}_{-0.13}$ & 25.32 & 2 \\[0.9mm] 
ZwCl1742	& $0.717^{+0.073}_{-0.053}$ & $ 232^{+ 46}_{- 38}$ & ($ 5.23^{+0.84}_{-0.73}$) & $ 6.88^{+ 3.06}_{- 1.96}$ & $1.68^{+0.22}_{-0.18}$ & $11.05^{+ 4.93}_{- 3.16}$ & $2.67^{+0.35}_{-0.28}$ & 12.42 & 11 \\[0.9mm] 
A2319	& $0.591^{+0.013}_{-0.012}$ & $ 285^{+ 15}_{- 14}$ & $ 8.80^{+0.50}_{-0.50}$ & $11.16^{+ 1.39}_{- 1.20}$ & $1.97^{+0.08}_{-0.07}$ & $18.07^{+ 2.12}_{- 2.06}$ & $3.16^{+0.11}_{-0.13}$ & 17.17 & 1 \\[0.9mm] 
A3695	& $0.642^{+0.259}_{-0.117}$ & $ 399^{+254}_{-149}$ & ($ 5.29^{+0.85}_{-0.74}$) & $ 5.57^{+ 5.29}_{- 2.16}$ & $1.57^{+0.39}_{-0.24}$ & $ 9.32^{+ 9.56}_{- 3.74}$ & $2.53^{+0.67}_{-0.40}$ & 11.12 & 11 \\[0.9mm] 
IIZw108	& $0.662^{+0.167}_{-0.097}$ & $ 365^{+159}_{-105}$ & ($ 3.44^{+0.55}_{-0.48}$) & $ 2.96^{+ 2.00}_{- 1.02}$ & $1.27^{+0.24}_{-0.16}$ & $ 5.04^{+ 3.60}_{- 1.80}$ & $2.06^{+0.40}_{-0.28}$ & 7.47 & 11 \\[0.9mm] 
A3822	& $0.639^{+0.150}_{-0.093}$ & $ 351^{+160}_{-111}$ & ($ 4.90^{+0.78}_{-0.69}$) & $ 4.97^{+ 3.30}_{- 1.75}$ & $1.51^{+0.28}_{-0.20}$ & $ 8.26^{+ 5.64}_{- 3.02}$ & $2.43^{+0.46}_{-0.34}$ & 10.29 & 11 \\[0.9mm] 
A3827	& $0.989^{+0.410}_{-0.192}$ & $ 593^{+248}_{-149}$ & ($ 7.08^{+1.13}_{-0.99}$) & $16.35^{+17.02}_{- 6.76}$ & $2.25^{+0.60}_{-0.37}$ & $27.44^{+29.53}_{-11.46}$ & $3.62^{+0.99}_{-0.60}$ & 22.44 & 11 \\[0.9mm] 
A3888	& $0.928^{+0.084}_{-0.066}$ & $ 401^{+ 46}_{- 40}$ & ($ 8.84^{+1.41}_{-1.24}$) & $22.00^{+ 9.28}_{- 6.28}$ & $2.48^{+0.31}_{-0.26}$ & $35.74^{+15.07}_{-10.38}$ & $3.96^{+0.49}_{-0.44}$ & 26.85 & 11 \\[0.9mm] 
A3921	& $0.762^{+0.036}_{-0.030}$ & $ 328^{+ 26}_{- 23}$ & $ 5.73^{+0.24}_{-0.23}$ & $ 8.46^{+ 1.13}_{- 0.96}$ & $1.80^{+0.08}_{-0.07}$ & $13.80^{+ 1.87}_{- 1.59}$ & $2.89^{+0.12}_{-0.12}$ & 14.37 & 3 \\[0.9mm] 
HCG94	& $0.514^{+0.007}_{-0.006}$ & $  86^{+  4}_{-  4}$ & $ 3.45^{+0.30}_{-0.30}$ & $ 2.28^{+ 0.36}_{- 0.34}$ & $1.17^{+0.06}_{-0.06}$ & $ 3.62^{+ 0.56}_{- 0.51}$ & $1.84^{+0.09}_{-0.09}$ & 5.90 & 6 \\[0.9mm] 
RXJ2344	& $0.807^{+0.033}_{-0.030}$ & $ 301^{+ 20}_{- 18}$ & ($ 4.73^{+0.76}_{-0.66}$) & $ 6.91^{+ 2.30}_{- 1.69}$ & $1.68^{+0.17}_{-0.14}$ & $11.27^{+ 3.74}_{- 2.80}$ & $2.69^{+0.27}_{-0.25}$ & 12.58 & 11 \\[0.9mm] 
\enddata

\tablecomments{
Column (1) lists the cluster name. Column (2)
gives the $\beta$ parameter value and the corresponding 68\,\% c.l.\
statistical uncertainty for two interesting parameters. Column (3)
gives the core radius in $\kpc$ and the corresponding
uncertainty. Column (4) lists the X-ray temperature along with its
error.
For some references the temperature uncertainty is quoted at the 90\,\%
confidence level and therefore represents a conservative error estimate.
Columns (5) and (7) give $\mtf$ and $\mtz$ and their 
uncertainties in units of $10^{14} \msu$,
calculated as described in Sect.~\ref{massd}.
Columns (6) and (8) list
$r_{500}$ and $r_{200}$ and their uncertainties in
$\mpc$.
Column (9) gives $\mab\equiv \mt(<\rab)$ in units of $10^{14} \msu$. 
Column (10) lists the
code for the temperature reference decoded below.
Temperatures for codes 1--7 have been determined with
\as , code 8 with \ro , code 9 with EXOSAT, code 10 with Einstein, and
code 11 with a \ro --\as\ $\lx$--$\tx$ relation. Temperatures for code
11 are enclosed in parentheses and the corresponding errors have been
calculated using the scatter in the $\lx$--$\tx$ relation.
}

\tablerefs{
(1) \citealt{mfs98}.
(2) \citealt{fmt98}.
(3) \citealt{w00}.
(4) \citealt{m97}.
(5) \citealt{irb01}.
(6) \citealt{frb00}.
(7) This work.
(8) \citealt{s96a}.
(9) \citealt{es91a}.
(10) \citealt{dsj93}.
(11) Estimated from the $\lx$--$\tx$ relation given by \citealt{m98}.
}

\end{deluxetable}

\section{Results}\label{resul}

In this Section it is shown that a tight correlation exists between the
gravitational cluster mass and the X-ray luminosity. This ensures that
\gcs\ is essentially selected by cluster mass. In the second part of
this Section the cluster mass function is presented, including the
proper treatment of the scatter in the $\lx$--$\mt$ relation.

\subsection{Mass--Luminosity Relation}\label{relat}

\begin{figure}[thbp]
\psfig{file=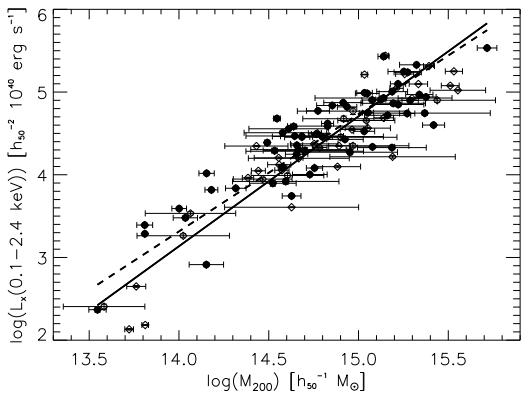,width=8cm,angle=0,clip=}
\caption{Gravitational mass--X-ray luminosity relation (solid line) for the
extended sample of 106 galaxy clusters. The dashed line gives the best
fit relation for the 63 clusters included in \gcs\ (filled circles only). The bisector fit
results are shown.
One-$\sigma$
statistical error bars are plotted for both axes, however, only the mass errors
are larger than the symbol sizes.}\label{mtlx}
\end{figure}

Since
the aim is the construction of a mass function from a flux-limited sample
it is now important to
test for a correlation between X-ray luminosity and gravitational mass. In
Fig.~\ref{mtlx} $\lx$, given in the \ro\ energy band, is plotted as a
function of $\mtz$, showing clearly the existence of a tight (linear
Pearson correlation coefficient $=$ 0.92) correlation, as expected.

To quantify the mass--luminosity relation, a linear regression
fit in log--log space has 
been performed. The method used allows for intrinsic scatter and errors in both
variables \citep{ab96}.
Tables~\ref{tab:lm1}--\ref{tab:lm5} in the appendix give the results for
different fit methods, where
minimization has been performed in vertical, horizontal, and
orthogonal direction, and the bisector result is given, which bisects the best
fit results of vertical and horizontal minimization.
The fits have been performed using the form
\begin{equation}
\log \left(\frac{\lx (\eb )}{\esl
}\right)=A+\alpha\log\left(\frac{\mtz}{\msu}\right)\,.
\label{eq:lmform}
\end{equation}
We find, as noted in general by previous authors (e.g., \citealt{ifa90}), that the
chosen fitting method has a significant influence on the best fit
parameter values.\footnote{This also implies that for a proper comparison of relations which have been
quantified by many different authors, e.g., the $\lx$--$\tx$ relation,
one and the same fitting statistic ought to be
used
(e.g., \citealt{wxf99}).}
In this work the appropriate relation for the application under
consideration is always indicated.

The difference between the fit results for 63 and 106
clusters may indicate a scale dependence of the $\lx$--$\mt$ relation,
since the difference is slightly larger than the uncertainty evaluated
with the bootstrap method. The small number of low luminosity clusters
in \gcs\ compared to the extended sample may be responsible for the
less steep relation obtained using the \gcs\ clusters only.
Note that only two out of the six clusters with $\lx<\esls$ are included
in \gcs .
To reliably detect any deviations from the power law shape of the
$\lx$--$\mt$ relation, however, more clusters with $\mt <
10^{14}\,h_{50}^{-1}\,M_{\odot}$ (and possibly $\mt >
3\times 10^{15}\,h_{50}^{-1}\,M_{\odot}$) need to be sampled. Such
work is in progress.
As will be seen later, in
the procedure used here for the comparison
of observed and predicted mass functions the precise
shape of the $\lx$--$\mt$ relation is not important.

When constructing the mass function the overall (measurement plus
intrinsic) scatter in the 
$\lx$--$\mt$ relation may become important (Sect.~\ref{funct}). After verifying that
the scatter is approximately Gaussian in log space the scatter has been
measured as given in Tab.~\ref{tab:scat} in the Appendix.
The scatter in $\log(\lx(\eb))$, $\log(\mtz)$, and orthogonal to the best fit line is
given by $\sll$, $\slm$, and $\sigma_{\log L/M}$, respectively.

\subsection{Mass Function}\label{funct}

The commonly used definition of the galaxy cluster mass function is analogous to the
definition of the luminosity function (e.g., \citealt{s76}):
the mass function, $\phi(M)$, denotes the number of
clusters, $N$, per unit comoving volume, $dV$, per unit mass in the interval
$[M,M+dM]$, i.e.\ $\phi(M)\equiv N(M)/(dV\,dM)\equiv
dn(M)/dM$.
Assuming constant density the
classical $\vmax$ estimator
\citep[e.g.,][]{s68,f76,bst88} can be used for estimation of
luminosity functions, i.e.\
$\hat\phi(L)=1/\Delta L\,\sum^{N}_{i=1}\,\frac{1}{\vmaxi}$. $\vmax$ is the
maximum comoving volume within which a cluster with given luminosity for a
given survey flux limit and sky coverage could have been detected.
As mentioned in Sect.~\ref{sample} the \gcs\ flux limit is constant over 99\,\% of
the covered area, which simplifies the calculation of $\vmax$.

In the previous Section it has been shown that the X-ray luminosity is
closely correlated with cluster mass. Therefore the $\vmax$ estimator can
also be applied to estimate the mass function; $\vmax$ then being a
function of mass.
We employ different methods to correct for the scatter
present in the $\lx$--$\mt$ relation. If $\vmax (\lx)$ is used instead
of $\vmax (\mt)\equiv \vmax (L(\mt))$, where $L(\mt)$ is
the luminosity estimated from the $\lx$--$\mt$ relation using the
determined cluster mass $\mt$, the scatter is automatically taken into
account. This method has been widely used in the construction of X-ray
temperature functions, recently, e.g., by \citet{h00}.
If $\vmax (\mt)$ is used and the utilized $\lx$--$\mt$ relation is
assumed to be the `true' relation then the scatter in this relation
has to be taken into account explicitly.
Therefore following the
method employed for the temperature function by
\citet{m98} and \citet{irb01} the mass function may also be estimated
by determining
$\vmax^\ast (\mt)$, where the measured scatter in $\log\lx$ is included.
Specifically we use 
\begin{eqnarray}
\vmax^\ast (\mt)\equiv \int^\infty_{-\infty}\vmax(L')\,(2\pi\sll^2)^{-1/2}
\nonumber \\
\times\ \exp\left({-\frac{(\log L'-(A+40)
-\alpha\log\mt)^2}{2\sll^2}}\right)\,d\log L'\,,
\label{eq:vmax}
\end{eqnarray}
where $A$ and $\alpha$ are the best fit parameter
values taken from the appropriate $\lx$--$\mt$ relation of the form
(\ref{eq:lmform}) and $\sll$ is the
corresponding measured
standard deviation in $\log \lx$ given in Tab.~\ref{tab:scat}.
However, we can also use the measured $\lx$--$\mt$ relation directly,
i.e.\ $\vmax(\mt)$, taking 
advantage of the fact that in our flux-limited sample there are fewer
low luminosity clusters for a given mass than high luminosity ones, which results in a slightly
increased normalization of the relation. Therefore using this relation
directly, the effect of the scatter and the resulting bias towards
higher luminosity clusters is already included and thus directly
accounted for.

The drawback of using $\vmax(\lx)$ is that a small number of clusters per
mass bin possibly does not represent the true scatter well. To
minimize this effect we use at least ten clusters per mass bin. The
drawback of using $\vmax^\ast(\mt)$ or $\vmax(\mt)$, as noted, e.g., by
\citet{m98} and \citet{ecf98}, is the
reliance on the validity of the measured relation over the
entire mass range.
The first method and the method that accounts for the scatter
explicitly (eq.~\ref{eq:vmax}) have been tested by using Monte Carlo simulations for a
precisely known $\lx$--$\tx$ relation
and scatter  and have been shown to give accurate estimates of
$\phi(T)$ for a large number of clusters
in the study of the \gcs\ temperature function by \citet{irb01}.

\begin{figure}[thp]
\psfig{file=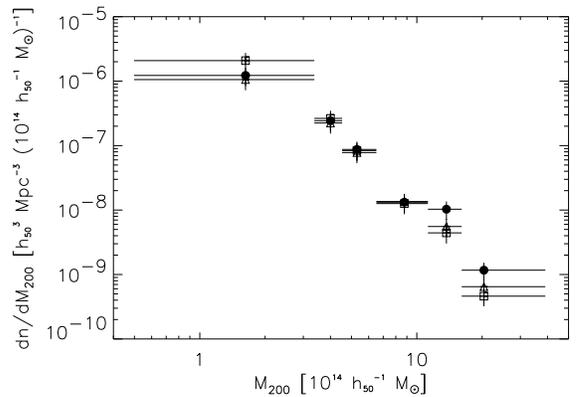,width=8cm,angle=0,clip=}
\caption{Gravitational mass functions for \gcs .
The mass function plotted with filled circles has been determined
using $\vmax(\lx)$, the ones with open symbols using $\vmax(\mt)$, where 
triangles correspond to the relation ($L\mid M$) and squares to ($M\mid
L$)
both for the flux-limited sample (see text Sect.~\ref{funct}). Vertical error bars
correspond to the formal 1-$\sigma$ Poisson errors, the horizontal
bars indicate the mass intervals covered. Each bin contains 10
clusters, apart from the highest mass bin, which contains 11
clusters. The highest and lowest mass clusters have been used to
calculate the highest and lowest mass intervals.}\label{vm_cp_ns}
\end{figure}

In Fig.~\ref{vm_cp_ns} \gcs\ mass functions are shown.
As expected the method
employing $\vmax(\lx)$ prompts a mass function exhibiting a larger
scatter, because in this case the scatter is accounted for by the
actual scatter of the ten or eleven clusters in each mass bin.
For comparison the two extreme mass functions calculated using $\vmax
(\mt)$ are shown. Extreme is meant in the sense of using the steepest (A)
and shallowest (B) $\lx$--$\mt$ relation for the \gcs\ sample, i.e.,
($M\mid L$) with $\alpha=1.538$ and ($L\mid M$) with $\alpha=1.310$
(Tab.~\ref{tab:lm2}). At the low mass end (A) predicts a lower
luminosity for a given mass than (B) resulting in a smaller $\vmax$ and
therefore a higher $dn/dM$. At the high mass side the effect is
opposite resulting in a lower $dn/dM$ for (A). The differences of these mass
functions to the mass function calculated using $\vmax (\lx)$ can be
understood in a similar way and are caused partly by the indication of a
deviation from a power law shape of the $\lx$--$\mt$ relation.
Using $\vmax^\ast(\mt)$ results in similar mass functions as shown for
the open symbols in Fig.~\ref{vm_cp_ns} but the points lie systematically
lower because the scatter is accounted for twice.
For the comparison of the
observational mass function to mass functions predicted by certain
cosmological models $\vmax(\lx)$ is used because it is independent of
the precise shape of the $\lx$--$\mt$ relation
and also because $\lx$ has a much smaller measurement uncertainty than
$\mt$. The influence of the choice of the $\vmax$ calculation on the
estimation of cosmological parameters is investigated in
Sect.~\ref{func_pred}. 

\section{Discussion}\label{discu}

A precise determination of distribution functions requires a high
sample completeness.
In Sect.~\ref{sample_c} several completeness tests
for \gcs\ are discussed, indicating a high completeness. The observed
$\lx$--$\mt$ relation is compared to expectations in
Sect.~\ref{relat_d} and possible applications are indicated. The
cluster mass function is compared to previous determinations and to
predictions of cosmological models in
Sect.~\ref{funct_d}.
The total gravitational mass contained in galaxy
clusters is compared to the total mass in the universe
in Sect.~\ref{func_dens}.

\subsection{Sample Completeness}\label{sample_c}

The sample completeness is important for the accuracy of the mass
function. 
The selection criteria detailed in Sect.~\ref{sample} are met by 63
clusters with mean redshift $\langle z \rangle
= 0.05$ and with two  clusters having $z>0.1$. 
The sample is constructed from surveys with much deeper flux limits
and high completenesses. A possible remaining incompleteness in these surveys is
likely to be present at low fluxes close to their flux limits, which
therefore would not effect \gcs .
Nevertheless
four completeness tests have been performed and are described in this
Section; they all
indicate a high completeness of \gcs . The $\log N$--$\log S$ and
$\lx$--$z$ diagram are compared to 
expectations, the luminosity function is compared to luminosity
functions of deeper surveys, and the $V/\vmax$ test is performed.

\begin{figure}[thbp]
\psfig{file=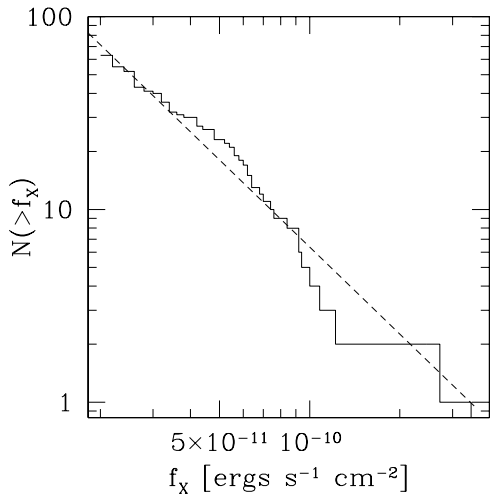,width=7cm,angle=0,clip=}
\caption{$\log N(>\fx)$--$\log \fx$ diagram. Fluxes are measured in the \ro\
energy band ($\eb$). The dashed line has a slope $-1.5$, expected for a uniform
distribution of clusters in a static Euclidean universe
(`three-halves-law'), the line is normalized to produce the same cluster number
at $\fx = 8\esc$.}\label{lnls}
\end{figure}
Figure~\ref{lnls} shows the integral number counts as a function of
X-ray flux (`$\log N$--$\log S$'). The slope in the $\log N$--$\log S$
diagram is very close to the value $-1.5$ expected in a static Euclidean
universe for uniformly distributed clusters.
Due to the small number of clusters (4) the
deviation is not significant for $\fx \gtrsim 1\escc$. 
Since the average redshift is
smallest for the highest fluxes large scale structure is
not completely washed out at the high flux end, therefore the slight bump visible around
$\fx \sim 6\esc$ in Fig.~\ref{lnls} suggests a deviation caused by cosmic
variance. The effect of an
expanding and finite universe on the $\log N$--$\log S$ -- flattening of the slope
towards low
fluxes -- is small for the redshift range covered by the sample. The
slope consistent with $-1.5$ towards the flux limit therefore
indicates a high
completeness of \gcs .

\begin{figure}[thp]
\psfig{file=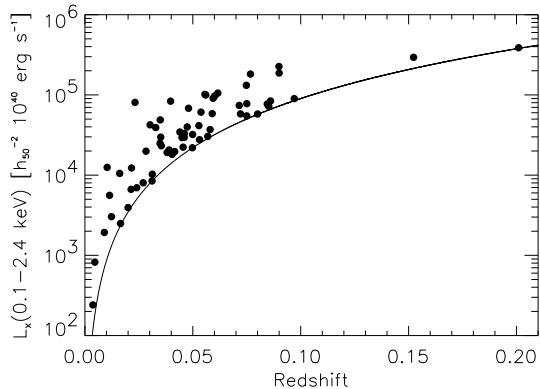,width=8cm,angle=0,clip=}
\caption{X-ray luminosity as a function of redshift. The flux limit is shown as
a solid line.}\label{lz}
\end{figure}

In Fig.~\ref{lz} the X-ray luminosity is plotted as a function of
redshift. The flux limit is shown as a solid line.\footnote{The correction
$K(\tg,z)$ for converting observer rest
frame luminosities to source rest frame luminosities depends on redshift
\emph{and}
source spectrum ($\tg$). For source rest frame luminosities it is therefore not
possible to plot the flux limit as one line in 2 dimensions ($\lx ,z$), but
rather as an
area in 3 dimensions ($\lx ,z,\tg$). For consistency we therefore give in this
2d plot the observer rest
frame luminosity (the correction is less than 6\,\% for 90\,\% of the clusters
anyway).} 
One notes the increase in rare luminous systems with increasing redshift
(volume).
Because of the seeming underdensity of clusters in the redshift range
$0.10<z<0.15$ a comparison with the expected number of
clusters as derived from $N$-body simulations has been performed.
An OCDM simulation,
carried out for analysis of the power spectral densities of REFLEX clusters
\citep{sbg00a}, adjusted to the \gcs\ survey volume in the southern
hemisphere (roughly half of the total volume sampled by \gcs ) has
been used.
The simulation yields 39 clusters while 33 \gcs\ clusters have been
detected in this region.
It is found that in fact not even one cluster with $z>0.1$ is expected for
this volume based on this
simulation and the \gcs\ subsample also does not contain any cluster with
a redshift larger than 0.1.
This is a further piece of evidence for the high completeness of the
sample. 

\begin{figure}[thbp]
\psfig{file=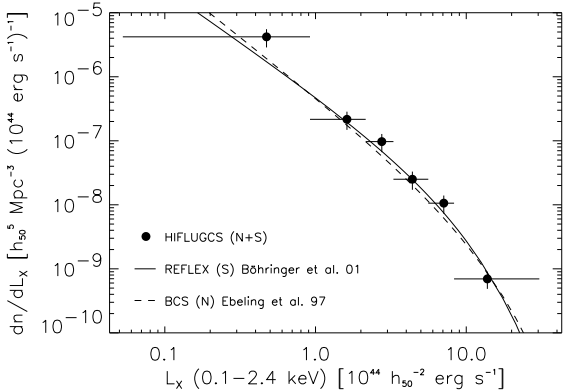,width=8cm,angle=0,clip=}
\caption{X-ray luminosity function for \gcs\ compared to surveys with deeper
flux limits in the northern (N) and southern (S) hemisphere. Vertical error bars
correspond to the formal 1-$\sigma$ Poisson errors (no
cosmic variance), the horizontal bars indicate the
luminosity intervals
covered.}\label{lxfunc}
\end{figure}

In Fig.~\ref{lxfunc} the \gcs\ X-ray luminosity function is compared to 
luminosity functions of larger surveys in the southern
(REFLEX,  \citealt{bcg01}) and 
northern (BCS, \citealt{eef97}) hemisphere. These surveys have much deeper flux
limits (Sect.~\ref{sample}) and contain many more clusters. Very good
agreement is found, which shows the high
completeness and homogeneous selection of \gcs .

The $V/\vmax$ test (e.g., \citealt{r68,s68,av80,p99}, Sect.~14.5) can
be used to asses a possible sample incompleteness.
Assuming a uniform distribution of clusters a value 1/2 is expected on
average.
For \gcs\ $\langle V/\vmax \rangle = 0.47\pm 0.04$, which is
consistent with the 
expectation and we interpret this result as a clear sign that \gcs\
covers a large enough volume for most of the $\lx$ range to be
representative of the local
universe with a high sample completeness. The local nature of \gcs\
becomes obvious by noting that the result of the comoving $V/\vmax$ test
is almost  identical to the result of the equivalent test assuming a
Euclidean and 
non expanding space, i.e., $\langle (\fx/\fxl)^{-3/2} \rangle = 0.46$.

\subsection{Mass--Luminosity Relation}\label{relat_d}

The close correlation between the X-ray luminosity and the
gravitational mass found in Sect.~\ref{relat} is not
surprising. Simple self similar scaling relations predict $\tg\propto
\mt \rch^{-1}$ and $\mt\propto \rch^3$, where $\rch$ is a
characteristic radius, e.g., the virial radius. Combined with bremsstrahlung
emission, $\lbol\propto\rog^2 T^{1/2} \rch^3$ ($\lx(\eb)\propto\rog^2
\rch^3$), the relation $\lbol\propto\fg^2 \mt^{4/3}$
($\lx(\eb)\propto\fg^2 \mt$) is predicted \citep[e.g.,][]{pe80}, where the gas fraction
$\fg\equiv\mg\mt^{-1}$ and $\lbol$ is the bolometric luminosity.

Observationally from the tight correlations between X-ray luminosity
and temperature (e.g., \citealt{m98}), and temperature and mass (e.g.,
\citealt{frb00}) a correlation between luminosity and mass clearly is
expected. Also correlations found between X-ray luminosity and galaxy
velocity dispersion (e.g., \citealt{es91b}) and X-ray luminosity and mean shear
strength from weak lensing studies (e.g., \citealt{sed97}) indicate a
correlation between $\lx$ and $\mt$.

 The X-ray luminosity has been compared directly to
gravitational mass estimates by
\citet{r98}, \citet{rb99b,rb00c}, \citet{s99}, \citet{jf99}, \citet{mlb99}, \citet{ef00},
and \citet{bg01},  where good correlations have been found in all of these
studies.

\begin{figure}[thbp]
\psfig{file=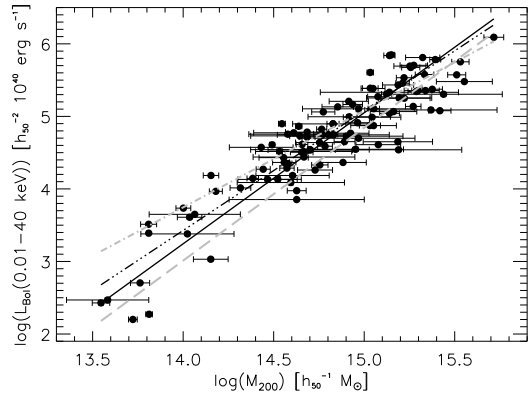,width=8cm,angle=0,clip=}
\caption{Gravitational mass--bolometric X-ray luminosity
relation compared to predicted relations. Shown are:
best fit relation for the extended sample (solid line),
best fit relation determined using \gcs\ (triple-dot-dashed line),
self-similar relation normalized by
simulations of \citet{nfw95} (dot-dashed line),
pre-heated relation given by
\citet{eh91}, using a normalization taken from the simulations of
pre-heated clusters by \citet{nfw95} (dashed line).}\label{mtlbol}
\end{figure}

In order to compare the empirical $\lx$--$\mt$ relation with predictions a
quasi bolometric luminosity, $\lbol$, has been calculated in the
source rest frame energy range
$\ebol$ (for the relevant range of cluster gas temperatures
at least 99\,\% of the flux is contained in this energy range).
In Fig.~\ref{mtlbol} this $\lbol$--$M_{200}$ relation
is compared to predicted relations. 
The solid line shows the best fit relation for the 106 clusters in the
extended sample and the 
triple-dot-dashed line shows the best fit relation determined using
\gcs . Here the bisector fit results have been used in order to treat
variables symmetrically, which is the appropriate method for a
comparison to theory \citep[e.g.,][]{ifa90}.
The dot-dashed line
shows the self-similar relation ($\lbol\propto M^{4/3}$) normalized by the
simulations of \citet{nfw95} and the dashed line shows the
`pre-heated' relation given by
\citet{eh91} ($\lbol\propto M^{11/6}$), using a normalization taken from the simulations of
pre-heated clusters by \citet{nfw95}. 
The idea of pre-heating is that the intracluster gas is not
cold initially, as in the self similar case, but is heated by some
form of non gravitational heat input, e.g., from supernovae or AGN, before
or during cluster formation. Assuming the central regions of all
clusters 
to have the same entropy yields the latter relationship.
Fig.~\ref{mtlbol} shows that measured and predicted
relations are in rough agreement,
the difference between the predicted relations being larger than the
difference to the observed relations.
Note, however, that the X-ray luminosity is one of the most uncertain
quantities to be derived from simulations. \citet{fwb99} recently
showed in a comparison of twelve different cosmological hydrodynamics
codes that a factor of two uncertainty is a realistic estimate of the
current accuracy. Including gas cooling in simulations worsens the
situation \citep[e.g.,][]{bpb01}.
The slopes of the observed relations are closer to the
pre-heated relation.
Observationally the effect of pre-heating can also result in a decrease of the gas
mass fraction for low temperature systems.
This has actually been observed for
the clusters in our sample \citep{r98,rb99c}. The possibility
that winds from, e.g., supernovae
-- originally invoked to explain the apparent low gas content of elliptical
galaxies \citep[e.g.,][]{mb71,l74} --  
pre-heat and dilute the central gas and thereby break
the self similarity has been pointed out by various authors (e.g.,
\citealt{k86}). Such a process would work most efficiently on the least massive
clusters (e.g., \citealt{w91,me97,pcn99}).
The $\lx$--$\mt$ relation and
other relevant relations between physical cluster parameters of the
extended sample have been discussed more thoroughly in this context in
\citet{r01}.

One may wonder about the origin of the scatter in the $\mt$--$\lx$
relation.
Due to the use of pointed observations for most of the clusters and the
local nature of \gcs\ the statistical errors of $\lx$ are negligible.
The logarithmic mean mass measurement uncertainty has been measured as
0.12. 
The overall scatter in log mass of the data points compared to the best
fit relation is larger and has been measured as 0.21
(Tab.~\ref{tab:scat}). This indicates a
possible contribution of intrinsic scatter to the overall scatter in the
$\mt$--$\lx$ relation. An obvious candidate to cause intrinsic scatter is the
central excess emission (central surface brightness exceeding single
$\beta$ model surface brightness) present in a number of clusters. This
excess emission may have its physical origin either in cooling flows
\citep[e.g.,][]{f94} or in the presence of cD galaxies
\citep[e.g.,][]{mef01}. A cooling flow analysis of the \gcs\ clusters
is in progress 
and first results indicate that indeed clusters with a high inferred mass
deposition rate lie on the high $\lx$ side of the $\mt$--$\lx$ relation
(Y. Chen et al., in preparation).

\begin{figure}[thbp]
\psfig{file=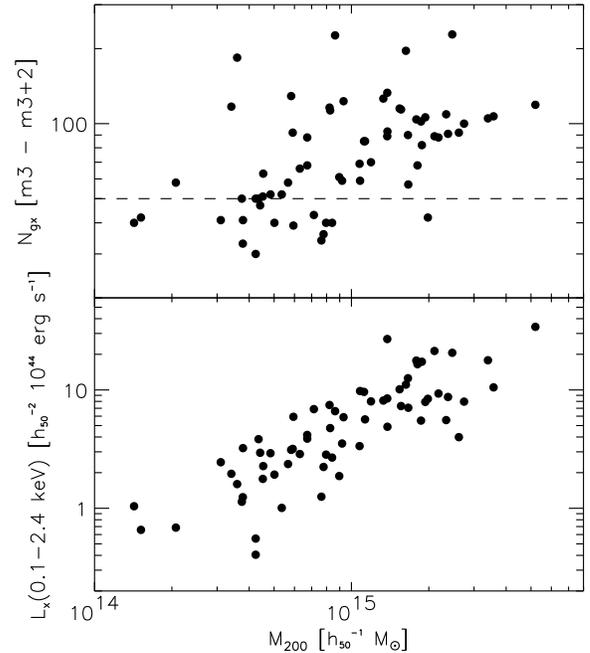,width=13cm,angle=0,clip=}
\caption{Measured number of cluster member galaxies, $N_{\rm gx}$, as
taken from \cite{aco89}
and X-ray luminosity of the same (66) clusters
as a function of the gravitational mass. Double clusters,
whose components have been treated separately here,
e.g., A3528n/s, are removed. Above the dashed
line all clusters have an Abell richness $R\ge 1$.}\label{nmlxm}
\end{figure}

In Fig.~\ref{nmlxm} the measured number of cluster member galaxies as taken from
\citet{aco89} is compared
to $\lx$ as gravitational mass tracer. It is clearly seen that a selection by
X-ray luminosity is much more efficient than a selection by Abell richness in
terms of mass.
Even though only the X-ray surface brightness profile and neither its
normalization nor the X-ray luminosity are directly used in the X-ray mass
determination via the hydrostatic equation, it is
nonetheless reassuring that a similar result is obtained when $\lx$
and richness are compared to masses estimated from optical velocity
dispersions \citep{bg01}.

A wide range
of possible applications becomes available with the quantification of the $\lx$--$\mt$
relation and its scatter. For large X-ray flux-limited cluster surveys, where
individual mass 
determinations are currently not feasible, luminosities can be directly
converted to masses. No combination of observations, simulations and theory is
then needed, like the frequently used approach of relating X-ray luminosities
to X-ray temperatures by an
observed relation, and converting X-ray temperatures to masses using a relation
where the slope is taken from theoretical arguments and the normalization from
hydrodynamical simulations (e.g., \citealt{mmd00}). The observational
$\lx$--$\mt$ relation
has first been applied directly in this sense in the power spectral analysis of REFLEX
clusters \citep{sbg00a}. An example of another direct application is given in
Sect.~\ref{lmtest}.
The $\lx$--$\mt$ relation may also be applied to
convert theoretical or simulated mass functions to luminosity functions for
comparison with observations of X-ray flux-limited samples, which is currently
being performed in the interpretation of the REFLEX luminosity function.

At this point it is important to note that even for the highest
redshift cluster in our sample ($z=0.2$) the dependence of the
observational determination of $\mt$ and $\lx$ on the chosen
cosmological model is very weak. For instance at $z=0.2$ the increase
in the luminosity distance, $D_L$, and the diameter distance,
$D_A$, is less than 5\,\% going from ($\om=1.0,\ol=0$) to
($\om=0.1,\ol=0$).
From (\ref{eq:mt}) one finds that
$\mt(<r)\propto r$ and therefore $\mt(<r)\propto D_A$, implying an
increase of $\mt$ by less than 5\,\% for the two models above.
For $\lx$ one has an increase of less than 10\,\%.
This means that the $\lx$--$\mt$ relation given here can be used unchanged for various
cosmological applications (unless redshift ranges are probed where
evolution becomes important, in this case a model dependent
redshift correction has to be introduced).
A similar calculation for $\vmax$ shows that for the
extreme case $\zmax=0.2$ the increase in $\vmax$ is less than
14\,\%, which is
less than the size of the Poissonian error bars in Fig.~\ref{PS0}.

More detailed investigations on the shape of the relation
are in progress and it is also envisaged to construct a
volume-limited sample, spanning a reasonably large range in
luminosity and mass, to test how much the $\lx$--$\mt$ relation given
here is affected by being estimated partly from a flux-limited sample.

\subsection{Mass Function}\label{funct_d}

\subsubsection{Comparison to previous estimates}\label{func_comp}

\begin{figure}[thbp]
\psfig{file=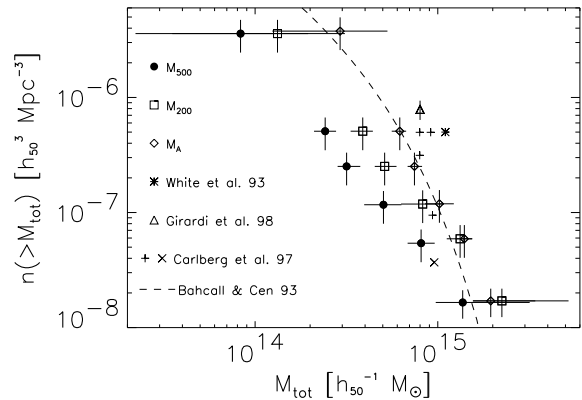,width=8.5cm,angle=0,clip=}
\caption{Cumulative gravitational mass functions for \gcs\ using three
different outer radii. Vertical error bars give the
Poissonian errors. Horizontal bars indicate the individual
bin sizes. Each bin contains 10 clusters, apart from the highest mass
bin, which contains 13 clusters. The abundances from previous works
are determined for
$\mab$.}\label{mfunc}
\end{figure}

\citet{bc93} give a mass function constructed a) from optically selected
clusters with masses determined from the galaxy richness and b) from the
cluster X-ray temperature function given by \citet{ha91}. Very good
agreement is found for masses determined within $\rab$
between the \citeauthor{bc93} and \gcs\ mass function
(Fig.~\ref{mfunc}).

\citet{wef93} constrain the cluster abundance by using published
values for the abundance and median velocity dispersion of richness
class $R\geq 1$ Abell clusters. It is not surprising that their density is
significantly higher than the \gcs\ density since they have 
intentionally used conservative mass estimates, which are overestimates
of the true cluster masses.

\citet{bgg93} and \citet{gbg98} have determined
the cluster mass function using optically selected cluster samples with masses
determined from published line-of-sight velocity dispersions of cluster galaxies. 
At the completeness limit given by \citet[triangle in
Fig.~\ref{mfunc}]{gbg98} the cluster density given by \citet{bgg93} is
about a factor of two higher than the density given by
\citet{gbg98}. The latter authors explain this by their on average
40\,\% smaller mass estimates due to an improved technique for removing
interlopers and the use of a surface-correction term in the virial theorem.
The value given by \citeauthor{gbg98} itself lies significantly higher
than the comparable \gcs\ density. The reason could lie in the fact
that their optically estimated masses are in general slightly larger
than the X-ray masses or that the external normalization for
$R\geq 1$ ($N_{\rm gx} \ge 50$) clusters which they applied to their mass function is
intrinsically higher than the normalization obtained through \gcs\
directly, or both. 
By comparing the mass estimates for a common subsample of 42 clusters
it has been found that the virial masses determined
by \citet{gbg98} are on average 25\,\% larger than the masses determined
in this work. This difference might be smaller if masses out to the Abell
radius were compared, since common radii would have been used in this case.
Increasing the X-ray masses artificially by 25\,\%, the diamonds
in Fig.~\ref{mfunc} shift towards higher masses, but the shift is too small
to account for the difference to the triangle.
The large scatter in the $N_{\rm gx}$--$\mt$ diagram (Fig.~\ref{nmlxm}) makes
a reliable estimate of a best fit relation between these two
quantities very
difficult. Nevertheless, in order to get a rough idea of the mass
within $\rab$ that
corresponds to $N_{\rm gx}=50$, we have performed fits using the minimization
methods outlined earlier and find $5.1\lesssim\mab (N_{\rm gx}=50)\lesssim8.8\times
10^{14}\,h_{50}^{-1}\,M_{\odot}$.
Note that this range is in agreement with the ranges
$5$--$8\times 10^{14}\,\msu$ and $5$--$7\times 10^{14}\,\msu$ 
given by \citet{bc93} and \citet{gbg98}, respectively, for $N_{\rm
gx}=50$. 
This mass range corresponds to a cumulative number 
density obtained through \gcs\ in the range $1.7\lesssim
n(>\mab)\lesssim 8.7\times 10^{-7}\,h_{50}^3\,\rm Mpc^{-3}$. The 
external density estimate applied to normalize the \citeauthor{gbg98}
mass function therefore is a factor 1.2--6.2 higher than the
corresponding estimate obtained here.
It is therefore concluded that
both effects (masses and normalization) are important but the latter
factor is responsible for a larger fraction of the
discrepancy. Assuming both normalizations to be determined from
samples that are highly complete and representative of the local
universe this may indicate that either X-ray and optical clusters
are drawn from
different populations or that projection effects (e.g., line of sight galaxy
overdensities, which do not form a bound structure in three
dimensions) possibly bias optically determined normalizations high.

\citet{gg00} recently extended the mass function to loose galaxy
groups, finding
$n(>1.8\,h_{50}^{-1}\times10^{13}\,M_{\odot})=1.6$--$2.4\times 10^{-4}
\,h_{50}^3\,\rm
Mpc^{-3}$, which is outside the mass range we can test. They find that
the group mass function can be described by a smooth extension of the
cluster mass function by \citet{gbg98}. Consistently this abundance is
higher than the abundance given by \citet{bc93} at that mass scale.

\cite{cmy97} have compiled and partially reestimated abundances for
local cluster samples \citep{ha91,mkd96,wef93,ecf96} for comparison with
higher redshift samples (the `$\times$' shows the density for a sample with
higher mean redshift and therefore it cannot be compared directly).
Note that \citet{bgc99} find that the reestimate of
the mass limit of the \citeauthor{mkd96} sample
by \citeauthor{cmy97} (the `+' at $n\approx 3\times 10^{-7} h_{50}^3 \rm Mpc^{-3}$)
may lead to an underestimated mass limit.
In general the comparison to the \gcs\ mass
function shows better agreement than the abundances given by
\citet{gbg98} and \citet{wef93}. 

The obvious importance of the definition of the cluster outer radius for the cluster
mass function can be directly appreciated by noting the large
differences between the mass functions determined for \gcs\ for
$\mtf$, $\mtz$, and $\mab$ in Fig.~\ref{mfunc}, especially towards
lower masses. Since for self similar clusters the mass scales with the
third power of the characteristic radius (Sect.~\ref{relat_d}),
determination of the mass within a characteristic overdensity is the
natural choice. We mainly
give the formally determined $\mab$ mass function for the comparison
with previous mass functions and recall again that
especially for the low mass systems the assumption of hydrostatic
equilibrium may not be justified out to $\rab$, and therefore our mass
estimates of $\mtf$ and $\mtz$ should be considered much
more precise than the estimates for $\mab$. 

\subsubsection{Comparison to predicted mass functions}\label{func_pred}

We use the standard formalism based on the Press--Schechter (PS)
prescription to predict cluster mass functions for given cosmological
models (see, e.g., \citealt{brt99}).
To allow easier comparison with the theoretical literature on this
subject in this paragraph $h_{100}=h_{50}/2$ is used.
The mass function is then given by
\begin{eqnarray}
\frac{dn(M)}{dM}=\sqrt{\frac{2}{\pi}}\,\frac{\bar\rho_0}{M}\,\frac{\dc(z)}{\sigma(M)^2}
\,\left\vert\frac{d\sigma(M)}{dM}\right\vert
\exp\left(-\frac{\dc(z)^2}{2\,\sigma(M)^2}\right)
\label{eq:ps}
\end{eqnarray}
(\citealt{ps74}, \citealt{bce91}; see, e.g., \citealt{sba00} for a compilation of
published extensions of the PS mass function).
Here $M$ represents the halo (cluster) virial mass and $\bar\rho_0 =
2.7755\times 10^{11}\,\om\,h_{100}^2\,M_\odot\,\rm
Mpc^{-3}$ is the present day mean matter density. The linear
overdensity computed at present $\dc(z)=\dcv(z)\,D(0)\,D(z)^{-1}$,
where the linear overdensity at the time of virialization, $\dcv(z)$,
is computed using the spherical collapse model
summarized in \citet{ks96}, for $\om =1$ using (A2) and for
$\om<1 \wedge \ok=0$ using (A6,7); the linear growth factor
$D(z)=2.5\,\om\,E(z)\int_z^\infty (1+z')\,E(z')^{-3}\,dz'$ and $E(z)$ has
been defined in Sect.~\ref{massd}. As mentioned earlier due to the
low redshift range spanned by \gcs , the effect of a redshift
correction is very small and we therefore set $z=0$ for all
calculations, unless noted otherwise.
The variance of the cosmic mass density fluctuations
\begin{equation}
\sigma(M)^2=\sigma_8^2\frac{\int_0^\infty
k^{2+n}\,T(k)^2\,\vert W(k\,R(M))\vert^2\,dk}{\int_0^\infty
k^{2+n}\,T(k)^2\,\vert W(k\,8\,h_{100}^{-1}\,{\rm Mpc})\vert^2\,dk}\,,
\label{eq:sigma}
\end{equation}
where $\sigma_8$ represents the amplitude of density fluctuations in
spheres of radius
$8\,h_{100}^{-1}\,{\rm Mpc}$.
Recent measurements of the cosmic microwave background (CMB)
anisotropies indicate that the primordial power spectral index,
$n$, has a value close to 1
\citep[e.g.,][]{bab00,jab01,phl01,wtz01,dab01} and is therefore set to 1,
unless noted otherwise.
For the transfer function we use the
fitting formula for Cold Dark Matter (CDM) cosmologies provided by
\citet{bbk86} for
$q(k)=k/(\Gamma\,h_{100}\,\rm Mpc^{-1})$
\begin{eqnarray}
T(k)\equiv T(q(k))=\ln(1+2.34q)/(2.34q)\nonumber\\
\times\, [1+3.89q+(16.1q)^2+(5.46q)^3+(6.71q)^4]^{-1/4}\,,
\label{eq:transf}
\end{eqnarray}
where the shape parameter is given by (modified to account for a small
normalized baryon density $\ob > 0$, \citealt{s95})
\begin{equation}
\Gamma = \om\,h_{100}\,\left(\frac{2.7\,\rm
K}{T_0}\right)^{2}\,\exp\, \left(-\ob-\sqrt{\frac{h_{100}}{0.5}}\,\frac{\ob}{\om}\right)\,.
\label{eq:gama}
\end{equation}
The temperature of the CMB
$T_0=2.726\,\rm K$ \citep{mcc94} and $\ob\,h_{100}^{2}=0.0193$
\citep{bt98}, for the latter equation and (\ref{eq:gama}) $h_{100}=0.71$ has
been used \citep{mhf00}. 
The comoving filter radius $R(M)=[3M/(4\pi\bar\rho_0)]^{1/3}$ for
the top hat filter function $W(x)=3\,(\sin x - x\cos x )/x^3$, which is
adopted in this analysis, because the \gcs\ masses have been
determined with a top hat filter, too.\footnote{This approach follows
the custom of disregarding 
the inconsistency of using top hat masses while the
PS mass function with the correct normalization has been derived for
the sharp $k$-space 
filter \citep{bce91}; see \citet{sba00} for a generalization to more
realistic filter functions.}
Since the PS recipe as outlined above assumes virial masses based on
the spherical collapse model we use $\mtz$ as approximation to the
virial mass (Sect.~\ref{massd}).

Similarly to the work of \citet{irb01} the statistical
uncertainty in the mass determination is incorporated
in the model mass function as
\begin{eqnarray}
\frac{d\tilde n(M)}{dM}\equiv \frac{1}{\vmax(M) }\,\int^\infty_{-\infty}\frac{dn(M')}{dM'}\,\vmax(M')\, \nonumber\\
\times\, (2\pi\, \bar\sigma_{\mt ,\log}^2)^{-1/2}\,\exp\left(\frac{-(\log M' -
\log M)^2}{2\,\bar\sigma_{\mt ,\log}^2}\right) d\log M'\,,
\label{eq:nschl}
\end{eqnarray}
where $\bar\sigma_{\mt ,\log} =
0.12$ represents the logarithmic mean
mass measurement uncertainty. Note that since \gcs\ is flux-limited and not
volume-limited the weighting has to be performed on the
mass distribution, $N(M)/dM$, rather than on the mass function
itself. The effect of this weighting on the model mass function is a
slight amplitude increase at the high mass end.

\begin{figure}[thbp]
\psfig{file=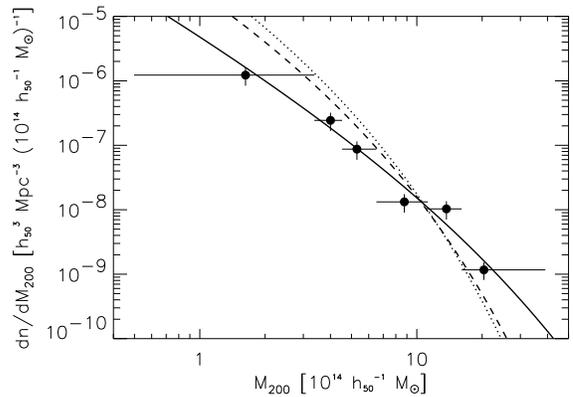,width=8cm,angle=0,clip=}
\caption{\gcs\ mass function compared to the best fit model mass function
with $\om=0.12$ and $\sigma_8=0.96$ (solid line). Also shown are the best fit model
mass functions for fixed $\om=0.5$ ($\Rightarrow \sigma_8=0.60$,
dashed line) and $\om=1.0$ ($\Rightarrow \sigma_8=0.46$, dotted
line).}\label{PS0}
\end{figure}

\begin{figure}[thbp]
\psfig{file=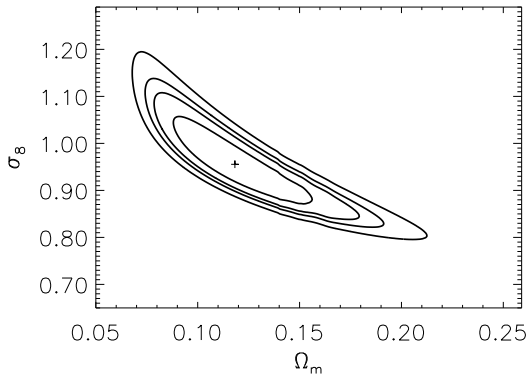,width=8cm,angle=0,clip=}
\caption{Statistical confidence contours for the $\chi^2$
procedure. The cross indicates the position of the minimum, $\chi^2_{\rm min}$.
Ellipses indicate the 68\,\%, 90\,\%, 95\,\%, and 99\,\% confidence levels
for two interesting parameters,
i.e.\ $\Delta\chi^2\equiv\chi^2-\chi^2_{\rm min}=$ 2.30, 4.61, 6.17, and 9.21,
respectively.
}\label{banana}
\end{figure}

For the modeling to be independent of the precise knowledge of the
$\lx$--$\mt$ relation the quantitative comparison has been performed
using a standard $\chi^2$ procedure on the
differential binned mass function given in Fig.~\ref{PS0} (rather than
using a maximum likelihood approach on the mass distribution).
After identifying the crude position where $\chi^2$ is minimal in a large
$\om$--$\sigma_8$ parameter space region
the $\chi^2$ values have been calculated in a fine grid of
200 by 200 $\om$--$\sigma_8$ values
in the range $0.05\le\om\le0.26$ and $0.65\le\sigma_8\le1.30$.
A flat cosmic geometry has been assumed, i.e.\ $\om+\ol=1$.
The cosmological
constant enters the calculation only through $\dc$, however, and therefore
has a negligible influence here. The minimum and statistical error ellipses for some
standard confidence levels (c.l.)\ are given in Fig.~\ref{banana}. The tight
constraints obtained show that with \gcs\ we can go beyond determining an
$\om$--$\sigma_8$ relation and put limits on $\om$ and $\sigma_8$
individually.  It is found that
\begin{equation}
\om=0.12^{+0.06}_{-0.04}\quad {\rm and}\quad  \sigma_8=0.96^{+0.15}_{-0.12}
\label{eq:omres}
\end{equation}
(90\,\% c.l.\ statistical uncertainty for two interesting parameters),
indicating a  relatively low value for the density parameter.
The large covered mass range, the specific region in
$\om$--$\sigma_8$ parameter space, and the assumption of CDM cosmological
models with
given $H_0$, $\ob$, $n$, and $T_0$ allow
to derive these tight
constraints from a local cluster sample.
For comparison for a given $\sigma_8$ value the
$\om$ value which minimizes $\chi^2$ is calculated.
In the interval shown in Fig.~\ref{banana}
these pairs can then
roughly be described by a straight line in log space given by 
\begin{equation}
\sigma_8=0.43\,\om^{-0.38}\,.
\label{eq:omsig}
\end{equation}
In
Fig.~\ref{PS0} we also plot the best fit model mass functions for
given $\om=0.5$ and $\om=1.0$ and one notes immediately
that these value pairs give a poorer description of  the shape of
the mass function.
The slight differences between the best fit values for $\sigma_8$
and the ones expected from (\ref{eq:omsig}) are caused by
the fact that the simple power law is only an approximation to the real shape of
the error ellipse and weakly by the extrapolation involved.
Previous estimates have
generally yielded a combination of slightly higher $\om$--$\sigma_8$
values (e.g., \citealt{p99}, Sect.~17.2). It has to be noted,
however, that for instance in the important work of \citet{wef93}, who
find $\sigma_8\approx 0.57\,\om^{-0.56}$, the authors explicitly
state that their estimates of
$\sigma_8$ are probably biased high due to their conservative mass
estimates.

\begin{figure}[thbp]
\psfig{file=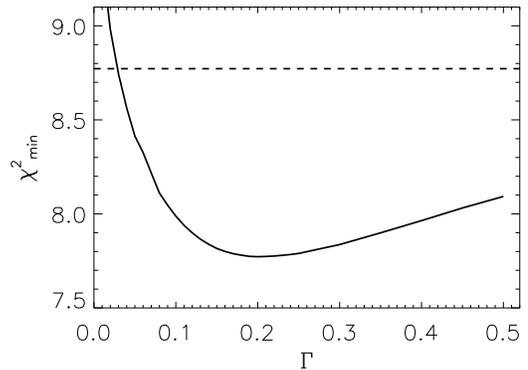,width=8cm,angle=0,clip=}
\caption{Minimal $\chi^2$ values calculated by varying $\om$ and
$\sigma_8$ for a variety of fixed values of $\Gamma$ instead of using
eq.~\ref{eq:gama}. The dashed line represents the 68\,\% c.l.\ statistical
uncertainty for one
interesting parameter. This line actually underestimates the uncertainty
since the procedure involves three interesting parameters. Obviously
$\Gamma$ cannot be constrained significantly.}\label{chigam}
\end{figure}

Before exploring sources of systematic errors let us evaluate our treatment
of the shape parameter $\Gamma$. By applying the \citet{bbk86} fitting
formula to calculate the transfer function we have adopted the assumption
of a CDM universe, i.e.\ $\om-\ob=\ocdm$. Within this model $\Gamma$ and
$\om$ are not independent. For a small $\ob/\om$ ratio $\Gamma$ is given by
(\ref{eq:gama}).
However, to test its influence we have set $\Gamma$ to a fixed value of 0.19 as
implied by Large Scale Structure (LSS) measurements of galaxies and clusters
\citep[e.g.,][see also the discussion at the end of this Section]{sjm01,sbg00a}.
The best fit values for $\om$ and $\sigma_8$ change significantly, the $\om$
value becoming smaller and the $\sigma_8$ value larger.
The $\om$--$\sigma_8$ relation changes slightly to about
$\sigma_8=0.40\,\om^{-0.46}$
for the $\om$ range [0.04--0.30] (note that for this range relation
(\ref{eq:omsig}) translates to about $\sigma_8=0.39\,\om^{-0.43}$, the steepening
being mainly due to a strong increase of $\sigma_8$ for the very lowest $\om$
values).
The error ellipse in general becomes much longer
when $\Gamma$ is set to a fixed value, preventing the possibility to obtain
tight individual constraints on $\om$ and $\sigma_8$,
even though $\om=1.0$ is still excluded at least at the 90\,\% c.l.\ for all
$\Gamma\ge 0.03$.
The decreased sensitivity is due to the loss of the amplification of a change of
$\om$ through $\Gamma$ (to first order $\om\propto\Gamma$, eq.~\ref{eq:gama}),
since both a larger (smaller) $\om$ and a larger (smaller) $\Gamma$ result in
steeper (less steep) model mass functions.
Having found a dependence of the results on $\Gamma$ we tried to constrain
$\Gamma$ directly from the observations, as done by previous authors
\citep[e.g.,][]{brt99}.
We have set $\Gamma$ to 31 different fixed values in the range
$0.01\le\Gamma\le0.50$ and calculated a $\chi^2_{\rm min}$ for each of
those by varying $\om$ and $\sigma_8$. The result is shown in
Fig.~\ref{chigam}. The conclusion is that the data do not contain enough
information to provide an independent constraint on $\Gamma$. Therefore we
have taken advantage of the dependence of $\Gamma$ on $\om$ within the CDM
framework and have continued to calculate $\Gamma$ with eq.~\ref{eq:gama}.
Notice that this procedure requires additional knowledge about $H_0$ and
$\ob$. The influence of the uncertainty of these two parameters is
tested along with various other possible systematic uncertainties in the
following. It is important to note that the best fit value for $\om$ is not
determined solely through (\ref{eq:gama}) but also directly through
$\bar\rho_0$ in (\ref{eq:ps}) and through the filter radius $R(M)$.

The quoted error ranges have been calculated from the $\chi^2$ procedure.
Most of the remaining part of this Section is devoted to an identification
and quantification of possible systematic uncertainties. The quantitative
results are summarized in Tab.~\ref{tab:sys}.

Due to the large given ranges of several orders of magnitude in mass and
especially density the $\chi^2$
values have been determined from comparison model/data naturally in
log space. However, we have verified that the same calculation in
linear space yields best fit values lying within the 68\,\% error ellipse.

In Sect.~\ref{funct} arguments have been given why we have used
$\vmax(\lx)$ for the determination of the mass function.
Nevertheless if $\vmax(\mt)$ is used instead (see Fig.~\ref{vm_cp_ns}),
consistent results are obtained. Since in this case we want to
estimate $L$ from $M$ the relation ($L\mid M$) is the appropriate one
\cite[e.g.,][]{ifa90}. Performing a fit to a mass function constructed
with $\vmax(\mt)$ results in best fit values $\om=0.14$ and
$\sigma_8=0.86$, which is consistent with the error range given in
(\ref{eq:omres}).

The data point in Fig.~\ref{PS0} that may be affected most by cosmic
variance is the one at the lowest mass, since the maximum search
volume is smallest for the clusters in this bin. We therefore tested
the sensitivity of the best fit results on this last point by ignoring
it. The decrease in the covered mass range of course increases the
resulting error ellipse, but the best fit values vary only within
the (smaller) 68\,\% error ellipse. It may be worth noting that
leaving out the highest mass bin or leaving out the highest \emph{and}
lowest mass bin changes the best fit values only within the 90\,\%
error ellipse.

Simulations seem to indicate that the mass density profile of virialized
halos follows the NFW profile \citep[but see, e.g.,
\citealt{kkb98}]{nfw96,nfw97}. As test therefore
gravitational masses have been recalculated assuming the dark matter
density to follow an NFW profile based on the measured values for $\beta$
and $\rc$. The approximation $\rs=\rc/0.22$ and $b=\beta/0.9$ given by
\citet{mss98} has been used combined with their eqs.~7 and
9. $r_{200}$ and $\mtz$ have been redetermined based on this model and the
resulting mass function has been compared to model mass functions.
On average it is found that the NFW masses are a factor of 0.90 lower, the
difference being smaller for high mass clusters than for low mass ones.
The
values for $\om$ and $\sigma_8$ for which $\chi^2$ is minimal lie within the
90\,\% statistical error ellipse shown in Fig.~\ref{banana}. Therefore the choice of
mass calculation does not affect the results significantly.

As shown in Sect.~\ref{massd} it is possible that the assumption
of isothermality leads to an overestimate of the cluster masses on
average. Therefore the robustness of the
results has been tested by multiplying the isothermal cluster
masses by 0.80 and
recalculating the minimum. As expected values for both $\om$ and
$\sigma_8$ are found to be lower. But the new minimum is contained
well within the error ranges given in (\ref{eq:omres}). On the other
hand in Sect.~\ref{func_comp} it has been shown by comparison with optical
virial mass measurements that masses could possibly be underestimated. After
increasing all cluster masses by 25\,\% a fit shows that both 
$\om$ and $\sigma_8$ become slightly larger, but again they lie well
within the range (\ref{eq:omres}).
Therefore these tests indicate that the constraints obtained here are
fairly insensitive against systematic uncertainties in the mass
measurements. Note that systematic effects that show a mass dependence have
a higher potential to affect the individual best fit values of $\om$ and
$\sigma_8$ than effects which simply shift all data points by about the
same amount in the same direction. However, no indications for a mass
dependence of the differences between the above mass estimations have been
found.

It has been mentioned that masses calculated within $r_{500}$ are less
affected by systematic uncertainties because less extrapolation beyond
$\rx$ is needed and because according to simulations the assumption of
hydrostatic equilibrium can be more safely applied. Nevertheless we have
used $\mtz$ for the comparison to predicted mass functions because it is a
better approximation to the cluster virial mass. However, for comparison
the best fit values have also been determined using $\mtf$, knowing that
virial masses are probably underestimated this way. The result is that the best
fit value for $\om$ changes only marginally. The best fit
$\sigma_8$ value is slightly lower than allowed by the statistical error.
On the other hand the spherical collapse model implies that one may
need to extrapolate even further than $r_{200}$ for low density universes.
For instance in a flat universe $\om=0.2$ implies that the virial radius
is close to $r_{87}$. Therefore a mass function has been constructed
using $M_{87}$ as cluster masses, knowing that these masses are likely to
be rather uncertain. Nevertheless the resulting best fit $\om$ value again
varies only insignificantly whereas the $\sigma_8$ value becomes significantly
larger.
Note that for both tests the best fit value for $\om$ hardly changes.

As mentioned in Sect.~\ref{massd} it has been shown that the small
range of low redshifts covered here ensures that no redshift
corrections need to be applied. Nevertheless we have tested whether or
not the best fit parameter values change if $\mtz$ is calculated using 
$\roc =\roc (z)$ for the extreme (strong evolution) case
($\om=1,\ol=0$), i.e., $\roc =
4.6975\times10^{-30}\,(z+1)^3\,\rm g\,cm^{-3}$ for each cluster
redshift. The model mass function is then calculated for the mean
redshift of \gcs , $\langle z\rangle=0.05$, using the formulae
outlined in the beginning of this Section. We have found that within
our grid the best fit values do not change at all and also the error
ellipses are almost not affected, thereby confirming that the
application of redshift corrections does not affect the results.

The value $H_0=71\,\rm km\,s^{-1}\,Mpc^{-1}$ has been adopted for the
calculation of model mass functions based on
the recent combination of constraints obtained using the Hubble Space
Telescope \citep{mhf00}. Setting the Hubble constant to their lower
limit, $H_0=65\,\rm km\,s^{-1}\,Mpc^{-1}$, does not affect the best fit
parameter values significantly. Using even $H_0=60\,\rm km\,s^{-1}\,Mpc^{-1}$
changes the results for $\om$ and $\sigma_8$ only well within the
68\,\% error ellipse.
In Tab.~\ref{tab:sys} results have also been listed for $H_0=50$ and
$80\,\rm km\,s^{-1}\,Mpc^{-1}$ showing that even variations within this
range change the $\om$ and $\sigma_8$ values only within their
statistical uncertainties given in (\ref{eq:omres}).
Therefore our constraints on these cosmological
parameters do not depend significantly on the specific choice of $H_0$.

A baryon density based on measurements of the primordial deuterium to
hydrogen ratio combined with standard nucleosynthesis theory has been used
in this work. The used value also agrees with recent CMB measurements
\citep[e.g.,][]{phl01}. However, we have also calculated the minimum for
the values $\ob\,h_{100}^{2}=0.0100$ and $0.0300$ finding that the position
of the minimum in each case does not deviate significantly from our
original result.

The primordial power spectral index has been set to $n=1$ based on
measurements of the CMB fluctuations. The influence of a change in this
parameter has been estimated by calculating the minimum for the values
$n=0.8$ and $n=1.2$. The differences in the best fit parameters are smaller
than the statistical uncertainties given in (\ref{eq:omres}).

Since we have found that for the estimate of the statistical errors we
need to explore ranges $\om < 0.1$ for $\ob\sim 0.04$ we regarded it
necessary to check
if the approximation to the transfer function as given in the
beginning of this Section is still applicable. 
Recently \citet{eh98} derived a fitting function, that includes, e.g.,
also the oscillations induced by the baryons, which gives a better
description of transfer functions computed with CMBFAST \citep{sz96} than
fitting functions for zero or small baryon contribution to the
total matter density derived previously. Therefore we incorporated
this improved version of an analytic transfer function in the $\chi^2$
procedure. We have found that within our grid
there is only a very weak shift of the minimum,
the choice of the \citet{bbk86} fitting function
combined with the shape parameter given by \citet{s95} therefore seems
to be accurate enough for our purposes. However, the confidence contours towards low
$\om$ are getting compressed when the \citet{eh98} fitting function is
used, thereby slightly decreasing the area of the error ellipse for a given
confidence level. Since we regard the latter statistical error
ellipse as more realistic, we show this one in Fig.~\ref{banana}.

\begin{deluxetable}{ccccc}
\tablecolumns{5} 
\tabletypesize{\footnotesize}
\tablecaption{Systematic uncertainties\label{tab:sys}}
\tablewidth{0pt}
\tablehead{
\colhead{$\om$}	& \colhead{$\sigma_8$} & \colhead{$\Delta \om$} &
\colhead{$\Delta \sigma_8$} & \colhead{Test}\\
}
\startdata
0.116	&	0.928	& $\pm 0.000$ & $-0.026$ & (1)\\
0.137	&	0.865	& $+0.021$ & $-0.094$ & (2)\\
0.150	&	0.897	& $+0.034$ & $-0.062$ & (3)\\
0.092	&	1.082	& $-0.024$ & $+0.124$ & (4)\\
0.083	&	1.050	& $-0.034$ & $+0.091$ & (5)\\
0.150	&	0.860	& $+0.034$ & $-0.099$ & (6)\\
0.107	&	0.907	& $-0.009$ & $-0.052$ & (7)\\
0.132	&	1.004	& $+0.016$ & $+0.046$ & (8)\\
0.106	&	0.826	& $-0.011$ & $-0.133$ & (9)\\
0.108	&	1.242	& $-0.008$ & $+0.281$ & (10)\\
0.116	&	0.959	& $\pm 0.000$ & $\pm 0.000$ & (11)\\
0.118	&	0.956	& $+0.002$ & $-0.003$ & (12)\\
0.153	&	0.861	& $+0.037$ & $-0.098$ & (13)\\
0.108	&	0.991	& $-0.008$ & $+0.033$ & (14)\\
0.106	&	1.001	& $-0.011$ & $+0.042$ & (15)\\
0.128	&	0.923	& $+0.012$ & $-0.036$ & (16)\\
0.148	&	0.874	& $+0.032$ & $-0.085$ & (17)\\
0.094	&	1.056	& $-0.022$ & $+0.098$ & (18)\\
\enddata

\tablecomments{See text for a more detailed description of the tests. The
differences to the best fit results have been been calculated with more
digits than given here.}

\tablerefs{
(1) $\chi^2$ calculated in linear space.
(2) $\vmax(\mt)$ used with relation ($L\mid M$) for \gcss\ (Tab.~\ref{tab:lm2}).
(3) Lowest mass bin ignored.
(4) Highest mass bin ignored.
(5) $r_{200}$ and $\mtz$ calculated assuming NFW profile.
(6) Universal mass function.
(7) $\mtz$ reduced by factor 0.80.
(8) $\mtz$ increased by factor 1.25.
(9) $\mtf$ used.
(10) $M_{87}$ used.
(11) Redshift correction applied for cluster mass determination and model
    mass function calculation.
(12) Transfer function calculated using \citeauthor{eh98} fitting formula.
(13) $h_{100}$ set to 0.5.
(14) $h_{100}$ set to 0.8.
(15) $\ob$ set to 0.0100\,$h_{100}^{-2}$.
(16) $\ob$ set to 0.0300\,$h_{100}^{-2}$.
(17) $n$ set to 0.8.
(18) $n$ set to 1.2.
}

\end{deluxetable}
We have also tested whether or not the recently found deviations of the PS
formalism compared to large $N$-body simulations
\citep[e.g.,][]{gbq99,jfw01} have a significant
influence on the results obtained here. We have compared the best fit
PS model ($\om=0.12$, $\sigma_8=0.96$) to the model obtained using the
`universal' mass function (fit to $N$-body simulations,
\citealt{jfw01}) for the same parameter values. These two models agree
well for $M\lesssim 10^{15}\,\msu$. The differences become larger than
the size of the Poissonian error bars (Fig.~\ref{PS0}) for $M\gtrsim
2\times 10^{15}\,\msu$, in the sense that the \citeauthor{jfw01} mass function
predicts higher cluster abundances than PS. For larger values of $\om$
the differences become comparable to the size of the error bars at
lower masses, e.g., for $\om =\sigma_8=0.5$ around $M\sim
5\times 10^{14}\,\msu$. To estimate the influence of these differences
on the best fit values derived using the PS mass function, we adjusted
the parameter values of the \citeauthor{jfw01} model to reproduce the PS mass
function, finding $\om =0.15$ and $\sigma_8=0.86$. The value for $\om$
becomes slightly larger but the combination of both values is still
contained within the 90\,\% error ellipse. We therefore
conclude that the differences between the model mass functions do not
significantly affect the interpretation of the \gcs\ mass
function. Moreover we regard this test as confirmation of the validity
of the PS mass function for the accuracy needed here.

Almost all identified systematics give rise to smaller uncertainties than the
statistical uncertainties (eq.~\ref{eq:omres}). However, it
is not impossible that several systematics combined have a significant
effect. In order to determine the highest $\om$ value allowed by a
conspiracy of all identified uncertainties, all positive $\Delta\om$ values
from Tab.~\ref{tab:sys} have been added to the positive statistical
uncertainty for $\om$ and this sum has been added to the $\om$
value which minimizes $\chi^2$ ($\om=0.12$). In this worst case procedure
the upper limit
\begin{equation}
\om<0.31
\label{eq:omup}
\end{equation}
has been found.

Using $\ob=0.19\pm 0.08\,h_{50}^{-3/2}\,\om$, as derived from the mean gas mass
fraction within $r_{200}$ for the 106 clusters in the enlarged sample,
results in model mass functions very similar to the ones calculated using the
baryon fraction given by \citet{bt98}. It is therefore not surprising
that the best fit values for $\om$ and $\sigma_8$ vary only well within the
68\,\% error ellipse if the former $\ob$ determination is used.
Moreover
it is worth noting that combining these two measurements of
$\ob$ gives further evidence for a low value for $\om$ by yielding an
estimate $\om=0.34^{+0.22}_{-0.10}$ using $H_0=71\,\rm
km\,s^{-1}\,Mpc^{-1}$ \citep{mhf00}, where the error has been determined
from the standard deviation of $\ob$ given above. This value for $\om$
is an upper limit since baryons, e.g., contained
in the cluster galaxies or forming part of the dark matter have been
neglected. A low value for
$\om$ has been indicated by
this method for smaller cluster samples by several works previously (e.g.,
\citealt{wf91,bhb92,b93,wne93,wf95,ef99}; but see \citealt{sb01}).
One has
to keep in mind, however,
that this estimate extrapolates the gas fraction from cluster scale to
cosmic scales. For the clusters in our
sample we have found that the gas fraction is not constant but varies with
radius and cluster mass \citep{rb99c}, therefore further
observational
tests of this assumption may be useful. 

\begin{deluxetable}{lllc}
\tablecolumns{4} 
\tabletypesize{\footnotesize}
\tablecaption{Examples of previous results for given $\om$\label{tab:sumcons}}
\tablewidth{0pt}
\tablehead{
\colhead{$\om$}	& \colhead{$\sigma_8$} & \colhead{Relation} & \colhead{Ref.}\\
}
\startdata
0.30	&	1.12	&	$\sigma_8 = 0.57\,\om^{-0.56}$	& 1\\
0.30	&	1.16	&	$\sigma_8 = 0.60\,\om^{-0.59+0.16\om-0.06\om^2}$	& 2\\
0.30	&	0.93	&	$\sigma_8 = 0.52\,\om^{-0.52+0.13\om}$	& 3\\
0.30	&	1.01	&	$\sigma_8 = 0.60\,\om^{-0.46+0.09\om}$	& 4\\
0.30	&	0.78	&		& 5\\
0.30	&	0.99	&	$\sigma_8 = 0.56\,\om^{-0.47}$	& 6\\
0.30	&	0.96	&	$\sigma_8 = 0.58\,\om^{-0.47+0.16\om}$	& 7\\
0.30	&	0.96	&		& 8\\
0.30	&	1.02	&	$\sigma_8 = 0.495\,\om^{-0.60}$	& 9\\
0.30	&	0.91	&	$\sigma_8 = 0.59\,\om^{-0.57+1.45\om-3.48\om^2+3.77\om^3-1.49\om^4}$	& 10\\
0.30	&	0.68\tablenotemark{a}	&	$\sigma_8 = 0.43\,\om^{-0.38}$	& 11\\
\enddata

\tablenotetext{a}{For consistency the $\sigma_8$ value has been calculated using the
relation; the directly determined best fit value for $\om=0.30$ is given by
$\sigma_8=0.72$.}

\tablerefs{
(1) \citealt{wef93}.
(2) \citealt{vl96}.
(3) \citealt{ecf96}.
(4) \citealt{gbg98}.
(5) \citealt{m98}.
(6) \citealt{vl99}.
(7) \citealt{brt99}.
(8) \citealt{bsb00}.
(9) \citealt{psw01}.
(10) \citealt{oa01}.
(11) This work.
}

\end{deluxetable}
Let us briefly discuss the role of $\Gamma$ again. The constraints
(\ref{eq:omres}) together with $h_{100}=0.71$, $\ob\,h_{100}^2=0.0193$,
and eq.~\ref{eq:gama} imply $\Gamma=0.06^{+0.04}_{-0.05}$, which is
significantly lower than values for $\Gamma$ determined through LSS studies of
galaxies (e.g., \citealt{sjm01} have found
$\Gamma=0.188\pm0.04$) or clusters (e.g., \citealt{sbg00a} have found
$\Gamma=0.195\pm0.055$). Adopting the worst case
upper limit on $\om$ found above yields an upper limit $\Gamma<0.18$
consistent with the LSS results. However, also many previous studies of
cluster abundances have yielded low values for $\Gamma$. For instance
\citet{ecf98} have found $\Gamma\approx 0.09\pm0.08$; \citet{h00} has found
$\Gamma\approx 0.05\pm0.22$; \citet{oa01} have found
$\Gamma=0.05\,\om^{-0.71+0.89\,\om}$ which, e.g., for $\om=0.3$ yields
$\Gamma\approx0.09$; judging from their Fig.~3 and their best fit values
for a spatially flat universe \citet{brt01} have found $\Gamma\approx 0.1$,
however, not significantly constrained. It is unclear whether these values
indicate a significant discrepancy but it is interesting to note that
cluster abundances seem to favor lower values for $\Gamma$ than LSS studies
do. If significant this could possibly imply either unidentified
systematic errors in one or both measurements -- from the cluster abundance side one
could for instance imagine a higher merger rate for more massive clusters
resulting in overestimated X-ray luminosities, temperatures, and
masses for the massive clusters resulting in
less steep $\lx$, $\tx$, and $\mt$ functions and therefore in
a decreased value for $\Gamma$
(and possibly $\om$) -- or that the description of
the power spectrum using a single $\Gamma$ parameter on all
scales is not accurate (e.g., \citealt{eh98,sa01}, Sect.~33.3.3).

Even though we made a conservative estimate
by neglecting the possible presence of gas temperature gradients,
previous estimates obtained from cluster abundances generally yielded
higher values for $\om$ and $\sigma_8$ \citep[e.g.,][]{wef93,gbg98}.
For the latter two works one could have expected this already from
Fig.~\ref{mfunc} and possible reasons
have been discussed in Sect.~\ref{func_comp}.
Recently a number of constraints on $\om$ and $\sigma_8$ have been obtained using
different kinds of distribution functions assuming relations
between the measured quantities and mass. Some of them incorporated
additional dynamical information from the evolution of the corresponding
distribution function. Especially intracluster gas
temperature functions have been constructed and used frequently. In
Tab.~\ref{tab:sumcons} some results, which allowed a simple comparison
for a given $\om=0.3$, are listed. If there was a choice, results assuming flat cosmological
models and isothermal clusters are quoted. This Table is not a complete summary of
recent results but 
merely shows that previous results yielded higher values for $\sigma_8$ than
obtained here from a newly constructed sample (even using $M_{87}$ instead
of $\mtz$ would increase the $\sigma_8$ value only up to 0.82).
The reasons may be manifold. Some have
been discussed
above. Other reasons may include different sources of temperature estimates.
For instance \citet{psw01} used cooling flow corrected temperatures given by
\citet{w00}, which are generally higher than temperatures derived previously.
Using theoretical/simulated $\mt$--$\tg$ relations
\citep[e.g.,][]{emn96,bn98} to connect observed temperature 
functions with theoretical mass functions will on average also result in higher values for
$\sigma_8$, since these relations have been shown to have higher normalizations than
observed X-ray mass--temperature relations
\citep[e.g.,][]{hms99,nmf00,frb00}. 
Assuming low temperature groups to be affected by non gravitational heat
input, temperature functions, unlike X-ray mass functions which incorporate
the additional information of the gas density profile, would show an
additional artificial increase towards low temperatures. If not taken into
account, e.g., by use of a modified $\mt$--$\tg$ relation, this effect
would result in artificially increased values for $\om$ (see
Fig.~\ref{PS0}).
Also the completeness and representativeness of the used cluster samples obviously
play an important role for the $\om$--$\sigma_8$ determination.
Here advantage has been taken of the currently best available local sample
in the sense of homogeneous X-ray selection, completeness, size, and availability of
high quality observations.
The results obtained here are, however, in good agreement with the
results from the power spectral analysis of the 452 REFLEX
clusters. \citet{sbg00a} find for a given $\Lambda$CDM model
($\om=0.3$) that $\sigma_8=0.7$ represents the data well, which is
very close to $\sigma_8 =0.68$ expected using relation (\ref{eq:omsig})
found here. Moreover the (1-$\sigma$) range $0.17\leq \om \leq 0.37$ (using
$h=0.71$ in their eq.\ 18) quoted for $\om$ directly is also
consistent with the 90\,\% range determined here. Furthermore
\citet{irb01}, who analyzed the \gcs\ temperature function using
temperatures from homogeneously reanalyzed \as\ data, find 
$\om=0.18_{-0.05}^{+0.08}$ and $\sigma_8=0.96_{-0.09}^{+0.11}$
(90\,\% c.l.\ statistical uncertainty; assuming an open cosmology) by
comparison with Press--Schechter 
models, which is in good agreement with our results.
Moreover, after this paper
had been submitted another paper appeared on astro-ph \citep{brt01}, where
the authors 
also find $\om$--$\sigma_8$ values that agree with values calculated using
relation (\ref{eq:omsig}) presented here. 

\subsubsection{Mass function estimated using $\lx$--$\mt$
relation}\label{lmtest}

\begin{figure}[thbp]
\psfig{file=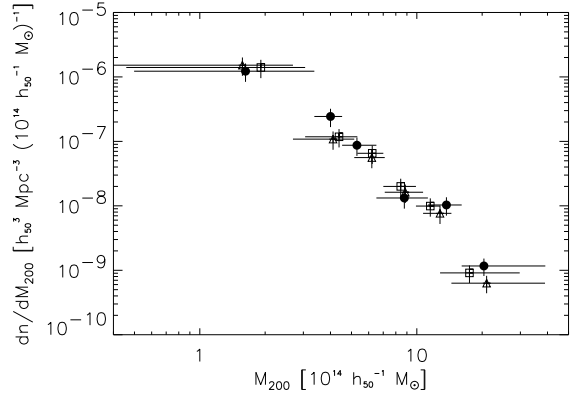,width=8cm,angle=0,clip=}
\caption{\gcs\ mass function (filled circles) compared to `mass'
functions estimated using measured luminosities and luminosity--mass
relations (open symbols). Squares have been calculated using the
($M\mid L$) relation and triangles using the ($L\mid M$) relation  for
\gcs\ clusters.}\label{mf_lm}
\end{figure}

To show consistency and the power of the $\lx$--$\mt$ relation, we have
also performed fits to `mass' functions, where masses have been
estimated from the measured X-ray luminosity.
Relations for the flux-limited sample have been used to get the best
mass estimate for the cluster luminosities included in \gcs . Mass
functions for the two extreme relations 
($M\mid L$) with $\alpha=1.538$ and ($L\mid M$) with $\alpha=1.310$
are shown in Fig.~\ref{mf_lm}. First of all one notes the fairly good
agreement between the three mass functions. In detail the differences
between the two mass functions estimated from different
luminosity--mass relations
can be understood by considering that
at the low luminosity end the steeper relation predicts a higher
mass for a given luminosity than the shallower relation, resulting in a
shift towards higher masses of the mass function. At the high mass side the effect is
opposite, resulting in a shift towards lower masses for the steeper
relation. On average the points for the steeper relation lie higher
which is caused by the fact that a steeper relation results in a
smaller $dM$ on average, which gives rise to an increased $dn/dM$.
The differences to the mass function calculated using the measured
masses are again understood by a similar comparison and are partially
caused by a
possible deviation of the shape of the $\lx$--$\mt$ relation from 
a pure power
law. Despite these small differences performing an actual
fit\footnote{For the fit the corresponding scatter in
$\log M$ for the two relations ($L\mid M$) and ($M\mid L$),
$\slm=0.22$ and $0.21$,
respectively (Tab.~\ref{tab:scat}), has replaced the mass
measurement error, $\bar\sigma_{\mt ,\log}=0.12$, in (\ref{eq:nschl}).}
results in the ($\om$, $\sigma_8$) values (0.14, 0.85) for ($L\mid M$)
and (0.22, 0.74) for ($M\mid L$). The first case is consistent with the error
range given in (\ref{eq:omres}) and in the second case the 90\,\%
statistical error ellipse overlaps with the 90\,\% ellipse in Fig.~\ref{banana}.
From the above and Fig.~\ref{PS0} it is clear that using steeper
luminosity--mass relations results in
higher values for $\om$ and lower values for $\sigma_8$.
Here we want to estimate $M$ from $L$ and therefore ($M\mid L$) is the
appropriate relation to use.
This test shows that with a comparatively easy to
obtain X-ray luminosity function of a statistical cluster sample and with
the knowledge of the empirical $\lx$--$\mt$ relation (even
if approximated as simple power law) and its scatter as presented here
useful constraints on cosmological parameters can be set by
construction of a `quasi mass function'.

\subsection{Total Gravitating Mass in Clusters}\label{func_dens}

\begin{figure}[thbp]
\psfig{file=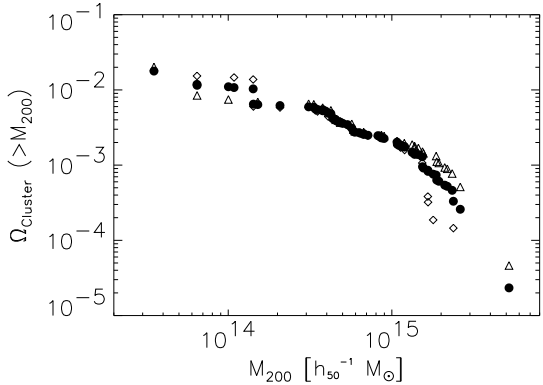,width=8cm,angle=0,clip=}
\caption{Mass density contained in galaxy clusters as a function of
minimum mass. Filled circles indicate the complete \gcs , open
triangles indicate the 34 clusters north of the galactic plane, and
open diamonds the
29 clusters at southern galactic latitudes included in \gcs .}\label{mdens}
\end{figure}

To estimate the fraction of the gravitational mass density relative to the
critical density contained in galaxy clusters, $\oc(>\mti) = 1/\roc
\int_{\mti}^\infty \mt\,\phi(\mt)\,d\mt$,
the individual cluster masses divided by the
corresponding maximum search volumes have been summed up for \gcs , i.e., $\oc
=1/\roc \sum_i \mtzi/\vmaxi$. Note that the determination of $\oc$ is
independent of the Hubble constant. The cumulative diagram for $\oc(>\mtz)$
is shown in Fig.~\ref{mdens}. In order to perform a conservative error estimate,
\gcs\ has been split into two parts with $\bii\ge +20\,\rm deg$ and $\bii\le
-20\,\rm deg$, and the results for these subsamples are also shown in the
Figure. This estimate is
conservative because \gcs\ is about twice as large as each
subsample. Taking the second and third lowest mass clusters together with
the maximum mass range given by their individual uncertainties, we obtain
\begin{equation}
\oc = 0.012^{+0.003}_{-0.004}
\label{eq:ocl}
\end{equation}
for masses larger than $6.4^{+0.7}_{-0.6}\times
10^{13}\,h_{50}^{-1}\,M_{\odot}$, i.e., the total gravitating mass
contained within the virial radius of clusters amounts only to 
$1.2^{+0.3}_{-0.4}$ percent of the total mass in a critical density
universe. Combined with our best estimate $\om=0.12$ this implies that
about 90\,\% of the total mass in the universe resides outside
virialized cluster regions above the given minimum mass. If galaxies
trace mass it also follows that by far most of the galaxies do not sit
in clusters.
This result is consistent with the general presumption
that clusters are rare objects, rare peaks in the density distribution
field.

Comparing the diagram to the 
mass fraction $\oc = 0.028^{+0.009}_{-0.008}$ in clusters with masses
larger than $2\times 10^{14}\,\msu$ given by \citet{fhp98},
based on the mass function determined by
\citet{bc93}, one finds that their estimate is a factor 4--5
higher. However, the \citeauthor{bc93} mass function is given for
$\mab$ and we get a consistent result if we calculate $\oc$
using our formally determined cluster masses within $\rab$. It needs
to be pointed out that at $\mab\sim 2\times 10^{14}\,\msu$ we find
that the typical virial radius is $\sim 1\,\mpc$ and a mass
determination at $3\,\mpc$ based on the assumption of virial
equilibrium may therefore be rather uncertain and possibly leading to
overestimates of $\oc$. This becomes more crucial if mass functions
for $\mab$ are extrapolated even down to galaxy masses. This way
\citet{fhp98} find $\og = 0.12\pm 0.02$ within the mass range $2\times
10^{12}$--$2\times 10^{14}\,\msu$, which, compared to our results
from the previous Section, would account already for almost all mass
in the universe.

\section{Summary}\label{conclu}
An X-ray selected and X-ray flux-limited sample
comprising the 63 X-ray brightest galaxy clusters in the
sky (excluding the galactic band, called \gcs ) has been constructed based on the
\ro\ All-Sky Survey.
The flux limit
has been set at $2\,\esc$ in the energy band $\eb$. It has been shown
that a high completeness is indicated by several tests. Due to the
high flux limit this sample can be used for a variety of
applications requiring a statistical cluster sample without any
corrections to the effective survey volume.

Mainly high quality pointed observations have been used to determine
fluxes and physical cluster parameters. It has been shown that a tight correlation
exists between the X-ray luminosity and the gravitational mass using
\gcs\ and an extended sample of 106 galaxy clusters. The
relation and its scatter have been quantified using different fitting
methods. A comparison to theoretical and numerical predictions shows
an overall agreement. This relation may be directly applied in large X-ray
cluster surveys or dark matter simulations for conversions between
X-ray luminosity and gravitating mass.

Using \gcs\ the gravitational mass function has been determined for
the mass interval $3.5\times 10^{13}< \mtz < 5.2\times
10^{15}\,\msu$. Comparison with Press--Schechter mass functions
assuming CDM power spectra
has yielded tight constraints on cosmological parameters. The
large covered mass range has allowed to put constraints on relevant
parameters individually. Specifically
we have found
$\om=0.12^{+0.06}_{-0.04}$ and $\sigma_8=0.96^{+0.15}_{-0.12}$ (90\,\%
c.l.\ statistical uncertainty). 
Various tests for
systematic uncertainties have been performed, including comparison of
the Press--Schechter mass function with the most recent results from
large $N$-body simulations, almost always yielding uncertainties smaller
than the statistical uncertainties.
Combining all identified systematic uncertainties in a worst case scenario
results in an upper limit $\om<0.31$.
For comparison we have
also determined the best fit $\om$ values for fixed $\sigma_8$ values
yielding the relation $\sigma_8=0.43\,\om^{-0.38}$.

The mass function has been integrated to obtain the fraction of the total
gravitating
mass in the universe contained in galaxy clusters. Normalized to the
critical density we have found
$\oc = 0.012^{+0.003}_{-0.004}$ for cluster masses larger than $6.4^{+0.7}_{-0.6}\times
10^{13}\,h_{50}^{-1}\,M_{\odot}$. With the value for $\om$ determined here
this implies that about 90\,\% of the mass
in the universe resides outside virialized cluster regions.

\acknowledgments

We are thankful for many enlightening discussions with Peter Schuecker, Yasushi Ikebe,
 and J\"org Retzlaff. The authors acknowledge the great benefit from the use
of unpublished catalogs from the REFLEX and NORAS teams.
Yasushi Ikebe has provided three cluster temperatures prior to
publication. Volker Mueller, J\"org Retzlaff and Peter
Schuecker have provided results from $N$-body simulations. 
Marisa Girardi has provided electronic tables of cluster masses 
based on galaxy velocity dispersions. Alexis Finoguenov has provided
electronic tables of partly unpublished X-ray mass measurements.
The BCES regression software has been provided by Michael Akritas, 
Tina Bird, and Matthew Bershady.
The code for calculation of the universal mass
function has been provided by
Adrian Jenkins. The code for calculation of the \citeauthor{eh98} transfer
function has been provided by these authors.
The authors thank the anonymous referee for useful comments on the manuscript. 
The University of Virginia has provided valuable computing resources.
Yutaka Fujita and Alexis Finoguenov have given helpful comments.
We acknowledge the dedicated work of the \ro\ and EXSAS hardware and
software teams.
We have made extensive use of the \ro\ Data Archive of the Max-Planck-Institut
f\"ur extraterrestrische Physik (MPE) at Garching, Germany, and also
of the NASA/IPAC Extragalactic Database (NED) 
which is operated by the Jet Propulsion Laboratory, Caltech, under contract
with the National Aeronautics and Space Administration.
This and related papers,
supplementary information, as well as electronic files of the Tables given in this paper
are available at:
\anchor{http://www.xray.mpe.mpg.de/~reiprich/}
{http://www.xray.mpe.mpg.de/$\sim$reiprich/}
and
\anchor{http://www.astro.virginia.edu/~thr4f/}
{http://www.astro.virginia.edu/$\sim$thr4f/}\ .



\appendix

\begin{deluxetable}{lcccc}
\tabletypesize{\footnotesize}
\tablecaption{Fit parameter values\label{tab:lm1}}
\tablewidth{0pt}
\tablehead{
\colhead{Fit}    &  {$\alpha$} &   \colhead{$\Delta\alpha$}   &
\colhead{$A$} & \colhead{$\Delta A$}
}
\startdata
BCES($L\mid M$) & 1.496 & 0.089 & -17.741 & 1.320 \\ 
bootstrap & 1.462 & 0.089 & -17.238 & 1.327 \\ 
BCES($M\mid L$) & 1.652 & 0.085 & -20.055 & 1.261 \\ 
bootstrap & 1.672 & 0.086 & -20.357 & 1.278 \\ 
BCES-Bisector & 1.571 & 0.083 & -18.857 & 1.237  \\ 
bootstrap & 1.562 & 0.083 & -18.717 & 1.239 \\ 
BCES-Orthogonal & 1.606 & 0.086 & -19.375 & 1.283 \\ 
bootstrap & 1.609 & 0.088 & -19.419 & 1.308 \\ 
 \enddata
\tablecomments{Best fit parameter values and standard deviations for
the extended sample (106
clusters) for a fit of the form given in eq.~\ref{eq:lmform}.
The rows denoted `bootstrap' give the results obtained for 10\,000
bootstrap resamplings.}
\end{deluxetable}
%
%
\begin{deluxetable}{lcccc}
\tabletypesize{\footnotesize}
\tablecaption{Fit parameter values\label{tab:lm2}}
\tablewidth{0pt}
\tablehead{
\colhead{Fit}    &  {$\alpha$} &   \colhead{$\Delta\alpha$}   &
\colhead{$A$} & \colhead{$\Delta A$}
}
\startdata
BCES($L\mid M$) & 1.310 & 0.103 & -14.935 & 1.526 \\
bootstrap & 1.256 & 0.103 & -14.146 & 1.531 \\
BCES($M\mid L$) & 1.538 & 0.105 & -18.320 & 1.568 \\
bootstrap & 1.584 & 0.113 & -18.995 & 1.681 \\
BCES-Bisector & 1.418 & 0.097 & -16.536 & 1.434 \\
bootstrap & 1.407 & 0.096 & -16.368 & 1.427 \\
BCES-Orthogonal & 1.460 & 0.105 & -17.157 & 1.559 \\
bootstrap & 1.468 & 0.110 & -17.274 & 1.633 \\
\enddata
\tablecomments{Same as Tab.~\ref{tab:lm1} but for the purely
flux-limited sample (63 clusters).}
\end{deluxetable}
%
%
\begin{deluxetable}{lcccc}
\tabletypesize{\footnotesize}
\tablecaption{Fit parameter values\label{tab:lm3}}
\tablewidth{0pt}
\tablehead{
\colhead{Fit}    &  {$\alpha$} &   \colhead{$\Delta\alpha$}   &
\colhead{$A$} & \colhead{$\Delta A$}
}
\startdata
BCES($L\mid M$) & 1.756 & 0.091 & -21.304 & 1.350 \\
bootstrap & 1.719 & 0.090 & -20.746 & 1.338 \\
BCES($M\mid L$) & 1.860 & 0.084 & -22.836 & 1.246 \\
bootstrap & 1.881 & 0.084 & -23.144 & 1.250 \\
BCES-Bisector & 1.807 & 0.084 & -22.053 & 1.251 \\
bootstrap & 1.797 & 0.084 & -21.899 & 1.241 \\
BCES-Orthogonal & 1.835 & 0.085 & -22.473 & 1.260 \\
bootstrap & 1.841 & 0.085 & -22.563 & 1.270 \\
\enddata
\tablecomments{Same as Tab.~\ref{tab:lm1} but for $\lbol$.}
\end{deluxetable}
%
%
\begin{deluxetable}{lcccc}
\tabletypesize{\footnotesize}
\tablecaption{Fit parameter values\label{tab:lm4}}
\tablewidth{0pt}
\tablehead{
\colhead{Fit}    &  {$\alpha$} &   \colhead{$\Delta\alpha$}   &
\colhead{$A$} & \colhead{$\Delta A$}
}
\startdata
BCES($L\mid M$) & 1.504 & 0.089 & -17.545 & 1.298 \\
bootstrap & 1.469 & 0.089 & -17.042 & 1.300 \\
BCES($M\mid L$) & 1.652 & 0.086 & -19.708 & 1.254 \\
bootstrap & 1.671 & 0.086 & -19.992 & 1.260 \\
BCES-Bisector & 1.575 & 0.084 & -18.590 & 1.228 \\
bootstrap & 1.565 & 0.083 & -18.445 & 1.224 \\
BCES-Orthogonal & 1.609 & 0.087 & -19.075 & 1.274 \\
bootstrap & 1.611 & 0.088 & -19.109 & 1.290 \\
\enddata
\tablecomments{Same as Tab.~\ref{tab:lm1} but for  $\mtf$.}
\end{deluxetable}
%
%
\begin{deluxetable}{lcccc}
\tabletypesize{\footnotesize}
\tablecaption{Fit parameter values\label{tab:lm5}}
\tablewidth{0pt}
\tablehead{
\colhead{Fit}    &  {$\alpha$} &   \colhead{$\Delta\alpha$}   &
\colhead{$A$} & \colhead{$\Delta A$}
}
\startdata
BCES($L\mid M$) & 2.683 & 0.215 & -35.675 & 3.219 \\ 
bootstrap & 2.606 & 0.218 & -34.525 & 3.265 \\ 
BCES($M\mid L$) & 2.488 & 0.127 & -32.761 & 1.902 \\ 
bootstrap & 2.520 & 0.130 & -33.239 & 1.946 \\ 
BCES-Bisector & 2.583 & 0.154 & -34.170 & 2.297 \\ 
bootstrap & 2.560 & 0.157 & -33.840 & 2.347 \\ 
BCES-Orthogonal & 2.513 & 0.128 & -33.137 & 1.915 \\ 
bootstrap & 2.530 & 0.132 & -33.388 & 1.975 \\ 
\enddata
\tablecomments{Same as Tab.~\ref{tab:lm1} but for  $\mab$ for
comparison (the same relative mass errors as for $\mtz$ have been assumed
here).} 
\end{deluxetable}
%
%
\begin{deluxetable}{lccc}
\tabletypesize{\footnotesize}
\tablecaption{Measured scatter\label{tab:scat}}
\tablewidth{0pt}
\tablehead{
\colhead{Scatter}    & \colhead{($L\mid M$)}  &   \colhead{($M\mid L$)}   &
\colhead{Bisector}
}
\startdata
\cutinhead{106 clusters included in the extended sample.} 
$\slm$ & 0.21 & 0.21 & 0.21 \\
$\sll$ & 0.31 & 0.34 & 0.32 \\
$\sigma_{\log L/M}$ & 0.17 & 0.18 & 0.17 \\
\cutinhead{63 clusters included in \gcss .} 
$\slm$ & 0.22 & 0.21 & 0.21 \\
$\sll$ & 0.29 & 0.32 & 0.30 \\
$\sigma_{\log L/M}$ & 0.18 & 0.18 & 0.17 \\
\enddata
\tablecomments{Scatter measured for different relations.}
\end{deluxetable}

\end{document}